\RequirePackage{lineno}
\documentclass[aps,prd,nofootinbib,showpacs,preprintnumbers,floatfix]{revtex4}
\usepackage{color}
\usepackage{graphicx}
\usepackage{amssymb}
\usepackage[english]{babel}
\def \simgt{\,\rlap{\lower 7.5 pt\hbox{$\mathchar \sim$}}\raise 3 pt \hbox{$>$}\,}
\def \simlt{\,\rlap{\lower 7.5 pt\hbox{$\mathchar \sim$}}\raise 3 pt \hbox{$<$}\,}

\def\lsim{\raise0.3ex\hbox{$<$\kern-0.75em\raise-1.1ex\hbox{$\sim$}}}
\def\gsim{\raise0.3ex\hbox{$>$\kern-0.75em\raise-1.1ex\hbox{$\sim$}}}

\begin{document}
\title{The chiral and deconfinement aspects of the QCD transition}
\author{A. Bazavov$^a$, 
T. Bhattacharya$^{\rm b}$, 
M. Cheng$^{\rm c}$, 
C. DeTar$^{\rm d}$, 
H.-T. Ding$^{\rm a}$, 
Steven Gottlieb$^{\rm e}$, 
R. Gupta$^{\rm b}$,\\
P. Hegde$^{\rm a}$, 
U.M. Heller$^{\rm f}$, 
F. Karsch$^{\rm a,g}$,
E. Laermann$^{\rm g}$, 
L. Levkova$^{\rm d}$, 
S. Mukherjee$^{\rm a}$, 
P. Petreczky$^{\rm a}$, \\
C. Schmidt$^{\rm g,h}$, 
R.A. Soltz$^{\rm c}$, 
W. Soeldner$^{\rm i}$,\\
R. Sugar$^{\rm j}$, 
D. Toussaint$^{\rm k}$, 
W. Unger$^{\rm l}$ and 
P. Vranas$^{\rm c}$
\\[1mm]
{\bf (HotQCD Collaboration)}
}
\affiliation{
$^{\rm a}$ Physics Department, Brookhaven National Laboratory, Upton, NY 11973, USA \\
$^{\rm b}$ Theoretical Division, Los Alamos National Laboratory, Los Alamos, NM 87545, USA\\
$^{\rm c}$ Physics Division, Lawrence Livermore National Laboratory, Livermore CA 94550, USA\\
$^{\rm d}$ Department of Physics and Astronomy, University of Utah, Salt Lake City, UT 84112, USA \\
$^{\rm e}$ Physics Department, Indiana University, Bloomington, IN 47405, USA\\
$^{\rm f}$ American Physical Society, One Research Road, Ridge, NY 11961, USA\\
$^{\rm g}$ Fakult\"at f\"ur Physik, Universit\"at Bielefeld, D-33615 Bielefeld, Germany\\
$^{\rm h}$ Frankfurt Institute for Advanced Studies, J.W.Goethe Universit\"at Frankfurt,\\D-60438 Frankfurt am Main, Germany \\
$^{\rm i}$ Institut f\"ur Theoretische Physik, Universit\"at Regensburg,
D-93040 Regensburg, Germany\\ 
$^{\rm j}$ Physics Department, University of California, Santa Barbara, CA 93106, USA\\
$^{\rm k}$ Physics Department, University of Arizona, Tucson, AZ 85721, USA\\
$^{\rm l}$ Institut f\"ur Theoretische Physik, ETH Z\"urich, CH-8093 Z\"urich,
Switzerland
}

\begin{abstract}
We present results on the chiral and deconfinement properties of the
QCD transition at finite temperature.  Calculations are performed with
$2+1$ flavors of quarks using the p4, asqtad and HISQ/tree
actions. Lattices with temporal extent $N_{\tau}=6$, $8$ and $12$ are
used to understand and control discretization errors and to reliably
extrapolate estimates obtained at finite lattice spacings to the
continuum limit.  The chiral transition temperature is defined in
terms of the phase transition in a theory with two massless flavors
and analyzed using $O(N)$ scaling fits to the chiral condensate and
susceptibility. We find consistent estimates from the HISQ/tree and
asqtad actions and our main result is $T_c=154 \pm 9$ MeV.
\vspace{0.2in}
\begin{center}
\bf{\today}
\end{center}
\end{abstract}

\vspace{0.2in}

\preprint{BNL-96539-2011-JA, LA-UR-11928}

\maketitle

\section{Introduction}

It was noted even before the advent of Quantum
Chromodynamics (QCD) as the underlying theory of strongly
interacting elementary particles, that nuclear matter cannot
exist as hadrons at an arbitrarily high temperature or density. The
existence of a limiting temperature was formulated in the
context of the Hagedorn resonance gas model \cite{Hagedorn:1965st}. This
phenomenon has been interpreted in the framework of QCD as a phase
transition~\cite{Cabibbo:1975ig} separating ordinary hadronic
matter from a new phase of strongly interacting matter --- the quark
gluon plasma \cite{Shuryak:1977ut}.  Today, the structure of the
QCD phase diagram and the transition temperature in the presence
of two light and a heavier strange quark is being investigated using 
high precision simulations of lattice QCD. 

Understanding the properties of strongly interacting matter at high
temperatures has been a central goal of numerical simulations of
lattice QCD ever since the first investigations of the
phase transition and the equation of state in a purely gluonic SU(2)
gauge theory \cite{McLerran:1981pb,Kuti:1980gh,Engels:1980ty}. Early work showed that chiral symmetry
and its spontaneous breaking at low temperatures play an important
role in understanding the phase diagram of strongly interacting
matter. Chiral symmetry breaking introduces a length scale, and the
possibility that it may be independent of deconfinement phenomena was
discussed \cite{Kogut:1982fn,Kogut:1982rt}.  Similarly, the consequences of the
existence of an exact global symmetry in the chiral limit of QCD, the
spontaneous breaking of this $O(4)$ symmetry, the influence of the
explicit breaking of the axial $U_A(1)$ symmetry and the presence of a
heavier strange quark for the QCD phase diagram were analyzed
\cite{Pisarski:1983ms}.  The possibilities that the strange quark mass could
be light enough to play a significant role in the QCD transition,
and/or an effective restoration of axial symmetry may trigger a first
order phase transition in QCD for even nonzero light quark masses were
also discussed. Neither of these situations seems to be realized
in QCD for physical light and strange quark masses.  Based on recent
high precision calculations, the QCD transition at nonzero temperature
and vanishing chemical potentials is observed to be an analytic
crossover \cite{Bernard:2004je,Cheng:2006qk,Aoki:2006we}.

In this paper we focus on understanding the universal properties of
the QCD phase transition in the chiral limit and extract the behavior
of QCD at physical quark masses using an $O(N)$ scaling analysis.  We
also calculate quantities that probe the deconfinement aspects of the
QCD transition: quark number susceptibilities and the renormalized
Polyakov loop.  It is expected that for temperatures below the
transition temperature these quantities should be well described by
the hadron resonance gas (HRG) model which is very successful in
describing thermodynamics at low temperature and the basic features of
the matter produced in relativistic heavy ion
collisions~\cite{BraunMunzinger:2003zd}. It is, therefore, interesting to quantify
the interplay between the universal properties of the chiral
transition and the physics of the HRG model. Some of these issues will
be addressed in a separate publication.

Investigations of QCD at finite temperature are carried out using a
number of different lattice formulations of the Dirac action.  While
studies based on the Wilson \cite{Ejiri:2009hq,Bornyakov:2009qh} or chiral fermion
formulations \cite{Cheng:2009be} are, at present, constrained to a regime of
moderately light quark masses ($m_l/m_s \gsim 0.2$), calculations
exploiting staggered fermion discretization schemes
\cite{Bernard:2004je,Aoki:2005vt,Cheng:2006qk,Bernard:2006nj,Cheng:2007jq,Bazavov:2009zn,Cheng:2009zi,Aoki:2006br,Aoki:2009sc,Borsanyi:2010bp}
can be performed with an almost realistic spectrum of dynamical light
and strange quarks. Today, high statistics calculations, performed at
a number of values of the lattice cutoff and quark masses, allow
for a detailed analysis of discretization errors and quark mass
effects.

Recent studies of QCD thermodynamics with two degenerate light quarks
and the heavier strange quark have been performed with several
staggered fermion actions that differ in the way improvements are
incorporated to reduce the effects of known sources of discretization
errors.  These include the asqtad, p4 and stout actions. The results
at $a\sim 0.1$ fm show differences not only in the determination of
relevant scales, such as the transition temperature, but also in the
temperature dependence of thermodynamic observables. Because estimates of all
observables ought to agree in the continuum limit, discretization errors 
and the dependence on light quark masses require
careful analysis. Furthermore, since QCD for physical quark masses
does not display a genuine phase transition, the definition of the
transition temperature itself requires care.  A proper definition of a
pseudocritical temperature should be related to the chiral phase
transition in the massless limit of QCD and reduce to it in that
limit. 
The aim of this paper is  to study chiral and deconfinement aspects of the QCD transition 
at sufficiently small lattice spacing as to give control over the continuum extrapolation,
and demonstrate the consistency of the results obtained with different actions.
Furthermore, for the first time we provide a determination of the chiral
transition temperature in the continuum limit that makes close connection with
the critical behavior of QCD for massless light quarks. 

This paper extends earlier calculations, performed with the asqtad and
p4 actions on lattices with temporal extent $N_\tau=8$ and light to
strange quark mass ratio $m_l/m_s=0.1$, in several ways.  We have
added calculations for $m_l/m_s=0.05$ on $N_\tau=8$ lattices and
performed new calculations with the asqtad action at smaller lattice
spacing, {\it i.e.}, for $N_\tau=12$ with $m_l/m_s = 0.05$. More
importantly, we have performed thermodynamic calculations with the
highly improved staggered quark action (HISQ)~\cite{Follana:2006rc} on
lattices with temporal extent $N_\tau =6$, $8$ and $12$ to quantify
discretization errors. Preliminary versions of the results given in
this paper have been presented in
Refs.~\cite{Detar:2007as,Gupta:2009tv,Bazavov:2009mi,Bazavov:2010sb,Bazavov:2010bx,Soldner:2010xk,Bazavov:2010pg,Bazavov:2011sd,Bazavov:2011jx}.

This paper is organized as follows. In the next section, we discuss
thermodynamics calculations with improved staggered fermion actions,
in particular, emphasizing the HISQ formulation which has been exploited in this
context for the first time. We analyze the so-called taste
symmetry violations in different improved staggered actions.  Details
of the simulation parameters used in our calculations are also
given in this section.  In Sec.~\ref{sec:observables}, we introduce the basic
observables used in the analysis of the QCD transition and discuss
their sensitivity to the expected critical behavior in the
chiral limit.  We present our numerical results for chiral
observables in Sec.~\ref{sec:chiral}.  In Sec.~\ref{sec:scaling}, we
discuss the universal properties of the chiral transition and the
determination of the pseudocritical temperature.  The deconfining
aspect of the QCD transition, which is reflected in the temperature
dependence of the quark number susceptibilities 
and the renormalized Polyakov loop, is discussed in Sec.~\ref{sec:deconf}.    
Finally, Sec.~\ref{sec:conclusions} contains our conclusions. Details of the 
simulations, data and analysis are given in the appendices.

\section{Thermodynamics with staggered fermion actions}
\label{sec:parameters}

\subsection{Staggered fermion actions}
All staggered discretization schemes suffer from the well known
fermion doubling problem, $i.e.$, a single staggered field describes
four copies of Dirac quarks. These extra degrees of freedom are
called taste and the full taste symmetry (degeneracy of the four
tastes) is realized only in the continuum limit.  At nonzero lattice
spacing $a$ taste symmetry is broken and only a taste non-singlet
axial $U(1)$ symmetry survives. Consequently, in the chiral limit
there is a single Goldstone meson, and the other 15 pseudoscalar
mesons have masses of order $\alpha_s a^2$.  For lattice spacings
accessible in current numerical studies, the effects of taste breaking
can produce significant distortions of the hadron spectrum.  In
thermodynamic calculations, these effects are expected to be most
significant at low temperatures where the equation of state is
governed by the spectrum of hadrons \cite{Karsch:2003vd}.  Current
lattice calculations show that the distortion of the spectrum accounts
for a large part of the deviations of the QCD equation of state from a
hadron resonance gas estimate~\cite{Karsch:2003vd,Huovinen:2009yb,Huovinen:2010tv,Huovinen:2011xc}.

Several improvements to the staggered fermion formulation have been
proposed to reduce the ${\cal O}(a^2)$ taste symmetry breaking
effects. These improvements involve using smeared gauge fields or the
so-called fat links~\cite{Blum:1996uf} by including paths up to length seven
in directions orthogonal to the link being fattened. With these
improvements it is possible to completely cancel taste symmetry
breaking effects at order $\alpha_s a^2$ \cite{Orginos:1999cr}.  Such an
action is called asqtad and has been studied
extensively~\cite{Bazavov:2009bb}.  Fat links are the sum of $SU(3)$
matrices corresponding to different paths on the hyper-cubic lattice,
and are not elements of the $SU(3)$ group.  It has been shown that
projecting the fat links back to $SU(3)$
\cite{Hasenfratz:2001hp,Hasenfratz:2002vv} or even to the $U(3)$ group
\cite{Hasenfratz:2007rf} greatly improves the taste symmetry. Projected fat
links are being used in simulations with the stout action~\cite{Aoki:2006br,Aoki:2009sc,Borsanyi:2010bp} 
and HISQ action
\cite{Follana:2006rc,Bazavov:2009wm,Bazavov:2010ru}. In this paper, we
confirm that this projection results in reductions of taste symmetry
violations and a much better reproduction of the physical hadron
spectrum in calculations starting at moderately coarse lattice
spacing, $i.e.$, $a \sim 0.15$ fm.

In the generation of background gauge configurations, the reduction of
the number of tastes from four to one for each flavor uses the
so-called rooting procedure, {\it i.e.}, the fermion determinant in
the QCD path integral is replaced by its fourth root. Effectively this
amounts to averaging over the non-degenerate spectra of mesons and
baryons, $e.g.$, over the non-degenerate spectrum of sixteen taste
pions. The validity of this procedure is still a subject of debate
\cite{Sharpe:2006re,Creutz:2007rk}.  (For a more detailed summary of
the issues, see Ref.~\cite{Bazavov:2009bb}.)  Reducing taste symmetry
violations is, in any case, important for making the rooted staggered
theory a good approximation to a single flavor physical theory.

\begin{figure}
\includegraphics[width=8cm]{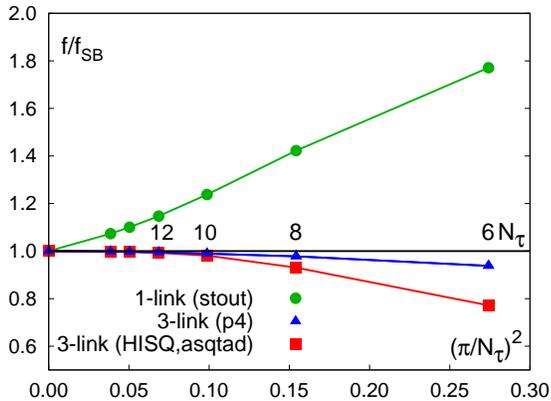}
\caption{The free energy density of an ideal quark gas calculated
for different values of the temporal extent $N_{\tau}$ 
divided by the corresponding value for $N_\tau = \infty$.}
\label{fig:pSB}
\end{figure}

In the context of thermodynamic calculations, it is also important to
control cutoff effects that manifest themselves as distortions of the
high temperature ideal gas and perturbative high temperature limits. 
To reduce these ${\cal O}(a^2)$ effects we use improved staggered
fermion actions that include three-link terms in the discretization of
partial derivatives in the Dirac action. These three-link terms remove
the tree-level ${\cal O}(a^2)$ discretization effects, which are the
dominant ones at high temperatures \cite{Heller:1999xz,Hegde:2008nx}, as can be
seen by considering the free energy density in the ideal gas limit
calculated on four dimensional lattices with varying temporal extent
$N_{\tau}$.  This free energy density of a quark gas divided by the
corresponding result in the continuum limit ($N_\tau\rightarrow
\infty$) is shown in Fig.~\ref{fig:pSB}. For the unimproved staggered
fermion action with 1-link discretization, there is a significant
cutoff dependence for $N_{\tau} < 16$. Including three-link terms in
the action (p4 and Naik) reduces cutoff effects to a few percent even
for $N_{\tau}=8$. The Naik action with straight three-link terms is
the building block for both the asqtad and HISQ actions; however,
projection of fat links to $U(3)$ to further reduce taste violations
is done only in the HISQ action.  The stout action, on the other hand,
uses just the standard 1-link discretization scheme with stout smeared
links that include projection to $SU(3)$.

The HISQ action improves both taste symmetry breaking
\cite{Follana:2006rc} and cutoff effects in the hadron spectrum which,
as mentioned above, are of particular relevance to thermodynamic
calculations at low temperatures. The construction of the projected
fat link action proceeds in three steps. In the first step, a fat7
link is constructed; {\it i.e.}, a fat link which includes all the
paths in orthogonal directions up to length seven. This step is common
to the asqtad action. In step two, the sum of the product of $SU(3)$
matrices along these paths is projected to $U(3)$. In the third step,
these projected fat links are used in the conventional asqtad Dirac
operator without tadpole improvement.  Thus, from the point of view of
reducing taste symmetry breaking at order ${\cal O}(\alpha_s a^2)$ the
asqtad and the HISQ actions are equivalent, but differ at higher
orders.  Unfortunately, these higher order terms are large in the
asqtad action as discussed in Sec.~\ref{subsec:tastebreaking} where we
show that the projection of fat links to $U(3)$ in the HISQ
formulation significantly reduces the distortion of the spectrum at
low temperatures. The straight three-link Naik term in the asqtad and
HISQ actions eliminates the tree-level ${\cal O}(a^2)$ discretization
effects, consequently, their behavior at high temperatures is
equivalent.

For the HISQ calculations presented here, we use a tree-level improved
Symanzik gauge action that is also common to the p4 and stout
formulations.  We refer to this combination of the gauge and 2+1 HISQ
quark actions as the HISQ/tree action to distinguish it from the HISQ
action used by the MILC collaboration in their large scale
zero-temperature 2+1+1 flavor simulations with a dynamical charm quark
\cite{Bazavov:2009wm,Bazavov:2010ru}.  In the 2+1+1 HISQ action, in
addition to the 1-loop tadpole improved version of the Symanzik gauge
action, the 1-loop and mass dependent corrections are included in the
Naik term for the charm quark.

\subsection{Lattice Parameters and Simulation Details}
\label{subsec:parameters}

A summary of the run parameters, statistics, and data for the p4, asqtad, and
HISQ/tree actions analyzed in this paper is given in
Appendix~\ref{sec:appendix1}.  We have previously presented the
equation of state and other thermodynamic quantities using the p4 and
asqtad actions on lattices with temporal extent $N_{\tau}=4$, 6, and 8
in Refs.~ \cite{Bernard:2006nj,Cheng:2007jq,Bazavov:2009zn,Cheng:2009zi}.  Here, we extend these 
studies in the following three ways:
\begin{itemize}
\item
Additional asqtad calculations for $N_{\tau}=8$ with 
$m_l/m_s =0.2$ and $0.05$. (See Table~\ref{tab:asqtad8_runs} in Appendix~\ref{sec:appendix1}.)
\item
New simulations with the asqtad action in the transition region on 
lattices with temporal extent $N_\tau=12$ and light to strange quark mass 
ratio $m_l/m_s=0.05$. (See Table~\ref{tab:asqtad12_runs} in Appendix~\ref{sec:appendix1}.)
\item
New results with the HISQ/tree action on $N_{\tau}=6$ lattices with
$m_l/m_s=0.2$, $0.05$ and $0.025$; on $N_{\tau}=8$ lattices with $m_l/m_s=0.05$
and $0.025$; and on $N_\tau=12$ lattices with $m_l/m_s=0.05$. (See 
Tables~\ref{tab:hisq_0.2ms_runs}, \ref{tab:hisq_0.05ms_runs} 
and \ref{tab:hisq_0.025ms_runs} in Appendix~\ref{sec:appendix1}).
\end{itemize}

As described in previous studies, the first step is to determine the
line of constant physics (LCP) by fixing the strange quark mass to its
physical value $m_s$ at each value of the gauge coupling $\beta$
~\cite{Bazavov:2009zn,Cheng:2009zi}\footnote{In the case of the p4 action and for
  gauge couplings $\beta$ close to the transition region, the bare
  quark masses were not varied with $\beta$ to facilitate the
  Ferrenberg-Swendsen re-weighting procedure.}.  In practice, we
tune $m_s$ until the mass of the fictitious
$\eta_{s\bar s}$ meson matches the lowest order chiral perturbation
theory estimate $m_{\eta_{s \bar s}}=\sqrt{2 m_K^2-m_{\pi}^2}$
\cite{Cheng:2007jq,Cheng:2009zi}.  Having fixed $m_s$, between one and three
values of the light quark mass, $m_l/m_s = 0.2$, $0.1$, $0.05$ and $0.025$ are
investigated at each $N_\tau$ and used to obtain estimates at the
physical point $m_l/m_s=0.037$ by a scaling analysis discussed in 
Sections~\ref{sec:observables} and \ref{sec:chiral}. 

All simulations use the rational hybrid Monte-Carlo (RHMC) algorithm
\cite{Clark:2004cp,Clark:2005sq}. The length of the RHMC trajectory is $0.5$ in molecular
dynamics (MD) time units (TU) for the p4 action and $1.0$ for the
asqtad action. The statistics, therefore, are given in terms of TUs in
Tables~\ref{tab:p4_runs}--\ref{tab:hisq_0.025ms_runs} in
Appendix~\ref{sec:appendix1}. For each value of the input parameters,
we accumulated several thousand TUs for zero-temperature ensembles and
over ten thousand TUs for finite temperature runs.  The RHMC algorithm
for the HISQ action is discussed in
Refs.~\cite{Bazavov:2010ru,Bazavov:2009jc}.  In the calculations with
the HISQ/tree action, the length of the RHMC trajectory is typically
one TU. For smaller values of $\beta$ (coarse lattices) we used
trajectories with length of $1/2$ and $1/3$ TU since frequent spikes
in the fermion force term \cite{Bazavov:2009jc} reduce the acceptance
rate for longer evolution times.

To control finite size effects, the ratio of spatial to temporal
lattice size is fixed at $N_\sigma/N_\tau =4$ in most of our finite
temperature simulations.  The exception is $N_\tau=6$ runs at
$m_l/m_s=0.2$, which were done on lattices of size $16^3\times
6$. Zero-temperature calculations have been performed for different
lattice volumes (see Table~\ref{tab:hisq_0.05ms_runs} in 
Appendix~\ref{sec:appendix1}) such that the spatial extent of the
lattice $L$ satisfies $L M_{\pi}>3$, except for the smallest lattice
spacings where $L M_{\pi} \simeq 2.6$.

The construction of renormalized finite temperature observables
requires performing additive and multiplicative renormalizations. We
implement these by subtracting corresponding estimates obtained at
zero-temperature. These matching zero-temperature calculations have
been performed at several values of the parameters and then fit by
smooth interpolating functions over the full range of temperatures
investigated.

In the following two subsections, we discuss the determination of the
lattice spacing and the LCP for the HISQ/tree action, and the effect of
taste symmetry breaking on the hadron spectrum. 

\subsection{The static potential and the determination of the lattice spacing}
\label{subsec:scalesetting}

The lattice spacing is determined using the 
parameters $r_0$ and $r_1$, which are fixed by  
the slope of the static quark anti-quark potential 
evaluated on zero-temperature lattices as~\cite{Sommer:1993ce}
\begin{equation}
\left( r^2 \frac{{\rm d}V_{\bar{q}q}(r)}{{\rm d}r} \right)_{r=r_0} =
1.65 \;\; , \;\;
\left( r^2 \frac{{\rm d}V_{\bar{q}q}(r)}{{\rm d}r} \right)_{r=r_1} =
1.0  \;\; ,
\label{r0r1}
\end{equation}
and set the scale for all thermodynamic observables discussed in this
work.  The calculation of the static potential, $r_0$ and $r_1$ for
the p4 action was discussed in Refs.~\cite{Cheng:2007jq,Cheng:2009zi}. In
particular, it was noticed that for the values of $\beta$ relevant for
the finite temperature crossover on $N_{\tau}=8$ lattices, the
parameter $r_0$ is the same, within statistical errors, for
$m_l=0.1m_s$ and $m_l=0.05m_s$. Therefore, we use the interpolation
formula for $r_0$ given in Ref.~\cite{Cheng:2007jq} to set the temperature
scale for the p4 data. The calculation of the static potential and
$r_1$ for the asqtad action was discussed in Ref.~\cite{Bazavov:2009bb}.
In Appendix~\ref{sec:appendix2}, we give further details on the
determination of $r_1$. Here we note that the statistical errors in
the $r_1/a$ determination are about $0.2\%$ for gauge couplings
relevant for the $N_{\tau}=8$ calculations and about $0.1\%$ for the
$N_{\tau}=12$ calculations. We also reevaluate systematic errors in
the determination of $r_1/a$ and find that these errors are smaller
than $1\%$ on $N_{\tau}=12$ and about $1\%$ on $N_{\tau}=8$
lattices. These uncertainties will impact the precision with which the
chiral transition temperature is estimated.

\begin{figure}
\includegraphics[width=0.4\textwidth]{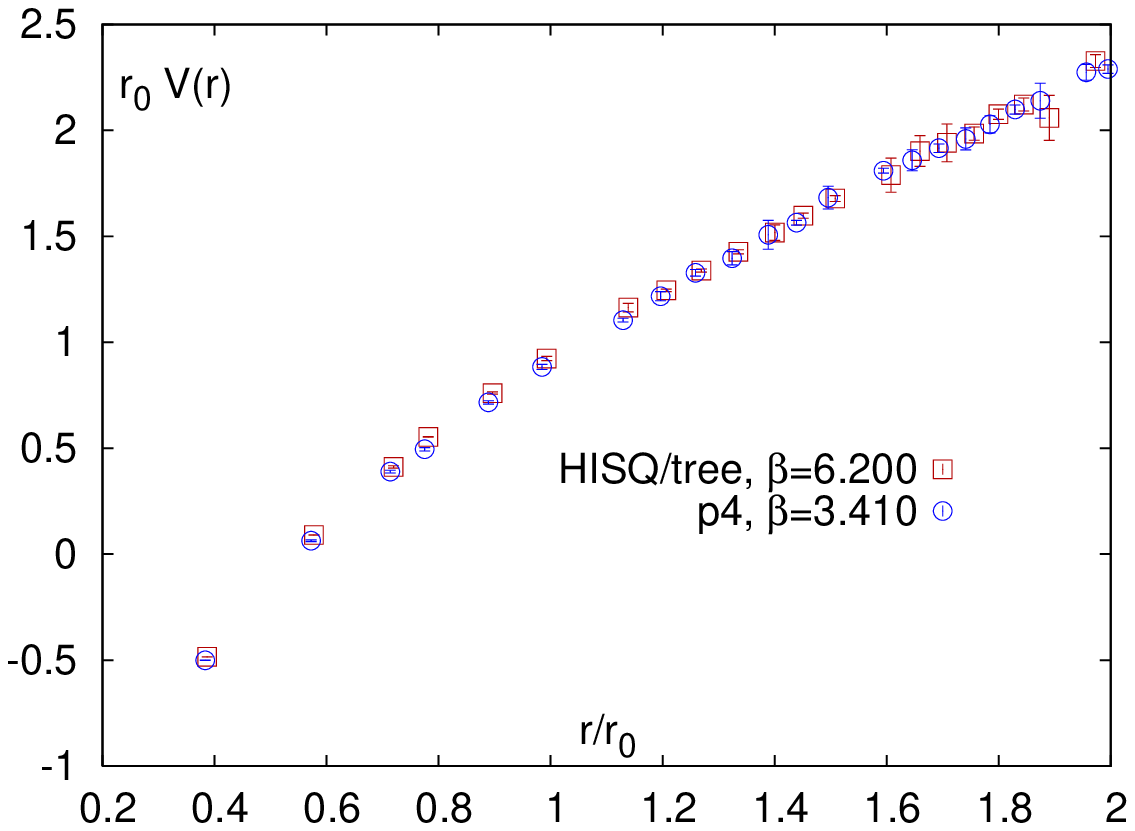}
\includegraphics[width=0.4\textwidth]{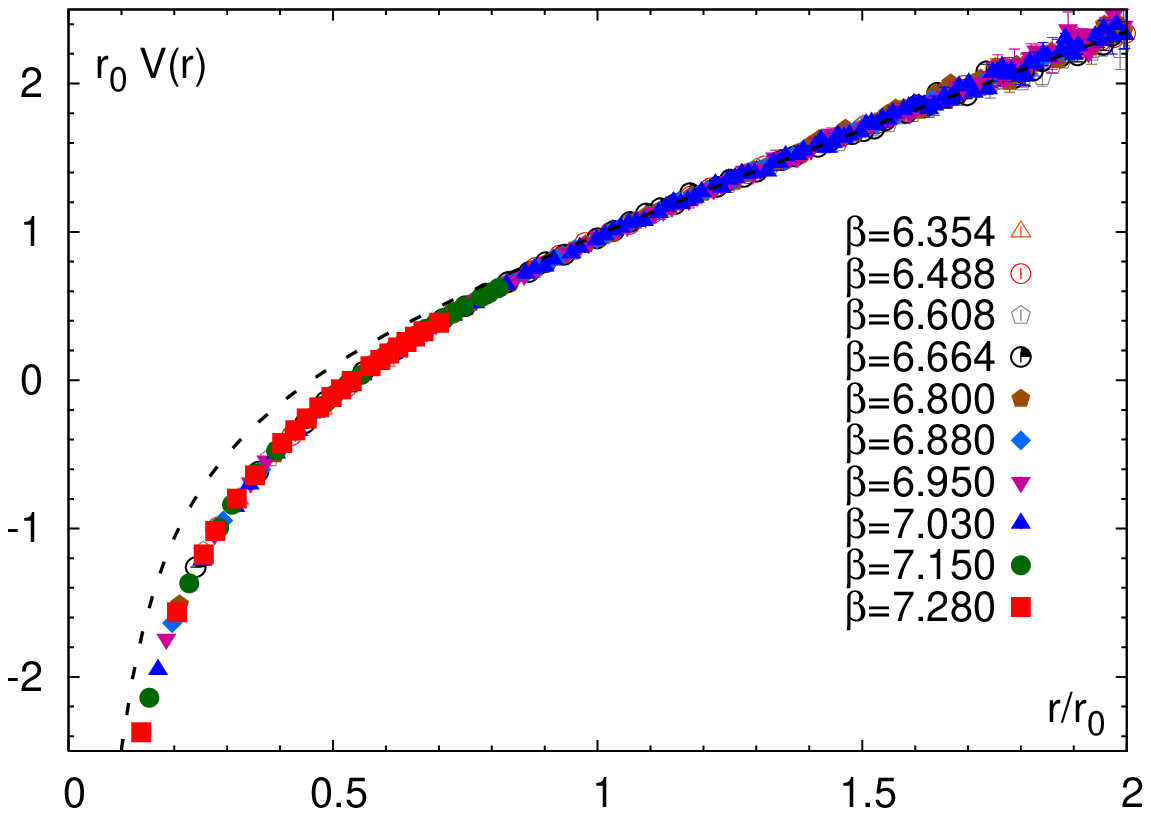}
\caption{The static potential calculated for the HISQ/tree action with $m_l=0.2m_s$ (left) and
  $m_l=0.05m_s$ (right) in units of $r_0$. In the plot on the left, we
  compare the HISQ/tree and p4 results obtained at a similar value of
  the lattice spacing. The dashed line in the plot on the right is the
  string potential $V_{string}(r)=-\pi/(12 r)+\sigma r$ matched to the
  data at $r/r_0=1.5$.}
\label{fig:pot}
\end{figure}

The static quark potential for the HISQ/tree action is calculated using the
correlation functions of temporal Wilson lines of different length
evaluated in the Coulomb gauge. The ratio of these correlators,
calculated for two different lengths, was fit to a constant plus
exponential function from which the static potential is extracted.  To
remove the additive UV divergence, we add a $\beta$-dependent constant
$c(\beta)$ defined by the requirement that the potential has the value
$0.954/r_0$ at $r=r_0$. This renormalization procedure is equivalent to
the normalization of the static potential to the string potential
$V_{string}(r)=-\pi/(12 r)+\sigma r$ at $r=1.5r_0$ \cite{Bazavov:2009zn}.  The
renormalized static potential, calculated for the HISQ/tree action for
$m_l/m_s= 0.05$, is shown in Fig.~\ref{fig:pot}(right) and we find no
significant dependence on $\beta$ (cutoff). We conclude that
discretization errors, including the effects of taste symmetry
violations, are much smaller in the static potential compared to other
hadronic observables.  Furthermore, for approximately the same value
of $r_0/a$, the static potentials calculated with the HISQ/tree and p4
actions agree within the statistical errors as shown in
Fig.~\ref{fig:pot}(left).

To determine the parameters $r_0$ and $r_1$, we fit the potential to a
functional form that includes Coulomb, linear, and constant terms~\cite{Aubin:2004wf,Bazavov:2009bb}:
\begin{equation}
V(r) = C + \frac{B}{r} + \sigma r + \lambda\left(\left.\frac{1}{r}\right|_{lat} - \frac{1}{r}\right) \, .
\label{eq:fitpot}
\end{equation}
In this Ansatz, the Coulomb part is corrected for tree-level lattice
artifacts by introducing a fourth parameter $\lambda$.  The term
proportional to $\lambda$ reduces systematic errors in the
determination of $r_0$ and $r_1$ due to the lack of rotational
symmetry on the lattice at distances comparable to the lattice
spacing.  The resulting fit has a $\chi^2/{\rm dof}$ close to unity for
$r/a>\sqrt{3}$, except for the coarsest lattices corresponding to the
transition region on $N_{\tau}=6$ lattices.  Consequently, we can
determine $r_0/a$ reliably for $\beta$ corresponding to the transition
region for $N_{\tau}=8$ and $N_{\tau}=12$ lattices.  On lattices
corresponding to the transition region for $N_{\tau}=6$ we use a three
parameter fit (Coulomb, linear and constant) with lattice distance
replaced by tree-level improved distance, $r \rightarrow r_I$.
Following Ref.~\cite{Necco:2001xg}, $r_I$ was determined from the Coulomb
potential on the lattice.  Neither the three-parameter nor the
four-parameter fit gives acceptable $\chi^2$; 
however, the difference in the $r_0$ values obtained from these
fits is of the order of the statistical errors. We use this difference
as an estimate of the systematic errors on coarse lattices.

One can, in principle, extract $r_0/a$ and $r_1/a$ by using any
functional form that fits the data in a limited range about these
points to calculate the derivatives defined in Eq.~(\ref{r0r1}). We
use the form given in Eq.~(\ref{eq:fitpot}), but perform separate fits
for extracting $r_0/a$ and $r_1/a$ for each ensemble. The fit range
about $r_0/a$ (or $r_1/a$) is varied keeping the maximum number of
points that yield $\chi^2/{\rm dof}\approx 1$.  The variation in the
estimates with the fit range is included in the estimate of the
systematic error.

The value of $r_1/a$ is more sensitive to the lattice artifacts in the
potential at short distances than $r_0/a$. Only for lattice spacings
corresponding to the transition region on $N_{\tau}=12$ lattices are
the lattice artifacts negligible. Again, we used the difference
between four and three parameter fits to estimate the systematic error
in $r_1/a$.

The lattice artifacts due to the lack of rotational symmetry play a
more pronounced role in the determination of $r_1/a$, and are
significantly larger for the HISQ/tree than for the asqtad
action. This is presumably due to the lack of tadpole improvement in
the gauge part of the HISQ/tree action. For this reason, we use $r_0$
on coarser lattices and $r_1$ on fine lattices and connect the two
using the continuum estimate of $r_0/r_1$ for estimating the
scale. Further details of this matching are given in
Appendix~\ref{sec:appendix2}.  As noted above, these effects are no
longer manifest for the crossover region for $N_{\tau}=12$ lattices.

\begin{figure}
\includegraphics[width=0.4\textwidth]{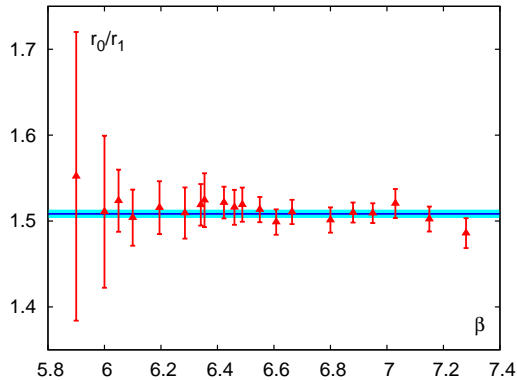}
\caption{The ratio $r_0/r_1$ for the HISQ/tree action. Fitting all the
  data at $\beta \geq 6.423$ by a constant gives $r_0/r_1 = 1.508(5) $
  as our best estimate of the continuum extrapolated value.}
\label{fig:r0r1_AB}
\end{figure}

Having calculated $r_0/a$, $r_1/a$ and $r_0/r_1$ at a number of values
of $\beta$, we estimate the continuum limit value for the ratio
$r_0/r_1$. In Fig.~\ref{fig:r0r1_AB} we plot the data for the
HISQ/tree action. It shows no significant variation with $\beta$, so
we make two constant fits to study the dependence on the range of
points included.  The first fit includes all points with $\beta \ge
6.423$ and the second with $\beta \ge 6.608$.  We take $r_0/r_1 =
1.508(5)$ from the first fit as our best estimate since it has a
better $\chi^2/dof = 0.32$, includes more points and matches the
estimate from a fit to all points.
This estimate is higher than the published MILC collaboration estimate
using the asqtad action: $r_0/r_1=1.474 (7)(18)$~\cite{Aubin:2004wf}.  A
more recent unpublished analysis, including data at smaller lattice
spacings, gives $r_0/r_1=1.50(1)$ for the asqtad action~\cite{Toussaint:private}, consistent
with the HISQ/tree estimate.  We will, therefore, quote $r_0/r_1 =
1.508(5)$ as our final estimate.

Finally, to extract the values of $r_1$ or $r_0$ in physical units, one
has to calculate these quantities in units of some observable with a 
precisely determined experimental value. We use the result $r_1 =
0.3106(8)(18)(4)$ fm obtained by the MILC collaboration using
$f_{\pi}$ to set the lattice spacing \cite{Bazavov:2010hj}. This estimate
is in good agreement with, and more precise than, the recent values
obtained by the HPQCD collaboration: $r_1=0.3091(44)$ fm using
bottomonium splitting, $r_1=0.3157(53)$ fm using the mass splitting of
$D_s$ and $\eta_c$ mesons, and $r_1=0.3148(28)(5)$ fm using
$f_{s\bar{s}}$, the decay constant of the fictitious pseudoscalar $s
\bar s$ meson~\cite{Davies:2009tsa}. To set the scale using $r_0$, we use the
above result for $r_0/r_1$ to convert from $r_1$ to $r_0$.  This gives
$r_0=0.468(4)$ fm, which is consistent with the estimates $r_0=0.462(11)(4)$ fm by the MILC
collaboration~\cite{Aubin:2004wf} and $r_0=0.469(7)$ fm by the HPQCD collaboration~\cite{Gray:2005ur}. 

There are two reasons why we prefer to use either $r_0$ or $r_1$ to
set the lattice scale. First, these are purely gluonic observables and
therefore not affected by taste symmetry breaking inherent in hadronic
probes. Second, as discussed above, $r_0$ (and $r_1$) do not show a
significant dependence on $m_l/m_s$ and thus one can extract reliable
estimates for the physical LCP from simulations at $m_l/m_s =
0.05$. Nevertheless, we will also analyze the data using $f_K$ to set
the scale and discuss its extraction in
Sec.~\ref{subsec:tastebreaking}.

\subsection{Hadron masses and taste symmetry violation}
\label{subsec:tastebreaking}

Precision calculations of the hadron spectrum have been carried out with the
asqtad action
in~Refs.~\cite{Aubin:2004wf,Bernard:2001av,Bazavov:2009bb}. Details of the
calculations of hadron correlators and hadron masses used in this
paper are given in Appendix~\ref{sec:appendix3}. For completeness,
we also list there the masses of baryons estimated at the same lattice
parameters.

As described in Sec.~\ref{subsec:parameters}, the strange quark mass
is fixed by setting the mass of the lightest $s \bar s$ pseudoscalar to
$\sqrt{2 M_K^2-M_{\pi}^2} = 686$ MeV~\cite{Davies:2009tsa}. Masses of all
other pseudoscalar mesons should then be constant along the lines of constant
physics defined by $m_l/m_s=0.2$ and $0.05$. Fits to the data give
\begin{eqnarray}
&
r_0 M_{\pi}=0.3813(12),~r_0 M_K=1.1956(33),~r_0 M_{\eta_{s \bar s}}=1.6488(46),~~ m_l=0.05m_s,\\
&
r_0 M_{\pi}=0.7373(14),~r_0 M_K=1.2581(23),~r_0 M_{\eta_{s \bar s}}=1.6206(30),~~ m_l=0.20m_s.
\end{eqnarray}
Using the value of $r_0$ determined in Sec.~\ref{subsec:scalesetting},
we find that the variation in $M_{\eta_{s \bar s}}$ over the range of $\beta$
values simulated on the LCP is up to $2\%$ for the HISQ/tree
action. We neglect the systematic effect introduced by this variation
in the rest of the paper as it is of the same order as the
statistical errors.  The LCP for the asqtad action, however,
corresponds to a strange quark mass that is about $20\%$ heavier than
the physical value. We will comment on how we account for this
deviation from the physical value in Sec.~\ref{sec:scaling}.

Lattice estimates of hadron masses should agree with the corresponding
experimental values in the continuum limit~\footnote{For the nucleon
  and $\Omega$-baryon this has been demonstrated in
  Ref.~\cite{Bazavov:2009bb}.}; however, at the finite lattice spacings
used in thermodynamic calculations there are significant
discretization errors.  In staggered formulations, all physical states
have taste partners with heavier masses that become degenerate only in
the continuum limit.  The breaking of the taste symmetry, therefore,
introduces additional discretization errors, in particular, in
thermodynamic observables at low temperatures where the degrees of
freedom are hadrons. These artifacts have been 
observed in the deviations between lattice results and the hadron
resonance gas model in the trace anomaly \cite{Bazavov:2009zn} and in
fluctuations of conserved charges \cite{Huovinen:2009yb}.  In this subsection, we
will quantify these taste symmetry violations in the asqtad, stout and
HISQ/tree actions and show that they are the smallest in the HISQ
action~\cite{Bazavov:2010ru}.

To discuss the effects of taste symmetry violations, we analyze all
sixteen pseudoscalar mesons that result from this four-fold doubling,
and are classified into eight multiplets with degenerate masses.
These are labeled by their taste index $\Gamma^F=\gamma_5,~\gamma_0
\gamma_5,~\gamma_i \gamma_5,~\gamma_0,~\gamma_i,~\gamma_i \gamma_0,~
\gamma_i \gamma_j,~1$ \cite{Lee:1999zxa}.  There is only one Goldstone
boson, $\Gamma^F=\gamma_5$, that is massless in the chiral limit and
the masses of the other fifteen pseudoscalar mesons vanish only in the
chiral and continuum limits.  The difference in the squared mass of
the non-Goldstone and Goldstone states, $M_\pi^2 -M_G^2$, is the
largest amongst mesons and their correlators have the best
statistical signal; therefore, it is a good measure of taste symmetry
violations. For different staggered actions, these violations, while
formally of order $\alpha_s^n a^2$, are large as discussed
below~\footnote{For the unimproved staggered fermion action as well as
  for the stout action the quadratic pseudoscalar meson splittings are
  formally of order $\alpha_s a^2$, while for the asqtad and the HISQ
  actions they are of order $\alpha_s^2 a^2$.  Projecting fat links to
  $U(3)$ reduces the coefficients of the $\alpha_s^n a^2$ taste
  violating terms.}.

\begin{figure}
\includegraphics[width=8cm]{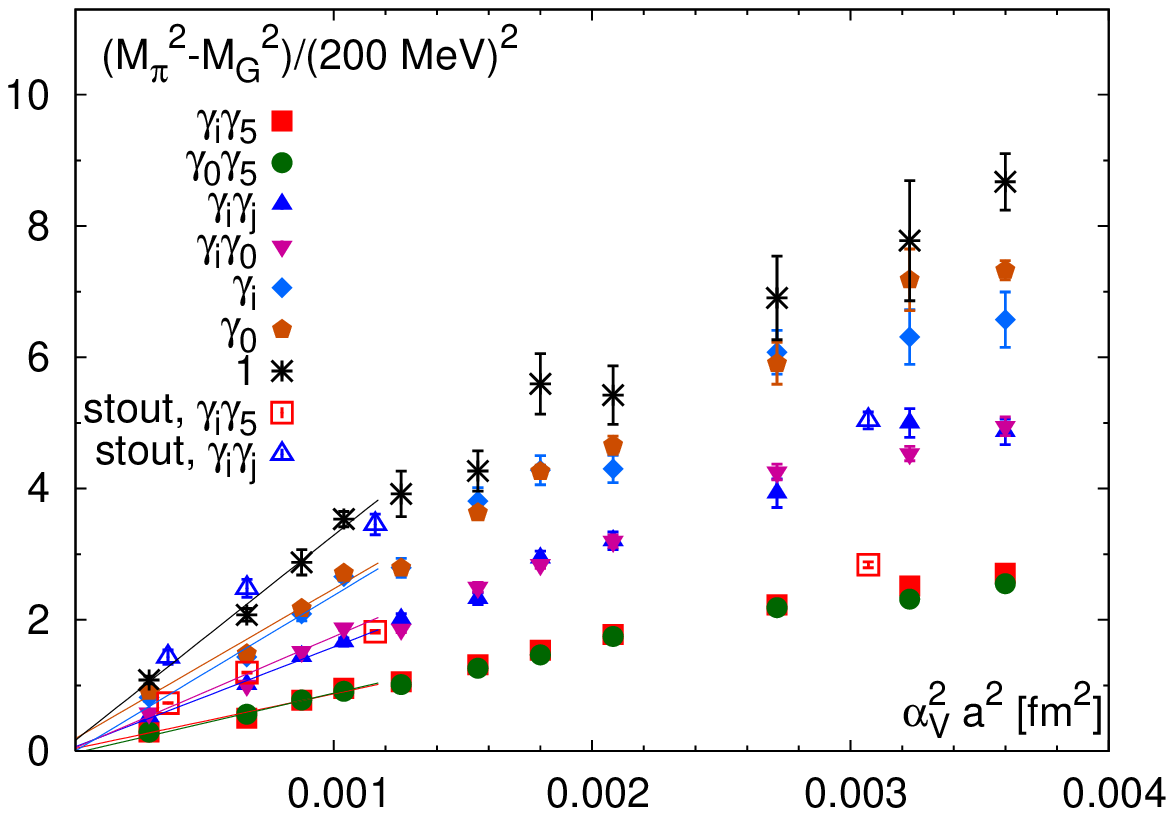}
\includegraphics[width=8cm]{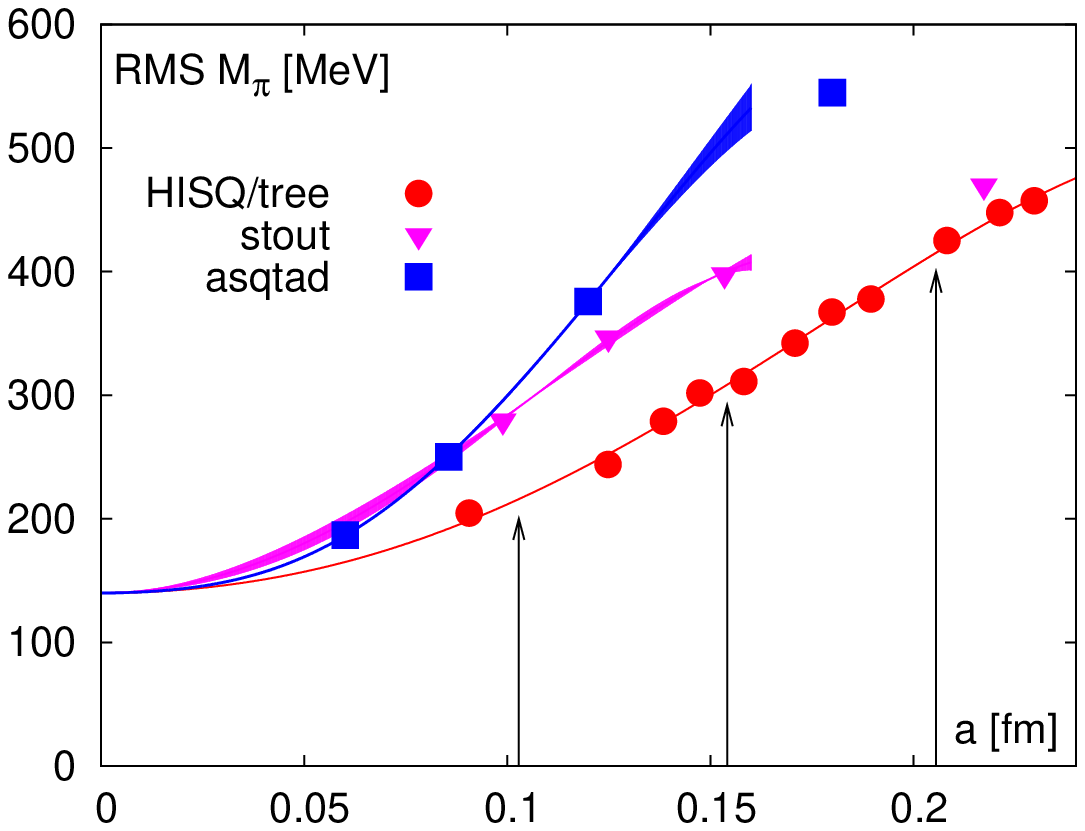}
\caption{The splitting $M_\pi^2 -M_G^2$ of pseudoscalar meson
  multiplets calculated with the HISQ/tree and stout actions as a
  function of $\alpha_V^2 a^2$ (left). The right panel shows the RMS
  pion mass with $M_G=140$ MeV as a function of the lattice spacing
  for the asqtad, stout and HISQ/tree actions. The band for the asqtad
  and stout actions shows the variation due to removing the fourth point at
  the largest $a$ in the fit. These fits become unreliable for $a \gsim 0.16$
  fm and are, therefore, truncated at $a=0.16$ fm.  The vertical arrows
  indicate the lattice spacing corresponding to $T \approx 160$ MeV
  for $N_{\tau}=6$, $8$ and $12$.}
\label{fig:ps_split}
\end{figure}

The taste splittings, $M_\pi^2 -M_G^2$, have been studied in detail
for the asqtad and p4 actions
\cite{Orginos:1999cr,Aubin:2004wf,Cheng:2006wj}. The conclusion is
that at a given $\beta$ they are, to a good approximation, independent
of the quark mass. Therefore, for the HISQ/tree action we calculate
them on $16^3 \times 32$ lattices with $m_l/m_s=0.2$ (see
Table~\ref{tab:hisq_0.2ms_runs}), and on four $32^4$ ensembles with
$m_l=0.05m_s$ for sea quarks and $m_l/m_s=0.2$ for valence
quarks. (The lattice parameters for these ensembles at $\beta=6.664$,
$6.8$, $6.95$, and $7.15$ are given in
Table~\ref{tab:hisq_0.05ms_runs}.)  The corresponding results, plotted
in Fig.~\ref{fig:ps_split}, show the expected $\alpha_s^2 a^2$
scaling, similar to that observed previously with the HISQ action in
the quenched approximation~\cite{Follana:2006rc} and in full QCD
calculations with four flavors \cite{Bazavov:2009wm,Bazavov:2010ru}.
In this analysis, following Ref.~\cite{Aubin:2004wf}, we use
$\alpha_V(q=3.33/a)$ from the potential as an estimate of
$\alpha_s$. Linear fits in $\alpha_s^2 a^2$ to the four points at the
smallest lattice spacings shown in Fig.~\ref{fig:ps_split}(left)
extrapolate to zero within errors in the continuum limit.  The data
also show the expected approximate degeneracies between the multiplets
that are related by the interchange $\gamma_i$ to $\gamma_0$ in the
definition of $\Gamma^F$ as predicted by staggered chiral perturbation
theory~\cite{Lee:1999zxa}.

The splittings for the stout action, taken from Ref.~\cite{Aoki:2009sc}, for
$\Gamma^F=\gamma_i \gamma_5$ and $\gamma_i \gamma_j$ are also shown in
Fig.~\ref{fig:ps_split} with open symbols. We find that they are
larger than those with the HISQ/tree action for comparable lattice
spacings.

To further quantify the magnitude of taste-symmetry violations, we
define, in MeV, the root mean square (RMS) pion mass as
\begin{equation}
  M_\pi^{RMS}= \sqrt{
  \frac{1}{16}\left(M_{\gamma_5}^2+M_{\gamma_0\gamma_5}^2
  +3M_{\gamma_i\gamma_5}^2+3M_{\gamma_i\gamma_j}^2
  +3M_{\gamma_i\gamma_0}^2+3M_{\gamma_i}^2
  +M_{\gamma_0}^2+M_{1}^2\right)} \, ,
\end{equation}
and plot the data in Fig.~\ref{fig:ps_split}(right) with $M_G$ tuned to
$140$ MeV.  The data for the asqtad and stout actions were taken from
Ref.~\cite{Aubin:2004wf} and Ref.~\cite{Borsanyi:2010bp}, respectively. As
expected, the RMS pion mass is the largest for the asqtad action and
smallest for the HISQ/tree action. However, for lattice spacing $a \sim
0.104$ fm, which corresponds to the transition region for $N_{\tau}=12$,
the RMS pion mass becomes comparable for the asqtad and stout actions.
The deviations from the physical mass, $M_\pi=140$ MeV, become 
significant above $a=0.08$ fm even for the HISQ/tree action. For
the lattice spacings $\sim 0.156$ fm ($a \sim 0.206$ fm), corresponding
to the transition region on $N_{\tau}=8$ ($N_{\tau}=6$) lattices,
the RMS mass is a factor of two (three) larger.  

\begin{figure}
\includegraphics[width=8cm]{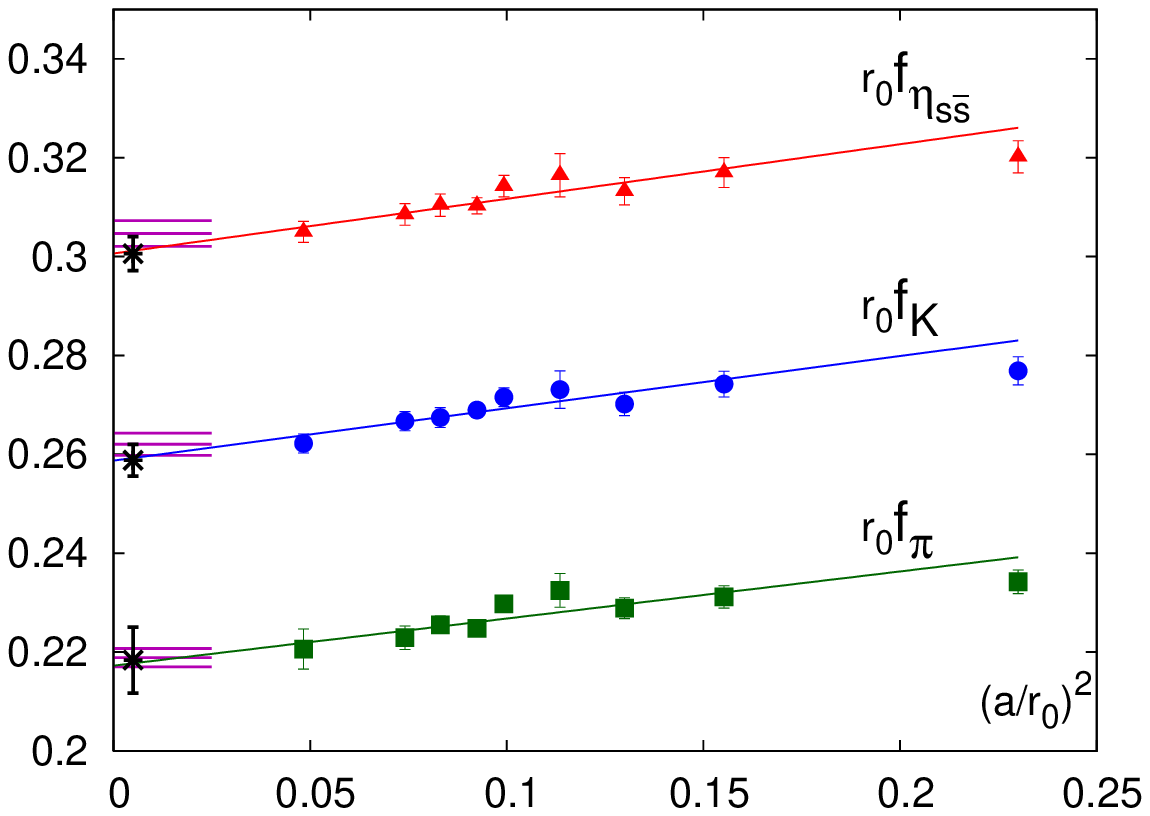} 
\includegraphics[width=8cm]{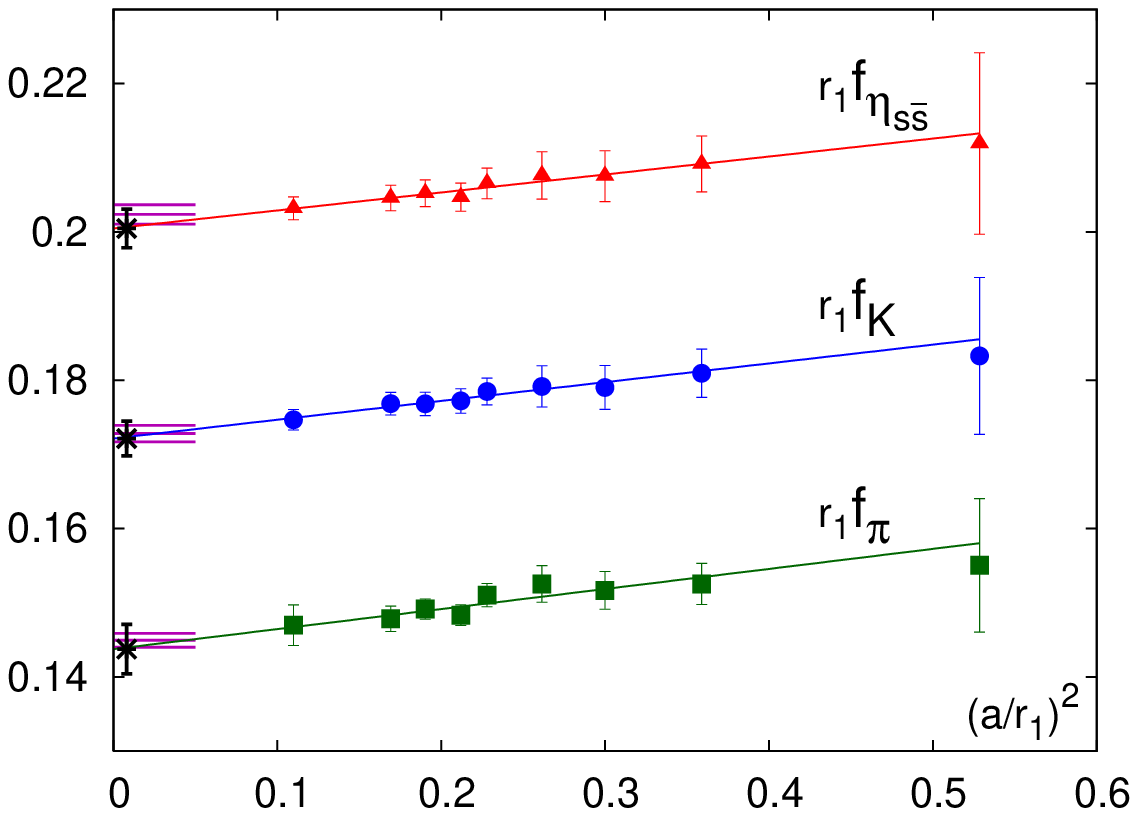}
\caption{The decay constants of $\eta_{s\bar s}$, K and $\pi$ mesons
  with the HISQ/tree action at $m_l = 0.05 m_s$ measured in units of
  $r_0$ (left) and $r_1$ (right) are shown as a function of the
  lattice spacings.  The black points along the y-axis are the result of a
  linear extrapolation to the continuum limit.  The experimental
  results are shown as horizontal bands along the y-axis with the
  width corresponding to the error in the determination of $r_0$ and
  $r_1$, respectively.  We use the HPQCD estimate for $f_{\eta_{s \bar
      s}}$~\cite{Davies:2009tsa} as the continuum value. }
\label{fig:fpi}
\end{figure}

Next, we analyze the HISQ/tree data for pion and kaon decay constants,
given in Appendix~\ref{sec:appendix3}, for $m_l/m_s =0.05$. We also
analyze the fictitious $\eta_{s\bar s}$ meson following
Ref.~\cite{Davies:2009tsa}.  In Fig.~\ref{fig:fpi}, we show our results in
units of $r_0$ and $r_1$ determined in Sec.~\ref{subsec:scalesetting}
as a function of the lattice spacing together with a continuum
extrapolation assuming linear dependence on $a^2$.  We vary the range
of the lattice spacings used in the fit and take the spread in the
extrapolated values as an estimate of the systematic errors.  These
extrapolated values agree with the experimental results within our
estimated errors (statistical and systematic errors are added in
quadrature) as also shown in Fig.~\ref{fig:fpi}. This consistency
justifies having used the continuum extrapolated value of $f_\pi r_1$ from
Ref.~\cite{Bazavov:2010hj} to convert $r_1$ to physical units as discussed
in Sec.~\ref{subsec:scalesetting}.  The deviation from the continuum
value in the region of the lattice spacings corresponding to our
finite temperature calculations is less than $8\%$ for all the decay
constants. We use these data to set the $f_K$ scale and analyze
thermodynamic quantities in terms of it and to make a direct
comparison with the stout action
data~\cite{Aoki:2006br,Aoki:2009sc,Borsanyi:2010bp}.

\begin{figure}
\includegraphics[width=8cm]{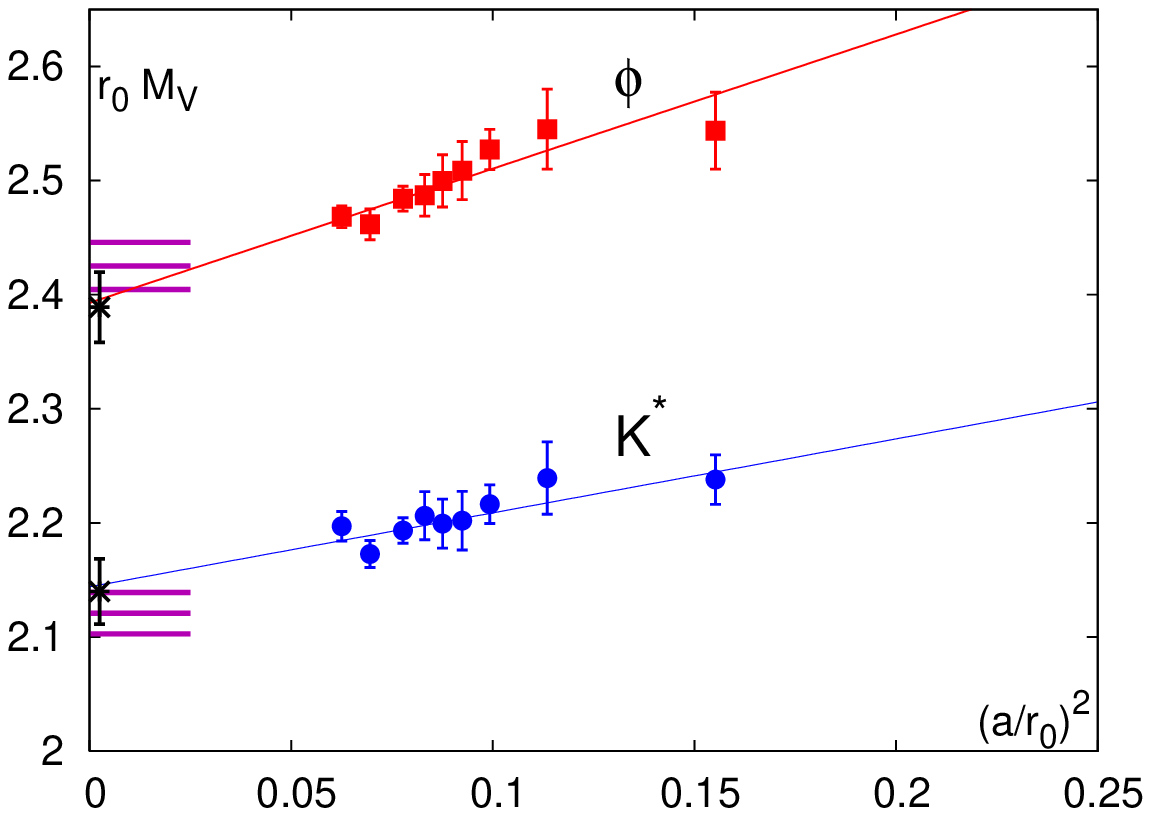}
\includegraphics[width=8cm]{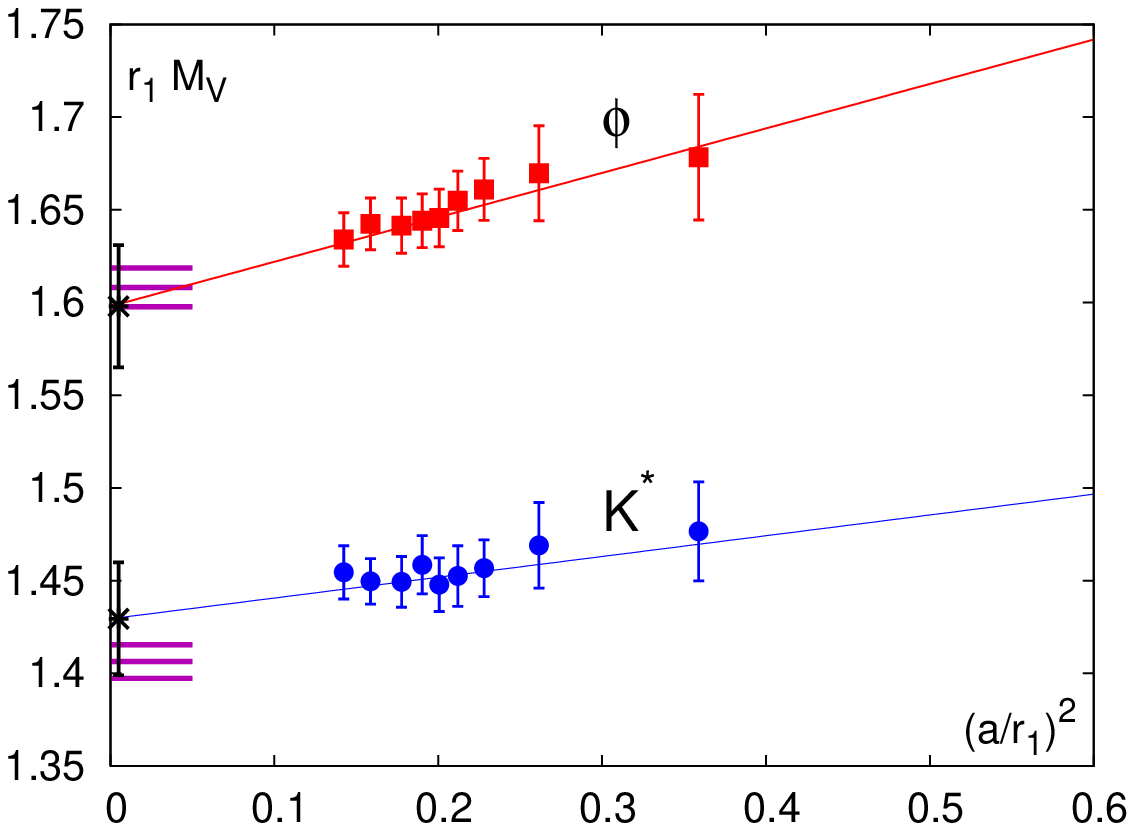}
\caption{The masses of the $\phi$ and $K^*$ mesons with the HISQ/tree
  action at $m_l = 0.05 m_s$ measured in units of $r_0$ (left) and
  $r_1$ (right) are shown as a function of the lattice spacing.  The
  lines show a linear continuum extrapolation and the black crosses
  denote the extrapolated values. The experimental results are shown as
  horizontal bands along the $y$-axis with the width corresponding to
  the error in the determination of $r_0$ and $r_1$, respectively.}
\label{fig:mVs}
\end{figure}

Finally, in Fig.~\ref{fig:mVs} we show the masses of $\phi$ and $K^*$
mesons given in Appendix~\ref{sec:appendix3} as a function of the
lattice spacing. (The rho meson correlators are very noisy, so we do
not present data for the rho mass.) Using extrapolations linear in
$a^2$ we obtain continuum estimates, and by varying the fit interval, 
we estimate the systematic errors and add these to the statistical
errors in quadrature. These estimates, in units of $r_0$ and $r_1$,
are plotted with the star symbol in Fig.~\ref{fig:mVs}. The
experimental values along with error estimates are shown as horizontal
bands and agree with lattice estimates, thereby providing an
independent check of the scale setting procedure. The slope of these
fits indicates that discretization errors are small and confirms the
findings in~\cite{Bazavov:2010ru} that taste symmetry
violations are much smaller in the HISQ/tree action compared to those
in the asqtad action. For the range of lattice spacings relevant for
the finite temperature transition region on $N_\tau=6$--12 lattices,
the discretization errors in the vector meson masses are less than
$5\%$.

\section{Universal scaling in the chiral limit and the QCD phase transition}
\label{sec:observables}

In the limit of vanishing light quark masses and for sufficiently large 
values of the strange quark mass, QCD is expected to undergo a second
order phase transition belonging to the universality class of three
dimensional $O(4)$ symmetric spin models \cite{Pisarski:1983ms}. Although
there remains the possibility that a fluctuation-induced first order
transition may appear at (very) small values of the quark mass, it 
seems that the QCD transition for physical values of the strange quark 
mass is, indeed, second order when the light quark masses are reduced
to zero. An additional complication in the analysis of the chiral 
phase transition in lattice calculations arises from the fact that the 
exact $O(4)$ symmetry is difficult to implement at nonzero values of 
the lattice spacing. Staggered fermions realize only a remnant of this
symmetry; the staggered fermion action has a global $O(2)$ symmetry.
The restoration of this symmetry at high temperatures 
is signaled by rapid changes in thermodynamic observables or peaks
in response functions, which define pseudocritical temperatures. For these
observables to be reliable indicators for the QCD transition, which becomes a
true phase transition only in the chiral limit, one must select observables
which, in the chiral limit, are dominated by contributions arising from 
the singular part of the QCD partition function $Z(V,T)$, or more precisely from the
free energy density, 
$f= - TV^{-1} \ln Z(V,T)$. A recent analysis
of scaling properties of the chiral condensate, performed with the p4 action
on coarse lattices, showed that critical behavior in the vicinity of the
chiral phase transition is well described by $O(N)$ scaling relations 
\cite{Ejiri:2009ac} which give a good description 
even in the physical quark mass regime. 

In the vicinity of the chiral phase transition, the free energy
density may be expressed as a sum of a singular and a regular
part,
\begin{equation}
f = -\frac{T}{V} \ln Z\equiv f_{sing}(t,h)+ f_{reg}(T,m_l,m_s) \; .
\label{free_energy}
\end{equation}
Here $t$ and $h$ are dimensionless couplings that control deviations from
criticality. They are related to the temperature $T$ and the light quark mass $m_l$,
which couples to the symmetry breaking (magnetic) field, as  
\begin{equation}
t = \frac{1}{t_0}\frac{T-T_c^0}{T_c^0} \quad , \quad 
h= \frac{1}{h_0} H \quad , \quad 
H= \frac{m_l}{m_s} \; ,
\label{reduced}
\end{equation}
where $T_c^0$ denotes the chiral phase transition temperature, 
{\it i.e.}, the transition temperature at $H=0$.  The scaling variables
$t$, $h$ are normalized by two parameters $t_0$ and $h_0$, which are
unique to QCD and similar to the low energy constants in the chiral
Lagrangian.  These need to be determined together with $T_c^0$. In the
continuum limit, all three parameters are uniquely defined, but depend
on the value of the strange quark mass.

The singular contribution to the free energy density is a homogeneous
function of the two variables $t$ and $h$. Its invariance under scale
transformations can be used to express it in terms of a single
scaling variable
\begin{equation}
z=t/h^{1/\beta\delta} = \frac{1}{t_0}\frac{T-T_c^0}{T_c^0} \left( \frac{h_0}{H} \right)^{1/\beta\delta}
 = \frac{1}{z_0}\frac{T-T_c^0}{T_c^0} \left( \frac{1}{H} \right)^{1/\beta\delta}
\label{eq:defz}
\end{equation}
where $\beta$ and $\delta$ are the critical exponents of the $O(N)$
universality class and $z_0 = t_0/h_0^{1/\beta\delta}$.
Thus, the dimensionless free energy density
$\tilde{f}\equiv f/T^4$ can be written as
\begin{equation}
\tilde{f}(T,m_l,m_s) = h^{1+1/\delta} f_s(z) + f_r(T,H,m_s) \; ,
\label{scaling}
\end{equation}
where the regular term $f_r$ gives rise to scaling violations. This regular term 
can be expanded in a Taylor series around $(t,h)=(0,0)$. In all
subsequent discussions, we analyze the data keeping $m_s$ in
Eq.~(\ref{scaling}) fixed at the physical value along the LCP. 
Therefore, the dependence on $m_s$ will, henceforth, be dropped.

We also note that the reduced temperature $t$ may depend on other
couplings in the QCD Lagrangian which do not explicitly break 
chiral symmetry. In particular, it depends on light and strange 
quark chemical potentials $\mu_q$, which in leading order enter only quadratically,
\begin{equation}
t = \frac{1}{t_0} \left( \frac{T-T_c^0}{T_c^0} + 
\sum_{q=l,s}\kappa_q\left(\frac{\mu_q}{T}\right)^2 +
\kappa_{ls} \frac{\mu_l}{T}\frac{\mu_s}{T} \right)
  \; .
\label{reduced2}
\end{equation}
Derivatives of the partition function with respect to $\mu_q$ are used
to define the quark number susceptibilities.

The above scaling form of the free energy density is the starting
point of a discussion of scaling properties of most observables used
to characterize the QCD phase transition. We will use this scaling
Ansatz to test to what extent various thermodynamic quantities remain
sensitive to universal features of the chiral phase transition at
nonzero quark masses when chiral symmetry is explicitly broken
and the singular behavior is replaced by a rapid crossover
characterized by pseudocritical temperatures (which we label $T_c$)
rather than a critical temperature.

A good probe of the chiral behavior is the {2-flavor} light quark
chiral condensate
\begin{equation}
\langle \bar{\psi}\psi \rangle_l^{n_f=2} = \frac{T}{V}
\frac{\partial \ln Z}{\partial m_l} \; . 
\label{chiral}
\end{equation}
Following the notation of Ref.~\cite{Ejiri:2009ac}, we introduce the dimensionless
order parameter $M_b$,
\begin{equation}
M_b \equiv \frac{m_s \langle \bar{\psi}\psi \rangle_l^{n_f=2}}{T^4} \; .
\label{order}
\end{equation}
Multiplication by the strange quark mass removes the need for
multiplicative renormalization constants; however, $M_b$ does require
additive renormalization. For a scaling analysis in $h$ at a fixed
value of the cutoff, this constant plays no role. Near $T_c^0$, $M_b$
is given by a scaling function $f_G(z)$
\begin{equation}
M_b(T,H) = h^{1/\delta} f_G(t/h^{1/\beta\delta}) + f_{M,reg}(T,H)  \; ,
\label{order_scaling}
\end{equation}
and a regular function  $f_{M,reg}(T,H)$ that gives rise to scaling violations. We 
consider only the leading order Taylor expansion of $f_{M,reg}(T,H)$ in $H$ and quadratic in $t$, 
\begin{eqnarray}
f_{M,reg}(T,H) &=& a_t(T)  H   \nonumber \\
&=& \left( a_0 + a_1 \frac{T-T_c^0}{T_c^0} + a_2 \left(\frac{T-T_c^0}{T_c^0} \right)^2 \right) H 
\label{eq:freg}
\end{eqnarray}
with parameters $a_0$, $a_1$ and $a_2$ to be determined.  The singular
function $f_G$ is well studied in three dimensional spin models and
has been parametrized for the $O(2)$ and $O(4)$ symmetry groups
\cite{Engels:2000xw,Toussaint:1996qr,Engels:1999wf,Engels:2001bq}.  Also, the exponents
$\beta$, $\gamma$, $\delta$ and $\nu$ used here are taken from Table 2
in Ref.~\cite{Engels:2001bq}.

Response functions, derived from the light 
quark chiral
condensate, are sensitive to critical behavior in the chiral limit. 
In particular, the derivative of $\langle \bar{\psi}\psi \rangle_l^{n_f=2}$ 
with respect to
the quark masses gives the chiral susceptibility
\begin{equation}
\chi_{m,l} =  \frac{\partial}{\partial m_l}  
\langle \bar{\psi}\psi \rangle_l^{n_f=2}  \;  .
\label{suscept}
\end{equation}
The scaling behavior of the light quark susceptibility, using Eq.~(\ref{order_scaling}), is 
\begin{eqnarray}
\frac{\chi_{m,l}}{T^2} &=& \frac{T^2}{m_s^2} 
\left( \frac{1}{h_0}
h^{1/\delta -1} f_\chi(z) + \frac{\partial f_{M,reg}(T,H)}{\partial H} 
\right) \; , \nonumber \\
&&{\rm with}\;\; f_{\chi}(z)=\frac{1}{\delta} [f_G(z)-\frac{z}{\beta} f_G'(z)].
\label{eq:chiralsuscept}
\end{eqnarray}
The function $f_\chi$ has a maximum at some value of the scaling
variable $z=z_p$. For small values of $h$ this defines the location of
the pseudocritical temperature $T_{c}$ as the maximum in the scaling
function $f_G(z)$.  Approaching the critical point along $h$ with $z$
fixed, $e.g.$, $z=0$ or $z=z_p$, $\chi_{m,l}$ diverges in the chiral
limit as
\begin{equation}
   \chi_{m,l} \sim m_l^{1/\delta - 1} \; .
\label{peaks}
\end{equation}

Similarly, the mixed susceptibility
\begin{equation}
\chi_{t,l} = -  \frac{T}{V} \frac{\partial^2}{\partial m_l\partial t} \ln Z \ ,
\label{chit}
\end{equation}
also has
a peak at some pseudocritical temperature and diverges in the chiral limit as 
\begin{eqnarray}
\chi_{t,l} \sim m_l^{(\beta -1)/\beta\delta} \;\; .
\label{peaks2}
\end{eqnarray}
One can calculate $\chi_{t,l}$ either by taking the derivative of
$\langle \bar{\psi}\psi \rangle$ with respect to $T$ or by taking the
second derivative with respect to $\mu_l$, {\it i.e.}, by calculating
the coefficient of the second order Taylor expansion for the chiral
condensate as a function of $\mu_l/T$ \cite{Kaczmarek:2011zz}.  The derivative
of $\langle \bar{\psi}\psi \rangle$ with respect to $T$ is the expectation value
of the chiral condensate times the energy density, which is difficult
to calculate in lattice simulations, as additional information on
temperature derivatives of temporal and spatial cutoff parameters is
needed.  Taylor expansion coefficients, on the other hand, are well
defined and have been calculated previously, although their calculation is
computationally intensive. This mixed susceptibility
has been used to determine the curvature
of the chiral transition line for small values of the baryon
chemical potential \cite{Kaczmarek:2011zz}.

Other thermodynamic observables analyzed in this paper are the 
light and strange quark number susceptibilities defined as
\begin{equation}
\frac{\chi_{q}}{T^2} = \frac{1}{VT^3}
\frac{\partial^2\ln Z}{\partial(\mu_{q}/T)^2} \; ,\;\; q=l,\; s \ . 
\label{chi_q}
\end{equation}
These are also sensitive to the singular part of the free energy since the 
reduced temperature $t$ depends on the quark chemical potentials
as indicated in Eq.~(\ref{reduced2}). 
However, unlike the temperature derivative of the chiral condensate,
{\it i.e.}, the mixed susceptibility $\chi_{t,l}$,
the temperature derivative of the light quark number susceptibility 
does not diverge in the chiral limit. Its slope at $T_c^0$ is given by 
\begin{equation}
\frac{\partial \chi_q}{\partial T} \sim c_r +A_{\pm} \left| 
\frac{T-T_c^0}{T_c^0} \right|^{-\alpha} \;\; ,
\label{chiq}
\end{equation}
and has the contribution $c_r$ from the regular part of the free energy, while
its variation with temperature is controlled by the singular part.
The critical exponent $\alpha$ is negative for QCD since the chiral
transition is expected to belong to the universality class of
three-dimensional $O(N)$ models. In short, while $\chi_q$ is sensitive to
the critical behavior, it does not diverge in the thermodynamic
limit. Consequently, it has been extremely difficult to extract reliable
information on $T_c^0$ or $T_c$ from scaling fits to $\chi_q$. Even in
high statistics $O(N)$ model calculations \cite{Cucchieri:2002hu} the
structure of the subleading term in Eq.~(\ref{chiq}) could only be
determined after using results for the dominant contribution $c_r$
extracted from other observables. We, therefore, consider quark number
susceptibilities as a good indicator of the transition in QCD, but not
useful for extracting precise values for the associated pseudocritical
temperature.

Finally, we consider the expectation value of the Polyakov loop $L$, 
\begin{equation}
L(\vec{x}) = \frac{1}{3} {\rm Tr} \prod_{x_0=1}^{N_\tau} U_0(x_0,\vec{x}) \; ,
\label{Ldef}
\end{equation}
which is the large distance limit of the static quark
correlation function,
\begin{equation}
L^2\equiv \lim_{|\vec{x}|\rightarrow\infty} 
\langle L(0) L^\dagger(\vec{x})\rangle \;\; .
\label{Polyakov}
\end{equation}
$L \equiv \langle L(\vec{x}) \rangle$ is a good order parameter for
deconfinement in the limit of infinitely heavy quarks. In that limit,
it can be related to the singular structure of the partition function
of the pure gauge theory and can be introduced as a symmetry breaking
field in the action. In QCD with light quarks, in particular in the
chiral limit, $L$ is no longer an order parameter due to the explicit
breaking of the $Z(3)$ center symmetry by the quark action. It does
not vanish for $T \le T_c^0$, but is determined by the value of the
free energy of a static quark $F_Q$ in the confined phase. This free
energy can be well approximated by the binding energy of the lightest
static-light meson, which is of order $\Lambda_{QCD}$. Similarly, $T_c
\sim \Lambda_{QCD}$; consequently $L \sim \exp(-F_Q/T) \sim 1/e$ is
not small in the confined phase.  The data for QCD with light quarks
show that $L$ varies significantly with temperature in the transition
region, reflecting the rapid change in screening properties of an
external color charge.  Thus, the Polyakov loop is sensitive to the
transition but it has no demonstrated relation to the singular part of
the QCD partition function. We, therefore, do not use it to determine
an associated pseudocritical temperature.

\section{Chiral observables}
\label{sec:chiral}

In this section, we present results for observables related to chiral
symmetry restoration at finite temperatures and discuss the cutoff
dependence of these quantities. To set the normalization of different
quantities we express them in terms of the staggered fermion matrix
$D_q=m_q \cdot 1 + D$ with $q=l,s$, as in
Ref. \cite{Bazavov:2009zn}.  In what follows, $\langle \bar \psi \psi
\rangle_{q,\tau}$ will denote the one-flavor chiral condensate, 
{\it  i.e.},
\begin{equation}
\langle \bar \psi \psi \rangle_{q,x}=\frac{1}{4} \frac{1}{N_{\sigma}^3 N_{\tau}} 
{\rm Tr} \langle D_q^{-1} \rangle,~~q=l,s \ ,
\end{equation}
where the subscript $x=\tau$ and $x=0$ will denote the expectation value
at finite and zero temperature, respectively.
The chiral susceptibility defined in Sec.~\ref{sec:observables} is the sum of connected 
and disconnected Feynman diagrams defined as 
\begin{eqnarray}
\chi_{m,l}(T)&=& {2}
\frac{\partial \langle \bar\psi \psi \rangle_{l,\tau}}{\partial m_l}
=\chi_{l, disc} + \chi_{l, con} \; , \\
\chi_{m,s}(T)&=& \frac{\partial \langle \bar\psi \psi \rangle_{s,\tau}}{\partial m_s}
=\chi_{s, disc} + \chi_{s, con} \; , 
\label{chi_tot}
\end{eqnarray}
with
\begin{eqnarray}
\chi_{q, disc} &=&
{{n_f^2} \over 16 N_{\sigma}^3 N_{\tau}} \left\{
\langle\bigl( {\rm Tr} D_q^{-1}\bigr)^2  \rangle -
\langle {\rm Tr} D_q^{-1}\rangle^2 \right\}
\label{chi_dis} \; , \\
\noalign{and}
\chi_{q, con} &=&  -
{{n_f} \over 4} {\rm Tr} \sum_x \langle \,D_q^{-1}(x,0) D_q^{-1}(0,x) \,\rangle \; ,~~~q=l,s.
\label{chi_con}
\end{eqnarray}
{Here $n_f=2$ for light quark susceptibilities, and $n_f=1$ for 
the strange quark susceptibilities.}
The disconnected part of the light quark susceptibility describes the
fluctuations in the light quark condensate and is directly analogous to
the fluctuations in the order parameter of an $O(N)$ spin model. The
second term ($\chi_{q,con}$) arises from the explicit quark mass
dependence of the chiral condensate and is the expectation value of
the volume integral of the correlation function of the (isovector)
scalar operator $\bar{\psi}\psi$.


\subsection{The chiral condensate}
\label{subsec:condensate}

\begin{figure}
\includegraphics[width=0.45\textwidth]{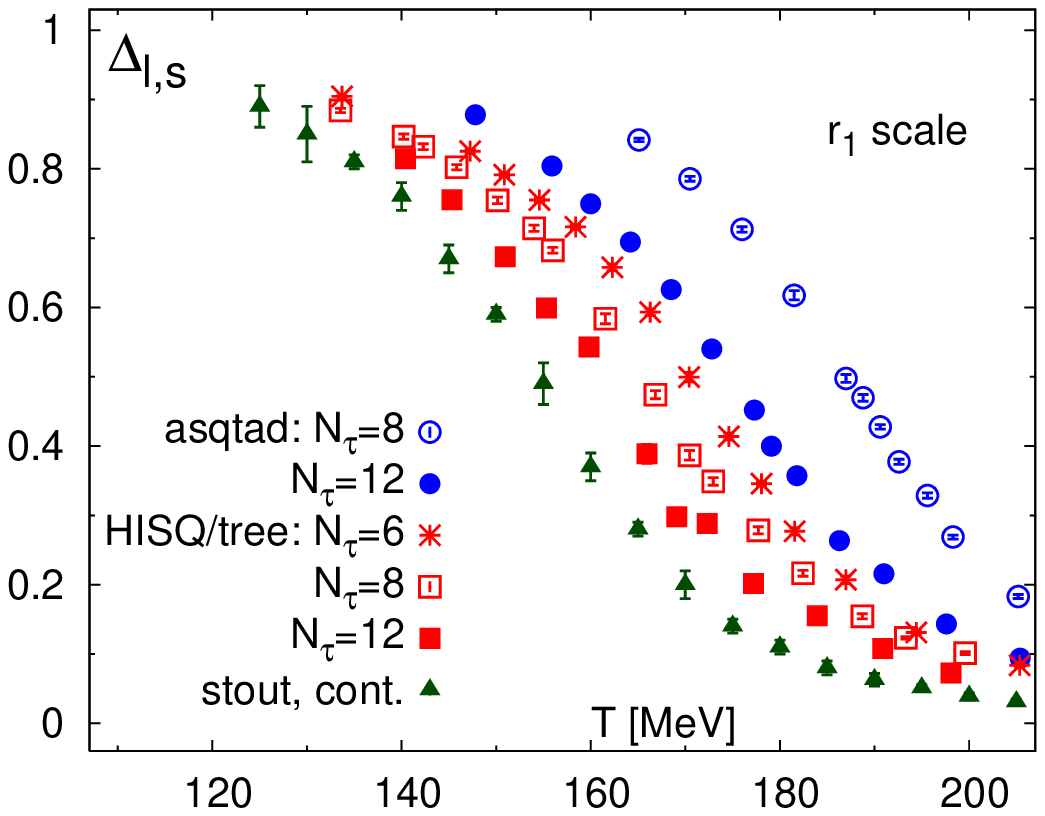}
\includegraphics[width=0.45\textwidth]{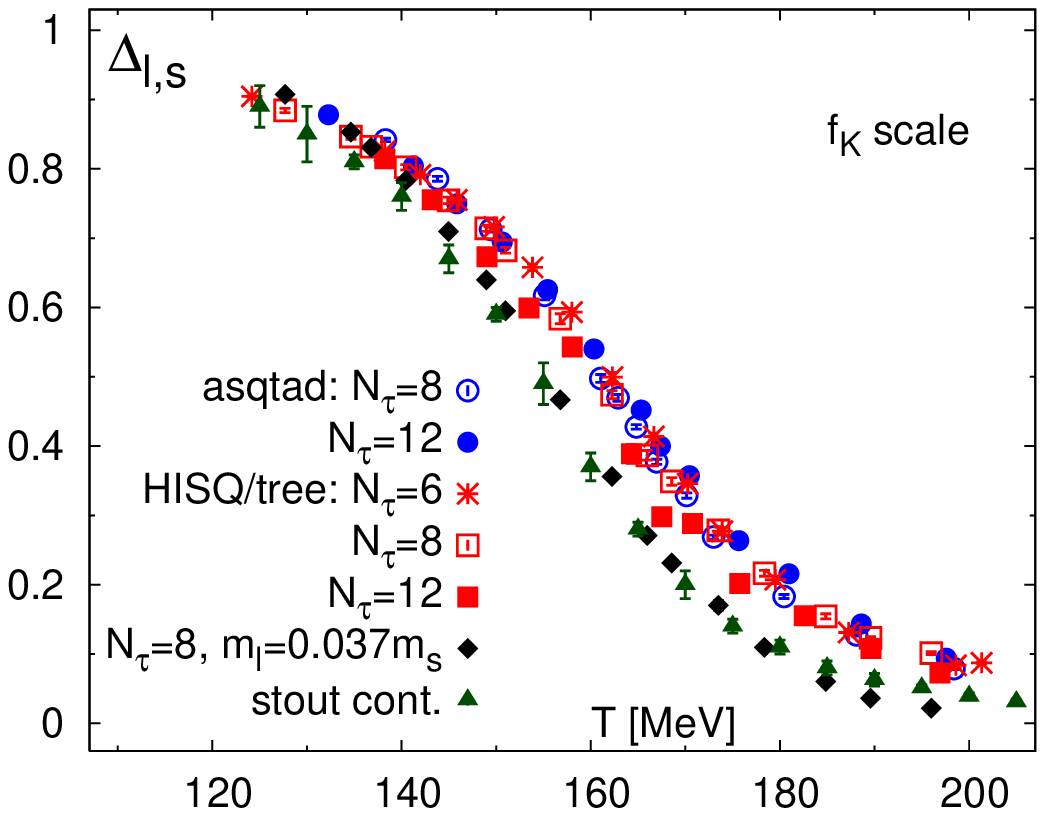}
\caption{The subtracted chiral condensate for the asqtad and HISQ/tree
  actions with $m_l=m_s/20$ is compared with the continuum extrapolated
  stout action results~\cite{Borsanyi:2010bp} (left panel).  The temperature
  $T$ is converted into physical units using $r_1$ in the left panel
  and $f_K$ in the right.  We find that the data collapse into a
  narrow band when $f_K$ is used to set the scale. 
  The black diamonds in the right panel show
  HISQ/tree results for $N_\tau=8$ lattices after an interpolation 
  to the physical light quark mass using the 
  $m_l/m_s = 0.05$ and $0.025$ data. }
\label{fig:pbp}
\end{figure}

The chiral condensate $\langle \bar{\psi} \psi \rangle$ requires both
multiplicative and additive renormalizations at finite quark
masses. The leading additive renormalization is proportional to
$(m_q/a^2)$.\footnote{There is also a logarithmic divergence 
proportional to $m_q^3$, which we neglect.} To remove these UV
divergences, we consider the
subtracted chiral condensate introduced in Ref.~\cite{Cheng:2007jq}, 
\begin{equation}
\Delta_{l,s}(T)=\frac{\langle \bar\psi \psi \rangle_{l,\tau}-\frac{m_l}{m_s} \langle \bar \psi \psi \rangle_{s,\tau}}
{\langle \bar \psi \psi \rangle_{l,0}-\frac{m_l}{m_s} \langle \bar \psi \psi \rangle_{s,0}}.
\end{equation}
Our results for the HISQ/tree and asqtad actions at $m_l=0.05m_s$ are
shown in Fig.~\ref{fig:pbp}(left) and compared to the continuum
estimate obtained with the stout action \cite{Borsanyi:2010bp}. The
temperature scale is set using $r_0$ and $r_1$ as discussed in
Section~\ref{subsec:scalesetting} and Appendix~\ref{sec:appendix2}.
The asqtad results obtained on $N_{\tau}=8$ lattices deviate
significantly from the stout results as observed previously
\cite{Bazavov:2009zn}. The new data show that these differences are much
smaller for $N_{\tau}=12$ ensembles.  More important, the
discretization effects and the differences from the stout continuum
results are much smaller for the HISQ/tree data.

In Fig.~\ref{fig:pbp}(right), we analyze the data for $\Delta_{l,s}$
using the kaon decay constant $f_K$ to set the lattice scale. For the
HISQ/tree action, we use the values of $f_K$ discussed in
Sec.~\ref{subsec:scalesetting}, while for the asqtad action, we use
$f_K$ from staggered chiral fits (see the discussion in appendix B).  We note that
for $m_l/m_s=1/20$ all the data obtained with the HISQ/tree and asqtad
actions on different $N_{\tau}$ lattices collapse into one curve,
indicating that $\Delta_{l,s}$ and $f_K$ have similar discretization
errors.  The remaining difference between the stout and our estimates,
as shown next, is due to the difference in the quark
masses---calculations with the stout action were done with
$m_l=0.037m_s$ whereas our calculations correspond to $m_l=0.05m_s$.

For a direct comparison with stout results, we extrapolate our
HISQ/tree data in the light quark mass.  This requires estimating the
quark mass dependence of the chiral condensate at both zero and
non-zero temperatures. For the $T=0$ data, we perform a linear
extrapolation in the quark mass using the HISQ/tree lattices at
$m_l=0.05m_s$ and $0.20m_s$. For the non-zero $T$ data, we use the
$O(N)$ scaling analysis, described in Sec.~\ref{sec:scaling}, which
gives a good description of the quark mass dependence in the
temperature interval $150 {\rm MeV} < T <200$ MeV.  The resulting $N_\tau=8$
HISQ/tree estimates at the physical quark mass are shown in
Fig.~\ref{fig:pbp}(right) as black diamonds and agree with the stout
action results~\cite{Aoki:2006br,Aoki:2009sc,Borsanyi:2010bp} plotted using
green triangles.

We can also remove the multiplicative renormalization factor in the
chiral condensate by considering the renormalization group invariant
quantity $r_1^4 m_s \langle \bar \psi \psi \rangle_l$, where $m_s$ is
the strange quark mass. The additive divergences can be removed by
subtracting the zero temperature analogue, {\it i.e.}, we consider the
quantity
\begin{equation}
\Delta_l^R=d+2 m_s r_1^4 ( \langle \bar \psi \psi \rangle_{l,\tau}-
\langle \bar \psi \psi \rangle_{l,0} ) \, .
\end{equation}
Note that $\Delta_l^R$ is very similar to the renormalized chiral
condensate $\langle \bar \psi \psi \rangle_R$ introduced in
Ref.~\cite{Borsanyi:2010bp}, but differs by the factor $(m_l/m_s)/(r_1^4
m_{\pi}^4)$ and $d$. A natural choice for $d$ is the value of the
chiral condensate in the zero light quark mass limit times $m_s
r_1^4$. In this limit $\Delta_l^R$ should vanish above the critical
temperature.  
To estimate $d$, we use the zero temperature estimate
$\langle \bar \psi \psi \rangle_l (\overline{MS},\ \mu=2{\rm GeV}) = 242(9)({}^{+5}_{-17})(4)\ {\rm MeV}^3$
determined in the chiral limit using $SU(2)$ staggered chiral perturbation
theory by the MILC collaboration~\cite{Bazavov:2009bb} and the corresponding strange
quark mass $m^{\overline{MS}}(\bar \mu=2{\rm GeV})=88(5)$ MeV. We get $d=0.0232244$.
\begin{figure}
\includegraphics[width=0.450\textwidth]{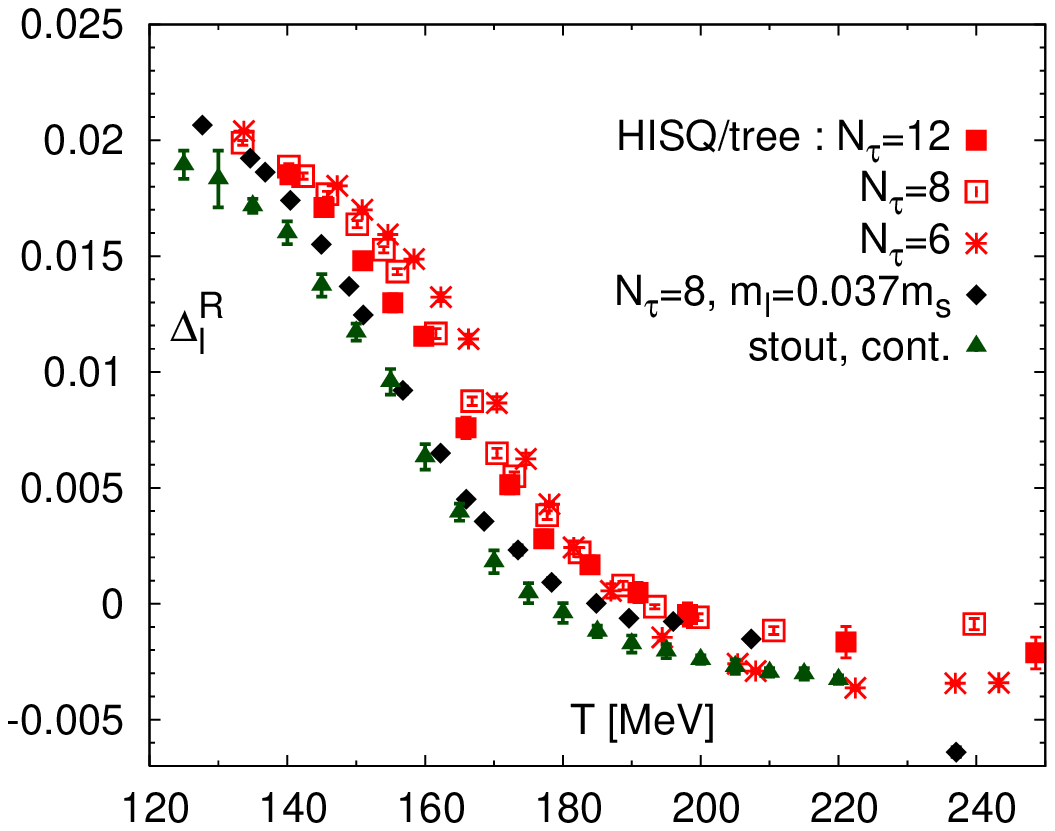}
\includegraphics[width=0.450\textwidth]{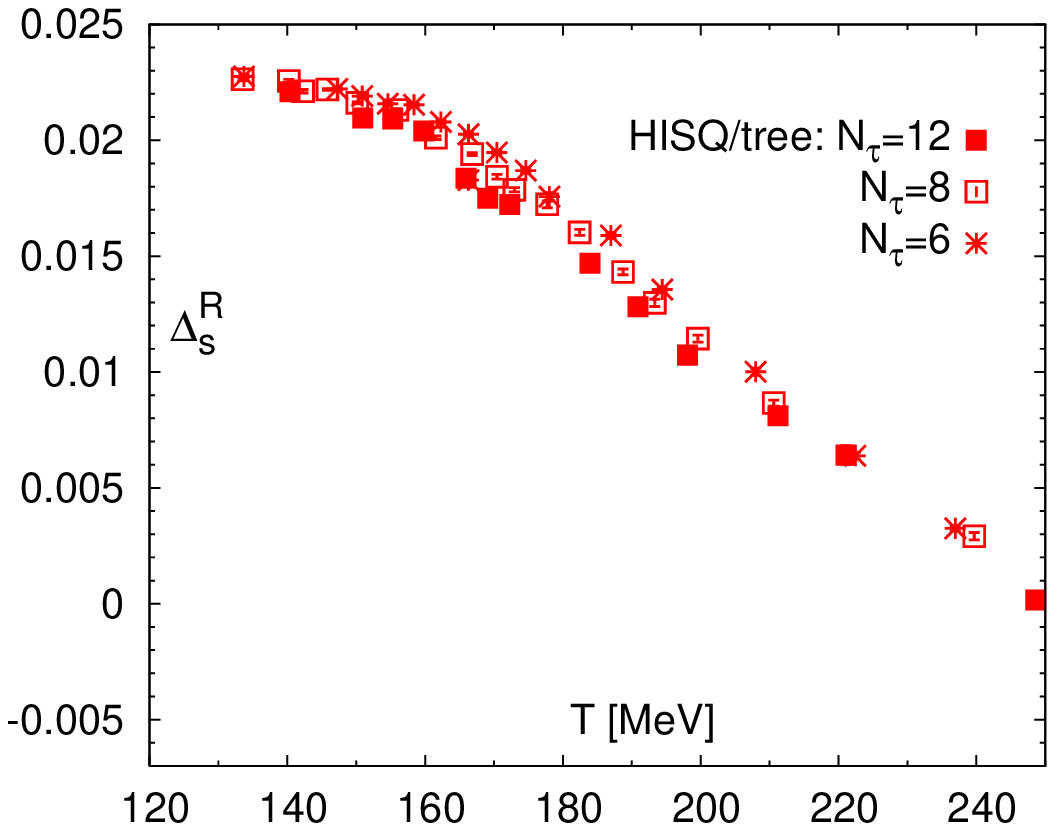}
\caption{The renormalized chiral condensate $\Delta_l^R$ for the
  HISQ/tree action with $m_l/m_s = 0.05$ is compared to the stout data.
  In the right panel, we show the renormalized strange quark
  condensate $\Delta_s^R$ for the HISQ/tree action. The temperature
  scale in both figures is set using $r_1$.  The black diamonds in the
  left panel show the $N_\tau=8$ HISQ/tree estimates using the $f_K$ scale and after an
  interpolation to the physical light quark mass $m_l/m_s = 0.037$
  as discussed in the text.
}
\label{fig:pbpR}
\end{figure}

We show $\Delta_l^R$ for the HISQ/tree action and the stout continuum
results in Fig.~\ref{fig:pbpR}(left)\footnote{We multiply the stout
  results by $(m_s/m_l)=27.3$ and by $r_1^4 m_{\pi}^4=0.0022275$. For
  the latter factor, we use the physical pion mass and the value of
  $r_1$ determined in~\cite{Bazavov:2010hj} and discussed in
  Sec.~\ref{subsec:scalesetting}.}.  To compare with the stout
continuum results, we need to extrapolate the HISQ/tree data both to
the continuum limit and to the physical quark mass.  To perform the
continuum extrapolation we convert $\Delta_l^R$ to the $f_K$ scale in
which discretization errors, as already noted for $\Delta_{l,s}$, are
small. We then interpolate these $N_\tau=8$ data at $m_l/m_s = 0.05$
and $0.025$ to the physical quark mass $m_l/m_s = 0.037$. These
estimates of the continuum HISQ/tree $\Delta_{l}^R$ are shown in
Fig.~\ref{fig:pbpR}(left) as black diamonds and are in agreement with
the stout results (green triangles)~\cite{Borsanyi:2010bp}.

Lastly, in Fig.~\ref{fig:pbpR}(right), we show the subtracted
renormalization group invariant quantity, $\Delta_s^R$, which is
related to the chiral symmetry restoration in the strange quark
sector.  We find a significant difference in the temperature
dependence between $\Delta_l^R$ and $\Delta_s^R$, with the latter
showing a gradual decrease rather than a crossover behavior.

\subsection{The chiral susceptibility}
\label{ssec:chiralsus}

As discussed in Sec.~\ref{sec:observables}, the chiral susceptibility
$\chi_{m,l}$ is a good probe of the chiral transition in QCD as it is
sensitive to the singular part of the free energy density.  It
diverges in the chiral limit, and the location of its maximum at
nonzero values of the quark mass defines a pseudocritical temperature
$T_c$ that approaches the chiral phase transition temperature $T_c^0$
as $m_l\rightarrow 0$.

For sufficiently small quark masses, the chiral susceptibility is
dominated by the disconnected part, therefore, $T_c$ can also be
defined as the location of the peak in the disconnected chiral
susceptibility defined in Eq.~(\ref{chi_dis}).  As we will show later,
$\chi_{q,disc}$ does not exhibit an additive ultraviolet divergence
but does require a multiplicative renormalization~\footnote{ It is
  easy to see that at leading order in perturbation theory, $i.e.$, in
  the free theory, the disconnected chiral susceptibility vanishes and
  thus is non-divergent. Our numerical results at zero temperature do
  not indicate any quadratic divergences in the disconnected chiral
  susceptibility, but logarithmic divergences are possible.}.

\subsubsection{Disconnected chiral susceptibility}
\label{ssec:chidisc}

\begin{figure}
\centering
\includegraphics[width=0.450\textwidth]{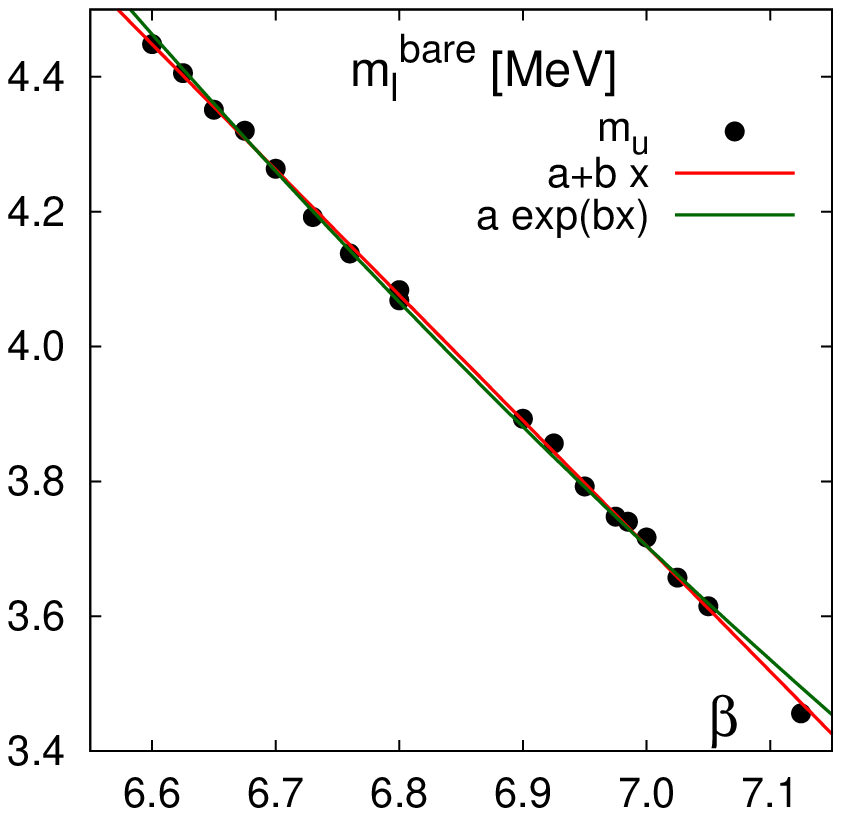}
\includegraphics[width=0.450\textwidth]{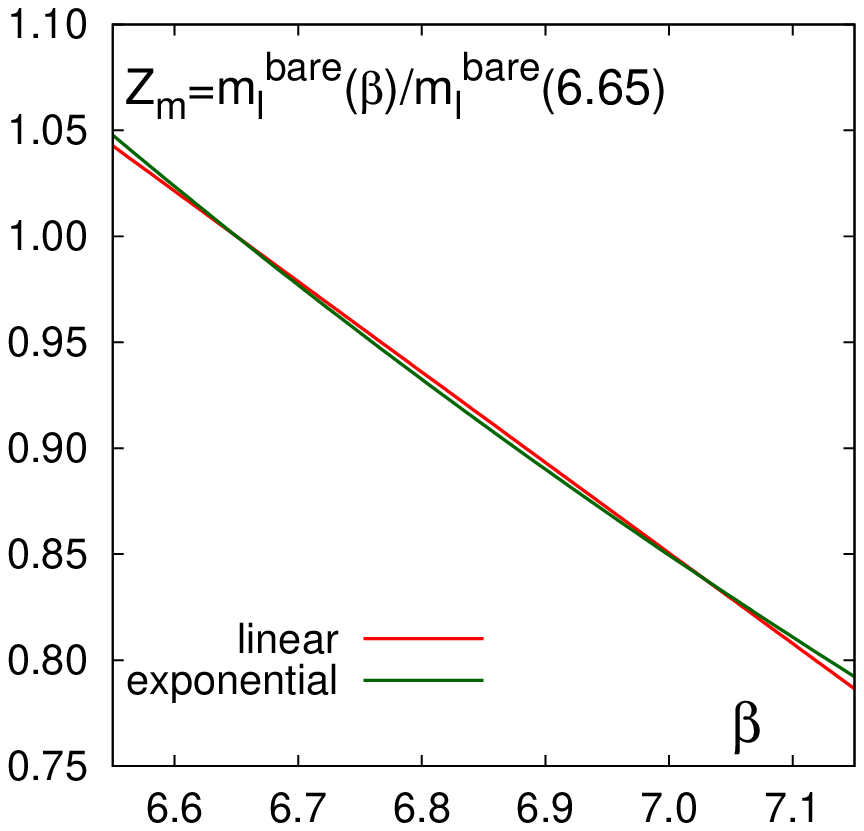}
\caption{Value of the bare light quark mass, in MeV using $r_1$ to set the scale, on the line of
  constant physics for the asqtad action {\it vs.}\ the lattice
  gauge coupling $\beta$.  The right hand part of the figure shows the
  change of the renormalization constants with $\beta$, {\it i.e.},
  with the cutoff, relative to the arbitrarily chosen renormalization
  point $\beta=6.65$.}
\label{fig:quarkmass}
\end{figure}

The multiplicative renormalization factors for the chiral condensate
and the chiral susceptibility can be deduced from an analysis of the
line of constant physics for the light quark masses, $m_l(\beta)$.
The values of the quark mass for the asqtad action, converted to physical units using
$r_1$, are shown in Fig.~\ref{fig:quarkmass}(left). The variation with
$\beta$ gives the scale dependent renormalization of the quark mass
(its reciprocal is the renormalization factor for the chiral
condensate).  What $m_l(\beta)$ does not fix is the renormalization
scale, which we choose to be $r_0/a=3.5$ (equivalently $r_1/a = 2.37$
or $a=0.134$ fm), and the ``scheme'', which we choose to be the asqtad
action.  For the asqtad action, this scale corresponds to the coupling
$\beta=6.65$ which is halfway between the peaks in the chiral
susceptibility on $N_\tau=8$ and $12$ lattices.  This specification,
$Z_m(asqtad) = 1$ at $r_0/a = 3.5$, is equivalent to choosing, for a
given action, the renormalization scale $\Lambda$ which controls the
variation of $Z_m$ with coupling $\beta$ as shown in
Fig.~\ref{fig:quarkmass}(right) for the asqtad action.

A similar calculation of $Z_m$ is performed for the p4 and HISQ/tree
actions. It is important to note that choosing the same reference
point $r_0/a=3.5$ and calculating $Z_m(\beta)$ for each of the actions 
leaves undetermined a relative renormalization factor between the
actions, $i.e.$, the relation between the corresponding $\Lambda$'s of the 
different schemes. This relative factor between any two actions is
also calculable and given by the ratio of the (bare) quark mass along
the physical LCP at $r_0/a=3.5$. At this scale our data give 
\begin{equation}
\frac{m({\rm asqtad})}{m({\rm HISQ/tree})} = 0.97828 \, .
\label{eq:scalerelation}
\end{equation}
Recall, however, that along the LCP the quark masses, $m_s$ and therefore $m_l$, for the
asqtad action are about $20\%$ heavier than the physical
values. Noting that the lattice scale at a given $\beta$ is set using
a quark-mass independent procedure, we correct $m({\rm asqtad})$ by the
factor $(M_\pi r_0)^2 |_{\rm HISQ/tree} / (M_\pi r_0)^2 |_{\rm asqtad}
$.  Then, at $r_0/a=3.5$
\begin{equation}
\frac{m({\rm asqtad})}{m({\rm HISQ/tree})} = 0.782 \qquad i.e. \qquad \frac{Z_m({\rm asqtad})}{Z_m({\rm HISQ/tree})} = 1.2786 \ .
\label{eq:Zmratio}
\end{equation}
Given $Z_m(\beta)$ we get $Z_{\bar{\psi}\psi} \equiv Z_S = 1/Z_m$ and $Z_{\chi} =
1/Z_m^2$. A similar calculation of $Z_m$ has been carried out for the
p4 action.

\begin{figure}
\includegraphics[viewport=18 5 279 243,width=0.45\textwidth]{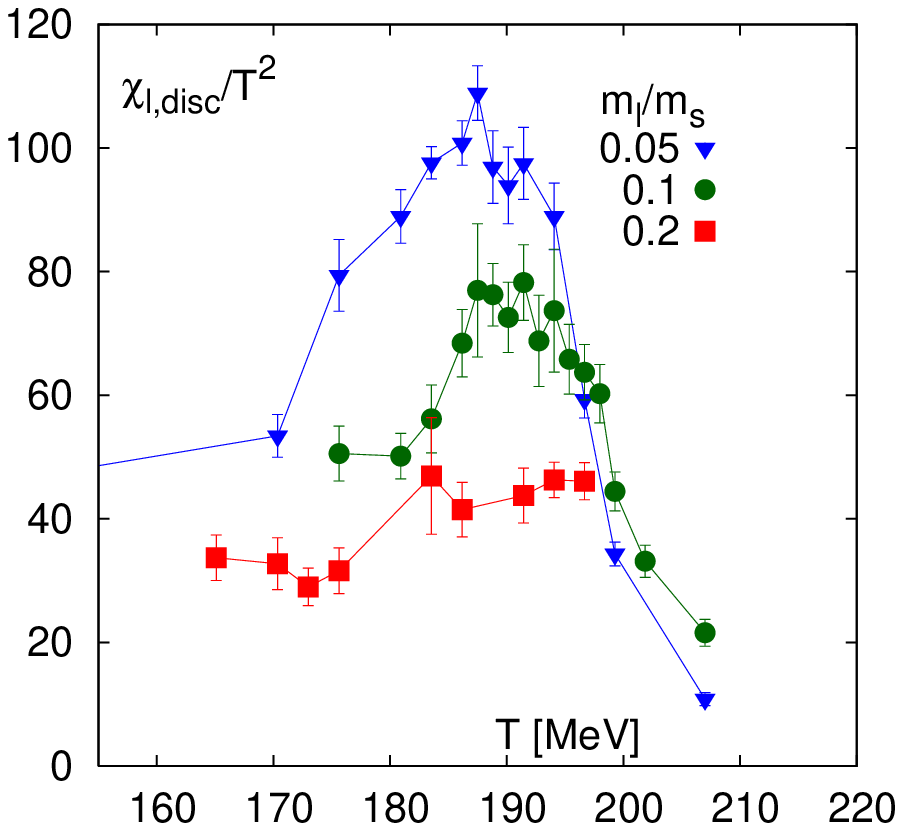}
\includegraphics[viewport=69 7 310 229,width=0.435\textwidth]{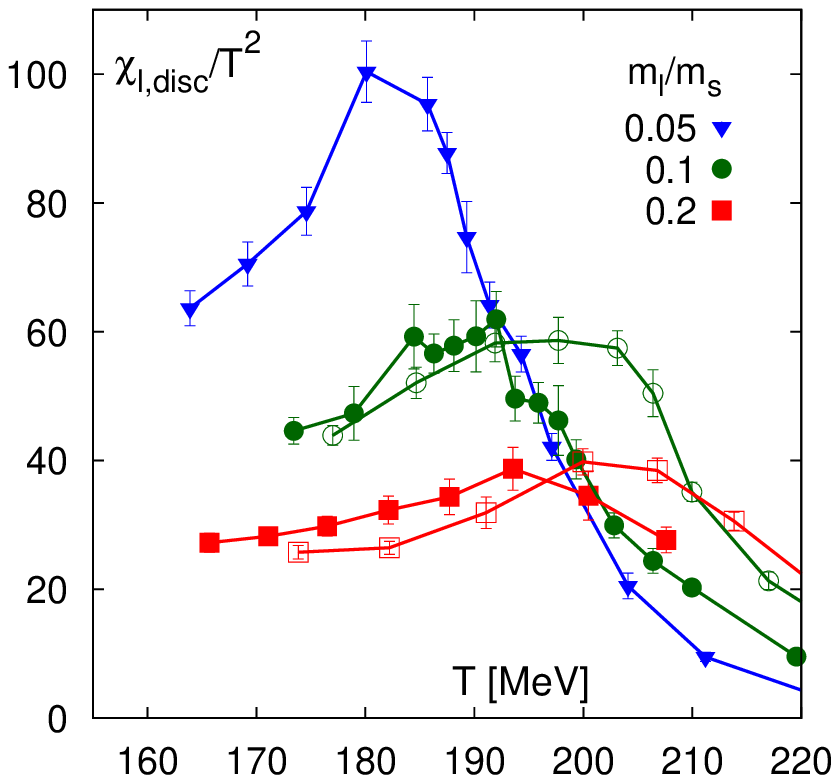}
\caption{The disconnected part of the chiral susceptibility, including
  multiplicative renormalization, calculated on $N_\tau=8$ lattices
  for the p4 (left) and asqtad (right) actions at three light quark
  masses. The figure on the right also shows asqtad data from
  $N_\tau=6$ lattices as open symbols. }
\label{fig:chi_disc}
\end{figure}

The systematics of the quark mass and cutoff dependence of the
disconnected part of the chiral susceptibility is analyzed in more
detail for the p4 and asqtad actions in Fig.~\ref{fig:chi_disc}.  The
data show a rapid rise in $\chi_{l,disc}/T^2$ with decreasing quark
mass at low temperatures and in the transition region. This mass
dependence can be traced back to the leading thermal correction to the
chiral condensate.  At finite temperature and for sufficiently small
quark masses, the chiral order parameter can be understood in terms of
the 3-dimensional $O(N)$ models.  A dimensional reduction is
applicable because the Goldstone modes are light in this region. Based
on the $O(N)$ model analysis, the quark mass dependence of the chiral
condensate is expected to have the form
\cite{Wallace:1975vi,Hasenfratz:1989pk,Smilga:1993in,Smilga:1995qf}
\begin{equation}
\langle \bar \psi \psi \rangle_l(T,m_l)=\langle \bar \psi \psi \rangle_l(0)+c_2(T) \sqrt{m_l}+... \ ,
\label{pbp_ml}
\end{equation}
as has been confirmed in numerical simulations with the p4 action on
$N_{\tau}=4$ lattices \cite{Ejiri:2009ac}.  Consequently for $T < T_c^0$,
there is a $m_l^{-1/2}$ singularity in the chiral susceptibility in
the limit of zero quark mass which explains the rise in
$\chi_{l,disc}/T^2$.  

A second feature of the data is shown in
Fig.~\ref{fig:chi_disc}(right) which compares data for the asqtad
action on lattices of different $N_\tau$ at $m_l/m_s = 0.2$ and
$0.1$. Open (filled) symbols denote data on $N_\tau=6$ ($N_\tau=8$)
lattices. The data show a shift towards smaller temperature values of
both the peak and the rapidly dropping high temperature part when the
lattice spacing is reduced.  The data also show that the variation of the
shape of the susceptibility above the peak is weakly dependent on the
quark mass. This is expected as $\chi_l$ is the derivative of the
chiral condensate with respect to the mass which, as shown in
Fig.~\ref{fig:pbpR}, is almost linear in the quark mass in this
temperature regime.  Thirdly, the data in
Fig.~\ref{fig:chi_disc}(right) show that the height of the
multiplicatively renormalized disconnected chiral susceptibility at
fixed $m_l/m_s$ is similar for $N_{\tau}=6$ and $N_{\tau}=8$ lattices.
This lack of increase in height with $N_\tau$ supports the hypothesis
that there are no remaining additive divergent contributions in the
disconnected part of the chiral susceptibility.

\begin{figure}
\includegraphics[width=8.5cm]{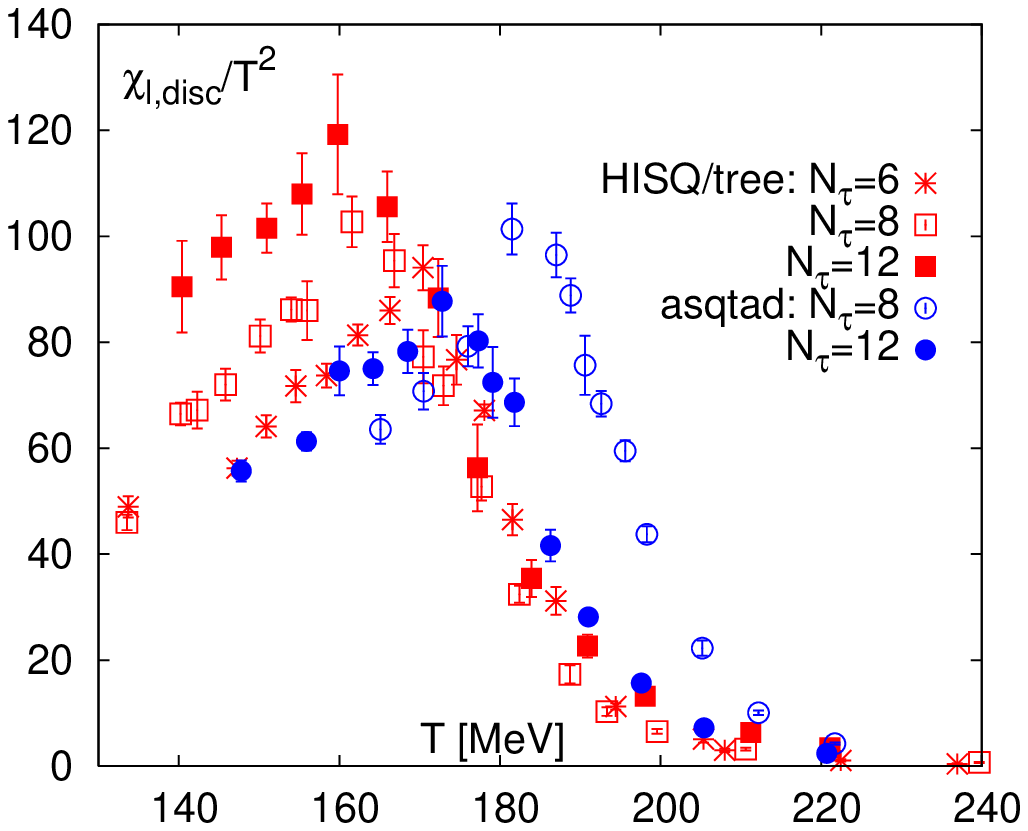}
\includegraphics[width=8.5cm]{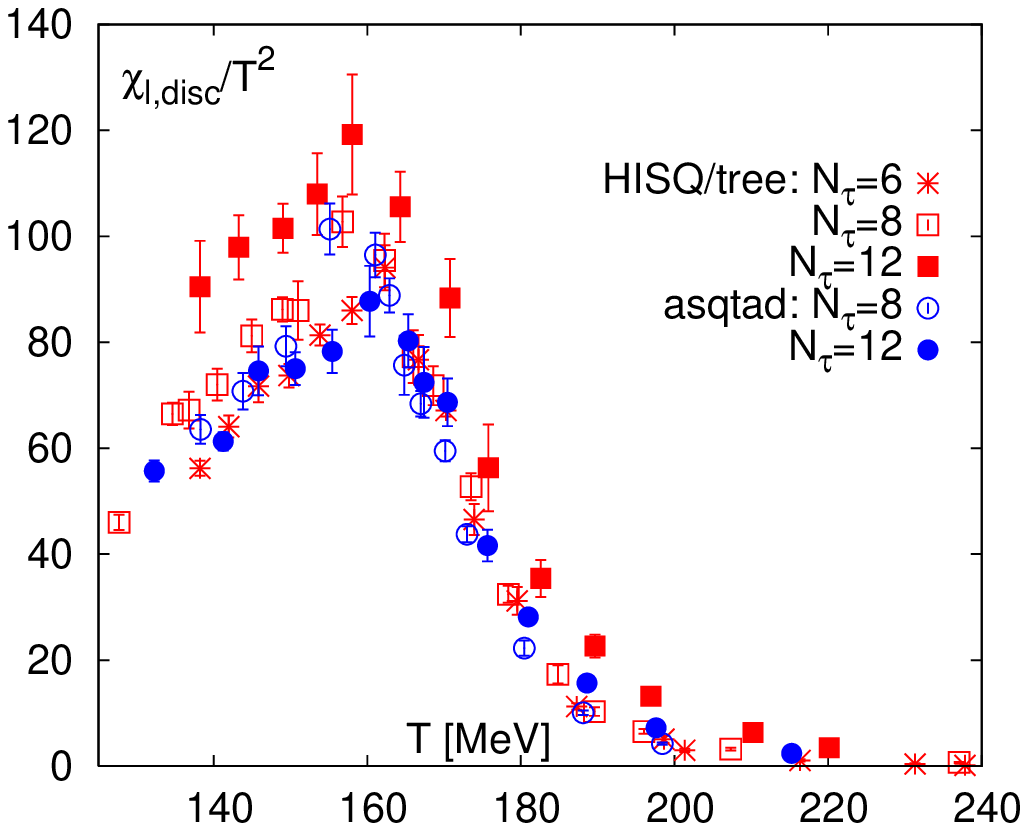}
\caption{ The disconnected part of the chiral susceptibility for the
  asqtad and HISQ/tree actions, including the multiplicative
  renormalization constant discussed in the text, is shown for
  $m_l=m_s/20$ and different $N_{\tau}$. In the right panel, the same
  data are plotted using $f_K$ to set the scale. Plotted this way they show much smaller
  variation with $N_\tau$.
}
\label{fig:chi_disc_asqtad_hisq}
\end{figure}

In Fig.~\ref{fig:chi_disc_asqtad_hisq}, we compare, for
$m_l/m_s=0.05$, the disconnected part of the chiral susceptibility
including the multiplicative renormalization factor $Z_\chi$. We note
three features in the data. First, the variation in the position of
the peak for the asqtad action is larger between $N_\tau=8$ and $12$ than for
the HISQ/tree action between $N_\tau=6$ and $8$. Second, the peak
height increases for the HISQ/tree data and decreases for the asqtad
data with $N_\tau$. Lastly, the agreement in the location of the peak
for the two actions and the data above the peak is much better when
$f_K$ is used to set the scale as shown in
Fig.~\ref{fig:chi_disc_asqtad_hisq}(right).  Note that the peak height
for the two actions is not expected to match since the quark masses on the LCP for the
asqtad data are about $20\%$ heavier than for the HISQ/tree data.

\subsubsection{Connected chiral susceptibilities}

The connected part of the chiral susceptibility, 
Eq.~(\ref{chi_con}), is the volume integral of the
scalar, flavor nonsinglet meson correlation function. 
At large distances, where its behavior is controlled 
by the lightest scalar, nonsinglet screening mass, the correlation function
drops exponentially as this state has a mass gap, $i.e.$, $\chi_{l,con}$ 
can diverge in the thermodynamic limit only if the finite 
temperature screening mass in this channel vanishes. This, in turn, would 
require the restoration of the $U_A(1)$ symmetry, which is not expected 
at the QCD transition temperature. In fact, the scalar screening masses
are known to develop a minimum at temperatures above, but close to, the
transition temperature \cite{Cheng:2010fe}. One therefore expects that even in
the chiral limit, the connected part of the chiral susceptibility will only 
exhibit a maximum above the chiral transition temperature. 

There are two subtle features of the connected part of the
susceptibility calculated at nonzero lattice spacings that require
further discussion.  First, taste symmetry violations in staggered
fermions introduce an additional divergence of the form
$a^2/\sqrt{m_l}$ for $T < T_c^0$. 
It also arises due to the long distance fluctuations of Goldstone pions, 
as explained  in Eq. (33), however, unlike
the divergence in the disconnected part which is physical, this term is 
proportional to the $O(a^2)$ taste breaking. Note that in the two-flavor theory 
there are no such divergences due to Goldstone modes in the continuum limit 
\cite{Smilga:1993in,Smilga:1995qf}.
Thus, we expect to observe a strong
quark mass dependence at low temperatures in $\chi_{l,con}$. Second,
there is a large reduction in the $U_A(1)$ symmetry breaking in the
transition region, consequently there will be a significant quark mass
dependence of scalar screening masses and of $\chi_{l,con}$.

\begin{figure}
\includegraphics[width=8.5cm]{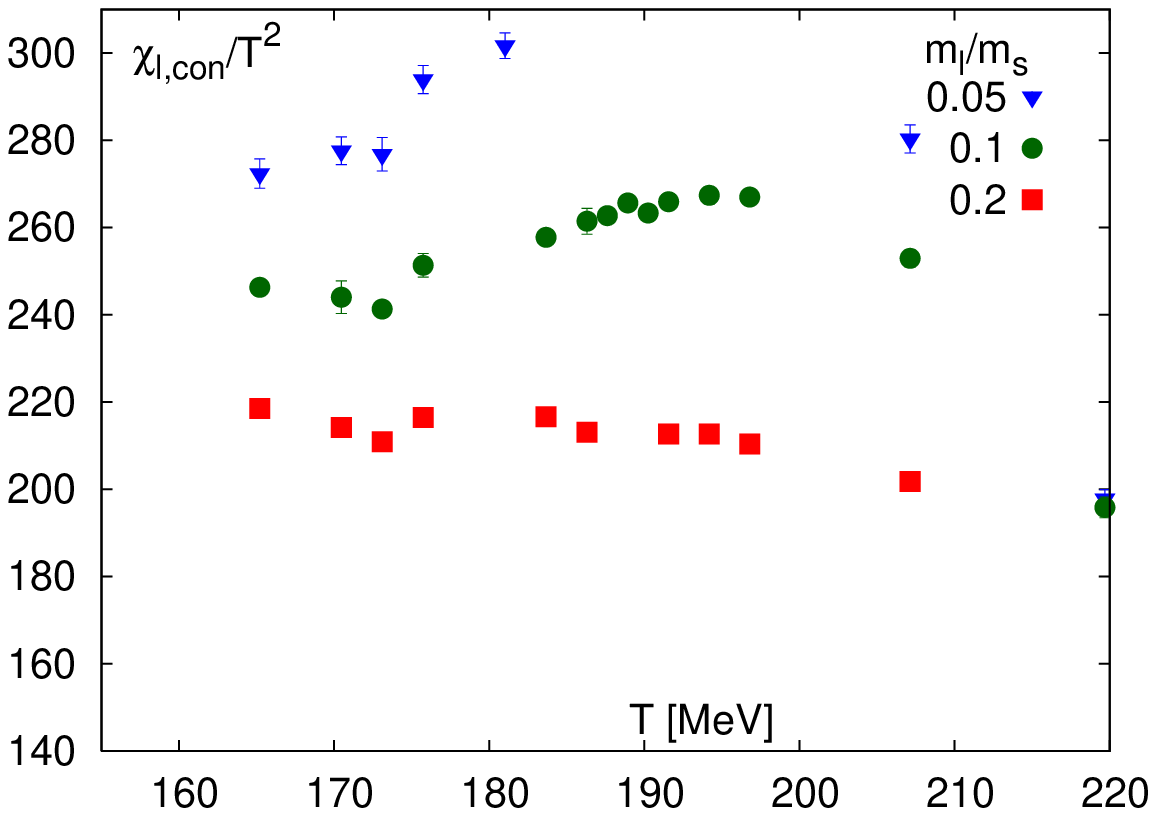}
\includegraphics[width=8.5cm]{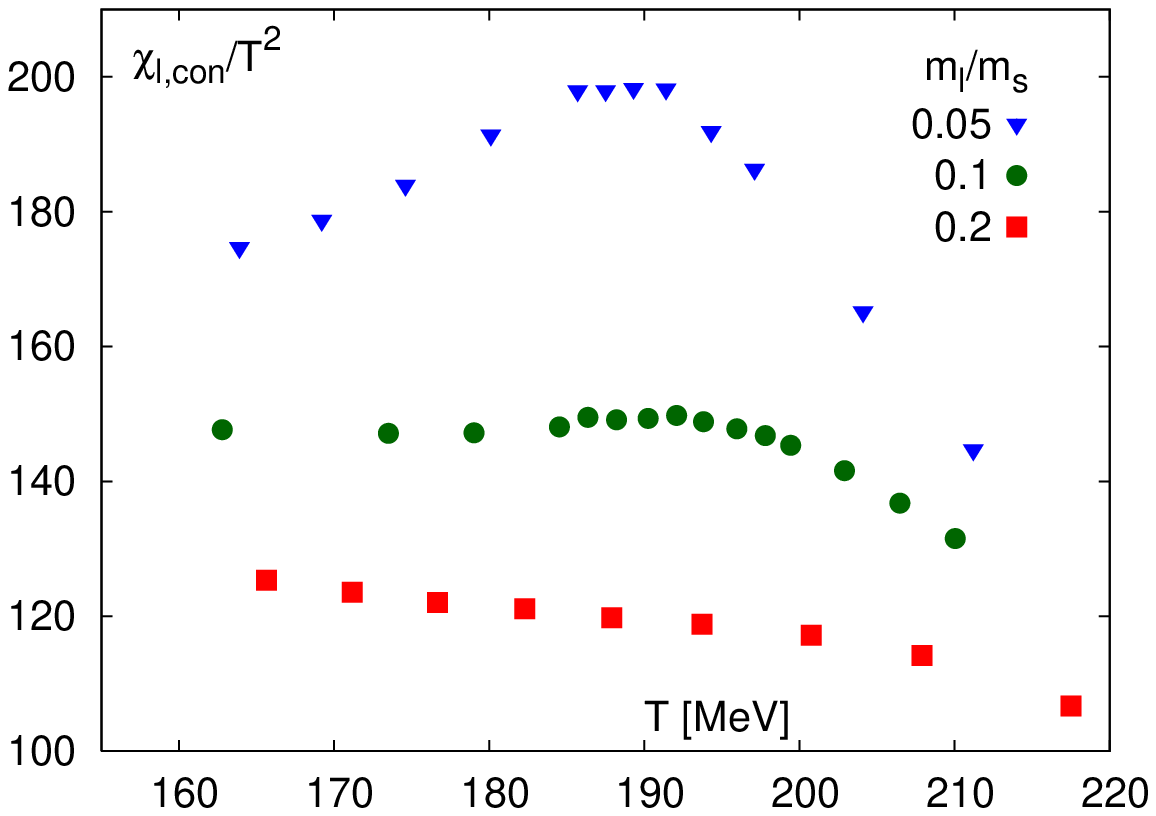}
\caption{The connected part of the chiral susceptibility for the p4
  (left) and asqtad (right) actions for different quark masses on
  $N_{\tau}=8$ lattices.  }
\label{fig:chi_conn_p4}
\end{figure}

We have calculated $\chi_{l,con}$ for
the p4, asqtad and HISQ/tree actions. Results at different light quark
masses from $N_\tau =8$ lattices are shown in
Fig.~\ref{fig:chi_conn_p4} for the p4 and asqtad actions with the
multiplicative renormalization performed in the same way as for
$\chi_{l,disc}$. A strong dependence on the quark mass is seen in
both the p4 and asqtad data.  This, as conjectured above, is due to a
combination of the artifacts that are due to taste symmetry breaking
and the variations with temperature of the scalar, flavor nonsinglet
screening mass at and above the crossover temperatures.

\begin{figure}
\includegraphics[width=8.5cm]{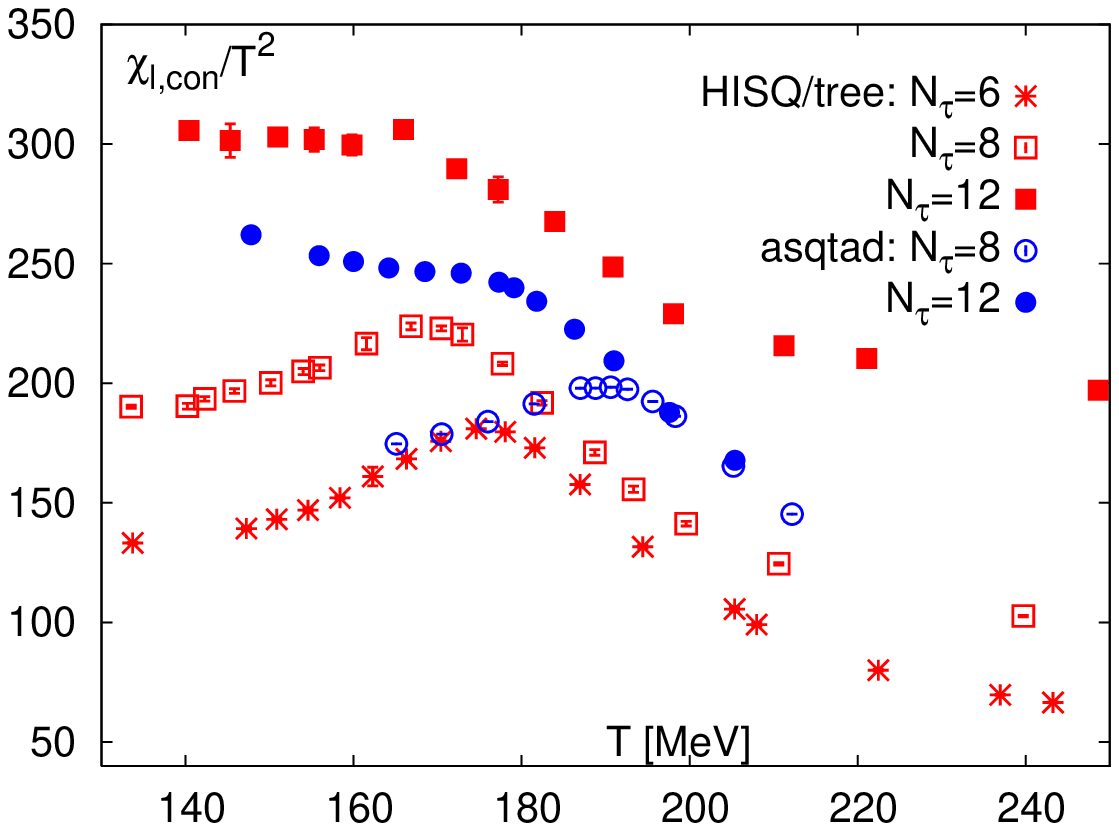}
\includegraphics[width=8.5cm]{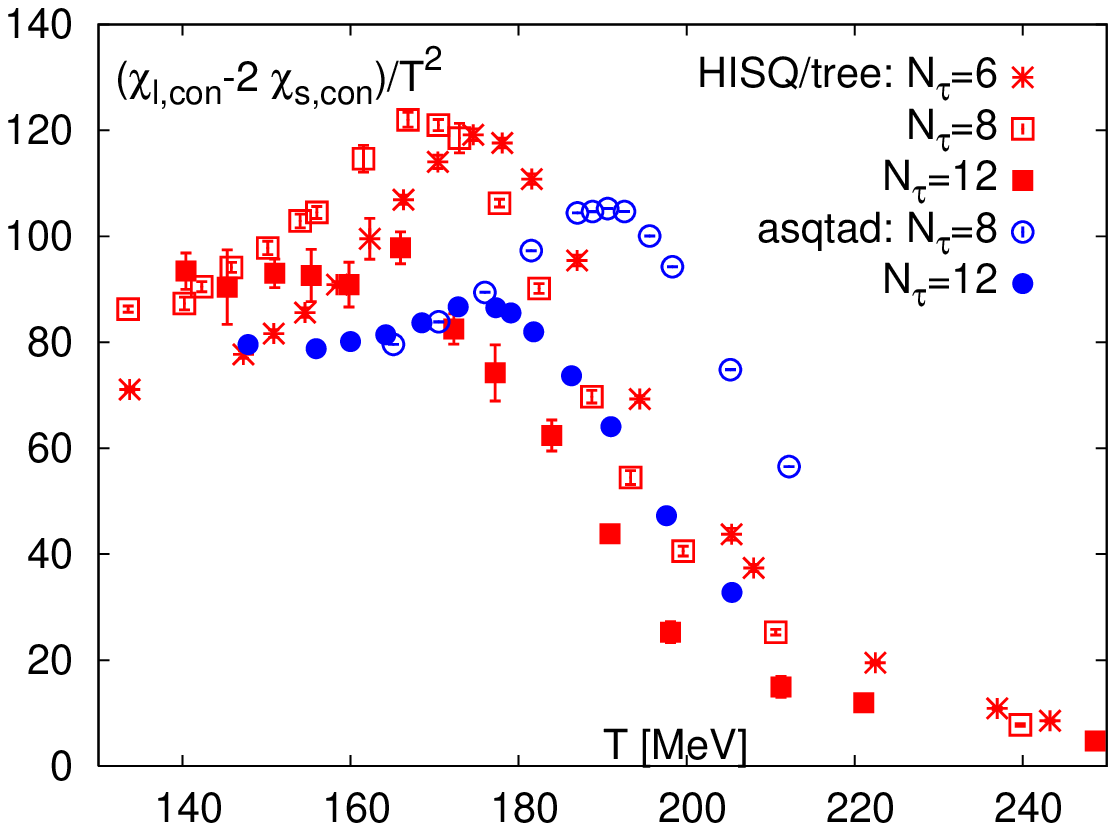}
\caption{The connected part of the chiral susceptibility for the
  asqtad and HISQ/tree actions on the LCP defined by $m_l=0.05m_s$.
  In the right hand figure, we show the difference of the light and
  strange quark connected susceptibilities in which the divergent
  additive artifact cancels.
}
\label{fig:chi_conn}
\end{figure}

In Fig.~\ref{fig:chi_conn}, we show the connected chiral
susceptibility for the asqtad and HISQ/tree actions at fixed $m_l =
0.05m_s$ for different $N_\tau$.  In Sec.~\ref{subsec:condensate}, we
noted the presence of an additive quadratic divergence, proportional
to $m_q/a^2$, in the chiral condensate, which will give rise to a
mass-independent quadratic divergence in the chiral susceptibility.
We find that the absolute value of the data grows with $N_{\tau}$ as 
expected.  Since this divergent contribution is the same for light
and strange susceptibilities, it can be eliminated by constructing
the difference $\chi_{l,con}-2\chi_{s,con}$.  The resulting data are
shown in Fig.~\ref{fig:chi_conn}(right), and we find that the peak
occurs at slightly higher $T$ as compared to the disconnected chiral
susceptibility shown in Fig.~\ref{fig:chi_disc_asqtad_hisq}.  Also, we
find that the height of the peak decreases with $N_\tau$ and the
position of the peak is shifted to smaller temperatures on decreasing
the lattice spacing, which is most evident when comparing the
$N_{\tau}=8$ and $N_{\tau}=12$ asqtad data.


\subsubsection{Renormalized two-flavor chiral susceptibility}

\begin{figure}
\includegraphics[width=8.5cm]{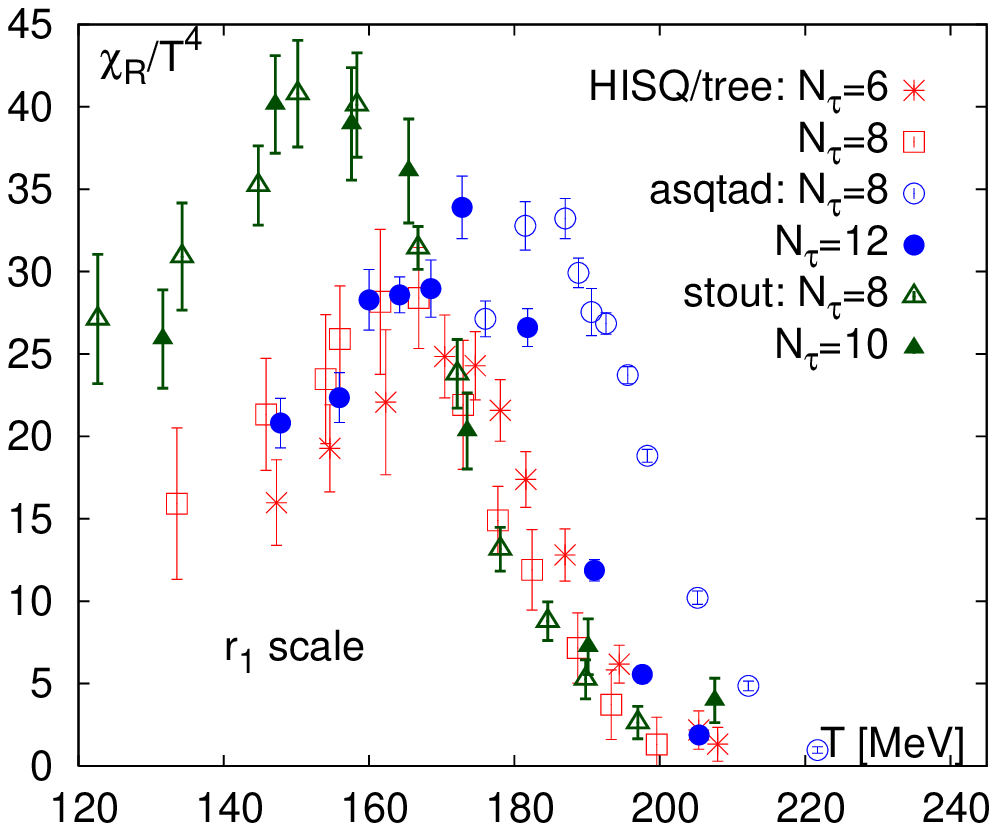}
\includegraphics[width=8.5cm]{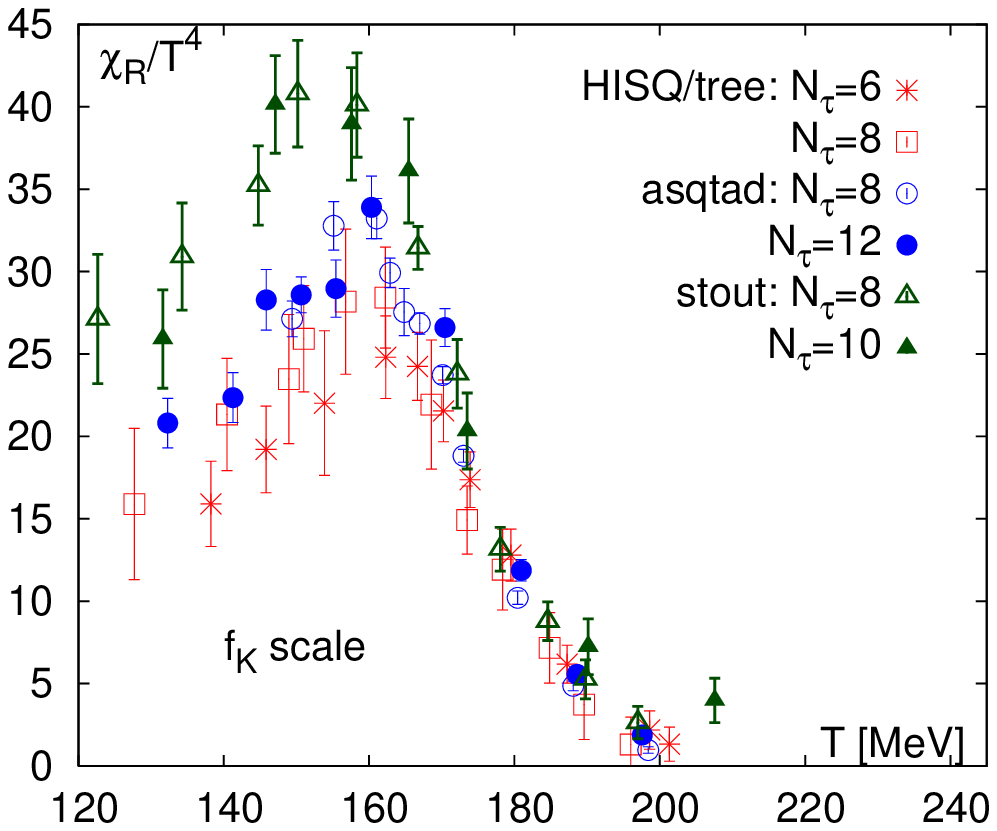}
\caption{The renormalized two-flavor chiral susceptibility $\chi_R$
  for the asqtad and HISQ/tree actions obtained at $m_l=0.05m_s$ and
  compared with the stout action results \cite{Aoki:2006br}.  The
  temperature scale is set using $r_1$ ($f_K$) in the left (right)
  panels.  }
\label{fig:chiR}
\end{figure}

Lastly, we compare our estimates for the two-flavor chiral
susceptibility, defined in Eqs.~(\ref{chi_dis}) and~(\ref{chi_con}), with
results obtained with the stout action \cite{Aoki:2006br}.  To remove
the additive ultraviolet divergence discussed above, we now subtract
the zero temperature rather than the strange quark chiral
susceptibility. Furthermore, to get rid of the multiplicative
renormalization, this combination is multiplied by $m_s^2$, $i.e.$, the
following quantity is considered 
\begin{equation}
\frac{\chi_R(T)}{T^4}=\frac{m_s^2}{T^4} \left( \chi_{m,l}(T)-\chi_{m,l}(T=0) \right).
\label{chiR}
\end{equation}
This construct has the advantage of being renormalization group invariant and, 
unlike the definition in Ref.~\cite{Aoki:2006br}, it does not vanish
in the chiral limit.  In Fig.~\ref{fig:chiR}(left), we show data for
the stout, HISQ/tree and asqtad actions with the scale set by
$r_1$. The stout data have been taken from Ref.~\cite{Aoki:2006br} and
multiplied by $(m_s/m_l)^2 = (27.3)^2$ to conform to
Eq.~(\ref{chiR})~\cite{Aoki:2006br,Aoki:2009sc}.  The HISQ/tree results on
$N_{\tau}=12$ lattices are not shown as the corresponding zero
temperature calculations are not yet complete.  We find that the large
difference between the continuum stout and $N_{\tau}=8$ asqtad results
is significantly reduced by $N_{\tau}=12$.  Second, the cutoff
dependence for the HISQ/tree data is much smaller than for the asqtad
data. A similar behavior was also observed in the case of the
chiral condensate, as discussed in Sec.~\ref{subsec:condensate}.

The cutoff dependence between the HISQ/tree, asqtad and continuum
stout data is significantly reduced when $f_K$ is used to set the
scale as shown in Fig.~\ref{fig:chiR}(right). The change in the HISQ/tree
data is small as the scales determined from $r_1$ and $f_K$ are
similar. The difference in the scales from the two observables is
larger for the asqtad action at these lattice spacings and using $f_K$
mostly shifts the $N_\tau=8$ data. The difference in the position of
the peak between the three actions also decreases, whereas the height
of the peak and the value in the low temperature region show
significant differences between the stout and HISQ/tree (or asqtad)
actions.  Since $\chi_R(T)/T^4$ is a renormalization group invariant
quantity, the only reason for the difference should be the different
values of the light quark mass: $m_l/m_s=0.05$ for the HISQ/tree
estimates versus $0.037$ for the stout data. $O(N)$ scaling, discussed
in Sec.~\ref{sec:observables}, suggests that the peak height should
scale as $h^{1/\delta - 1} \sim m_l^{-0.8}$.  Applying this factor to
the stout data reduces the peak height from $\sim 40$ to $\sim 31.5$.
Similarly, two corrections need to be applied to the asqtad data.
First, a factor of $1/1.44$ to undo the multiplication by a heavier
$m_s^2$ and the second, a multiplication by a factor of $1.2$ to scale
the susceptibility to the common light quark mass. After making these
adjustments to normalize all three data sets to $m_l/m_s=0.05$, we
find that the HISQ/tree, asqtad and stout data are consistent.

\section{\boldmath $O(N)$ scaling and the chiral transition temperature}
\label{sec:scaling}

\subsection{The transition temperature using the p4 action}
\label{ssec:scalingp4}

In this section, we use the universal properties of the chiral
transition to define the transition temperature and its quark mass
dependence for sufficiently small quark masses, as discussed in
Sec.~\ref{sec:observables}.  The scaling analysis of the chiral
condensate leads to a parameter free prediction for the shape and
magnitude of the chiral susceptibility. In the vicinity of the chiral
limit, the peak in the chiral susceptibility corresponds to the peak in
the scaling function $f_{\chi}(z)$ and the quark mass dependence of
the pseudocritical temperature $T_c$ is controlled entirely by the
universal $O(N)$ scaling behavior.  Keeping just the leading term proportional to 
$a_1$ in the regular part, the position of the peak in
$\chi_{m,l}$ is determined from Eq.~(\ref{eq:chiralsuscept}) using
\begin{eqnarray}
\frac{\partial }{\partial T} \left(\frac{m_s^2\ \chi_{m,l}(t,h)}{T^4} \right) &=&
\frac{1}{h_0 t_0 T_c^0} h^{1/\delta -1 -1/\beta\delta} \frac{\rm d}{ {\rm d} z}f_\chi(z)
+ \frac{a_1}{T_c^0} = 0 \; ,
\label{fchi2T}
  \end{eqnarray}
which, for zero scaling violation term, $i.e.$, $a_1=0$, gives the
position of the peak in the scaling function $f_\chi$ at $z = z_p$
(see Sec.~\ref{sec:observables}).  The strange quark mass on the
left hand side is included only for consistency as the derivative is
taken keeping it constant. For small light quark masses, we can expand
$f_\chi(z)$ around $z_p$:
\begin{equation}
f_\chi(z)= f_\chi(z_p)+A_p (z-z_p)^2 \, .
\label{eq:fchiexpand}
\end{equation}
In this approximation,
the location of the maximum in the chiral susceptibility varies as 
\begin{equation}
z = z_p -\frac{a_1 t_0 h_0}{2A_p} h^{1- 1/\delta  + 1/\beta\delta} \; ,
\label{zero2}
\end{equation}
and the variation of the pseudocritical temperature as a function of the quark mass is given by 
\begin{eqnarray}
&
T_c(H) = T_c^0+T_c^0 \frac{z_p}{z_0} H^{1/\beta\delta} \left(
1 -{\displaystyle\frac{a_1} {2A_p z_p z_0 h_0^{- 1/\delta}}} \, H^{1- 1/\delta  + 1/\beta\delta}  \right)
\nonumber \\
&
= T_c^0+ T_c^0 \frac{z_p}{z_0} H^{1/\beta\delta} \left(
1 -{\displaystyle\frac{a_1 t_0^\beta} {2A_p z_p {z_0}^{1-\beta} }} \, H^{1- 1/\delta  + 1/\beta\delta}  \right)
\; .
\label{zero4}
\end{eqnarray}
Recall that $T_c^0$ is the transition temperature in the chiral
limit. Thus, to determine the pseudocritical temperatures $T_c(H)$, we
need to perform fits to the chiral condensate $M_b$, defined in
Eqs.~(\ref{order}) and (\ref{order_scaling}), to determine the
parameters $T_c^0$, $z_0$, $t_0$, $a_0$, $a_1$ and $a_2$ in the scaling and
regular terms. Theoretically, one expects the $O(4)$ Ansatz to
describe the critical behavior in the continuum limit; however, 
$O(2)$ scaling is more appropriate for calculations with
staggered fermions at non-zero values of the cutoff. Fits are,
therefore, performed using both the $O(2)$ and $O(4)$ scaling
functions in all cases.

\begin{figure}[t]
\begin{center}
\includegraphics[width=8cm]{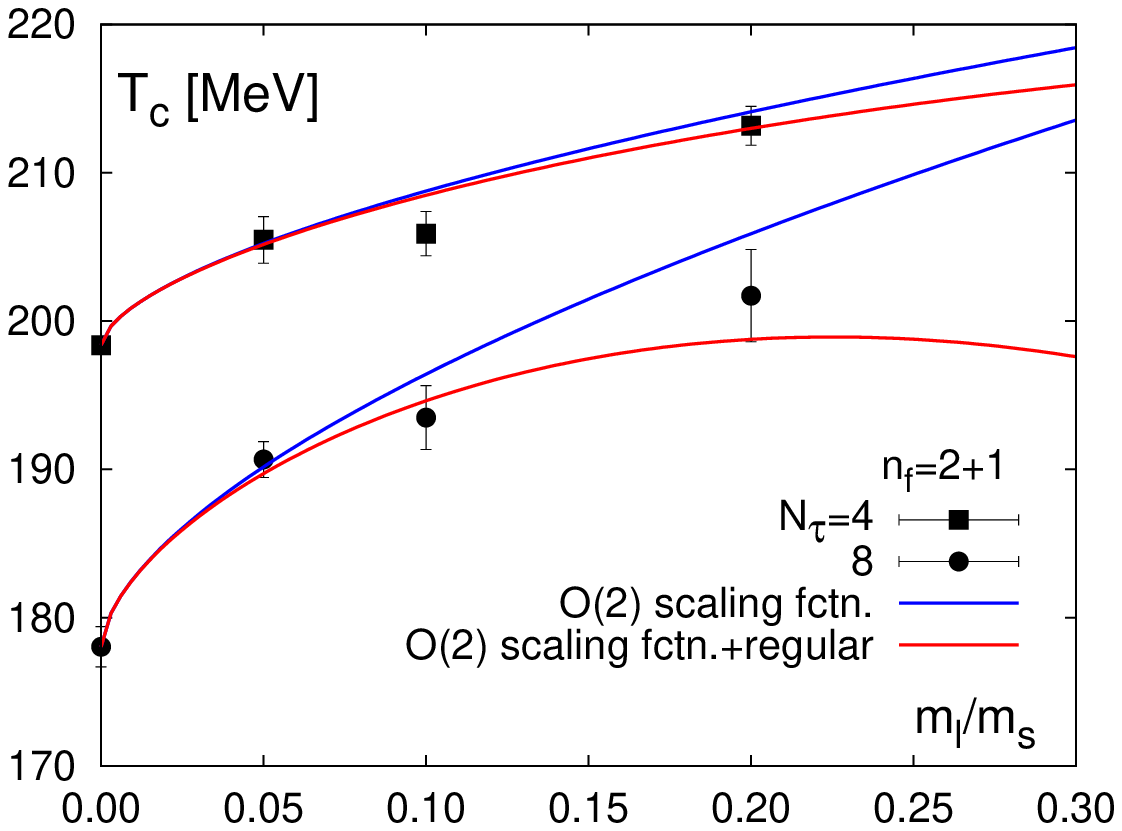}
\includegraphics[width=8cm]{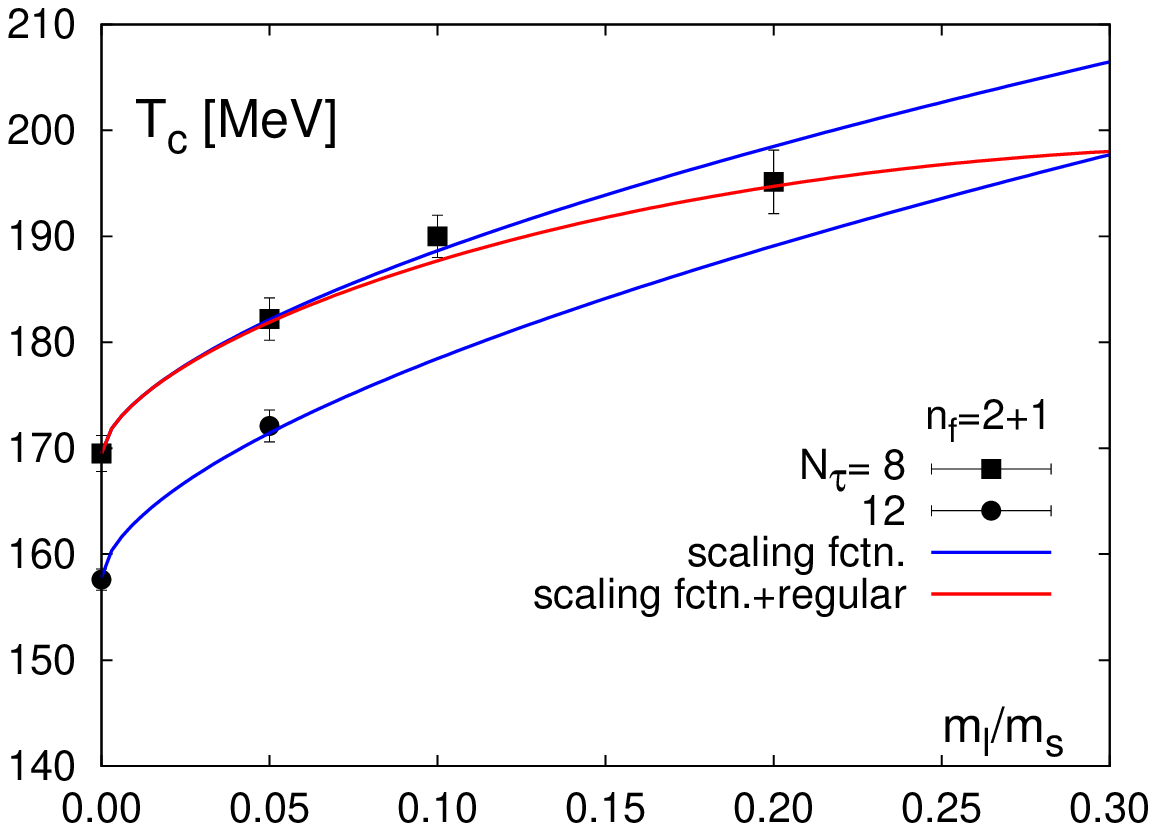}
\end{center}
\caption{
\label{fig:scaling_p4} Estimates of pseudocritical temperature determined
  from the peak in the disconnected susceptibility with the p4 action
  on lattices of temporal extent $N_\tau=4$ and $8$ (left).  Results
  for $m_s/m_l < 0.05$ will be presented elsewhere. Curves show
  results obtained from the $O(2)$ scaling fits to the chiral
  condensate without (blue) and with (red) scaling violation terms
  included. The right hand figure shows the corresponding analysis for
  the asqtad action.  }
\end{figure}

Performing universal $O(N)$ scaling fits to extract reliable estimates
of the chiral transition temperature as a function of the light quark
masses requires making choices for the range of data points
(temperature values) to include and the number of terms needed to
model the regular part.  To test these issues numerically, we use the
extensive data with the p4 action on lattices with $N_\tau = 4$ and
$8$ and light quark masses down to
$m_l/m_s=1/80$~\cite{Ejiri:2009ac,Kaczmarek:2011zz,rbcbi_future}. In
Fig.~\ref{fig:scaling_p4}(left), we compare scaling results for
$T_c(H)$ as a function of the light quark mass, $H=m_l/m_s$, with direct 
determination of the peak in the disconnected part of the chiral
susceptibility obtained using Gaussian and cubic fits to the data in
the peak region.  We find that the scaling fits reproduce the quark
mass dependence of $T_c(H)$ for $m_l/m_s\lsim 1/20$, which covers the
physical point $m_l/m_s\simeq 1/27$. For larger quark masses, scaling
violation terms have to be taken into account.  The range of temperature values 
included in these fits (defining the scaling window) is 15--20 MeV. 

Having demonstrated that the scaling analysis works for the p4 action, 
we apply this approach to the asqtad and HISQ/tree
actions even though the range of quark masses explored is smaller
and the coverage of the transition region (values of $T$ simulated) is
not as dense, so a larger $T$ range will be used.  

\subsection{Scaling analysis for the asqtad and HISQ/tree action}
\label{ssec:scalingHISQ}

The scaling analysis of the HISQ/tree ($N_\tau =6,\ 8$ and $12$
lattices) and asqtad actions ($N_\tau=8$ and $12$ lattices) was
performed in steps due to the limited number of $m_l/m_s$ and
temperature values simulated compared to the p4 action discussed in
Sec.~\ref{ssec:scalingp4}.  In the case of the HISQ/tree action, having
data at $m_l/m_s=1/20$ and $1/40$ on $N_{\tau}=6$ and $N_{\tau}=8$
lattices allowed us to test the range of validity of the scaling fits
in the region bracketing the physical light quark
mass. The scaling Ansatz included both the
singular and regular parts defined in Eqs.~(\ref{eq:freg}) 
and~(\ref{fchi2T}) in terms of the six parameters $T_c^0$, $z_0$, $t_0$,
$a_0$, $a_1$ and $a_2$. We find that a best fit to the
$m_l/m_s=1/40$ data for $M_b$ also fits the $m_l/m_s=1/20$ data; in
addition, the chiral susceptibility derived from this fit matches the
measurements at $m_l/m_s=1/20$.  We, therefore, conclude that the scaling 
extends to $m_l/m_s=1/20$. 

To get the value of $T_c$, we then carried out a simultaneous fit to
the chiral condensate at the two values of $m_l/m_s$ and its
derivative, the total chiral susceptibility, at $m_l/m_s=1/20$. (The
calculation of the connected part of the susceptibility at
$m_l/m_s=1/40$ is not complete, so we could not include these data.)
The range of temperature values selected for the three data sets was
independently adjusted to minimize the $\chi^2/{\rm dof}$ and, at the
same time, include as many points as possible in the region of the
peak in the susceptibility. We also varied the relative weight
assigned to points along the two LCPs and for our final fits decreased
the weight given to the heavier $m_l/m_s=1/20$ points by a factor of
sixteen in the calculation of the $\chi^2$. (This was done because the
scaling Ansatz with a truncated regular piece is expected to get
progressively worse as the mass increases, whereas the statistical
errors in $M_b$ are roughly a factor of two smaller at the heavier
mass.) In all these fits we find that while the height of the peak in
the susceptibility is sensitive to what points are included in the
fits and the relative weighting of the points along the two LCP, the
location of the peak does not vary by more than $0.3$ MeV for any
reasonable set of choices.  Repeating the analysis with just the
$m_l/m_s=1/40$ or the $m_l/m_s=1/20$ data did not change the estimates
of $T_c$ significantly. We therefore conclude that $m_l/m_s=1/20$ lies
within the range of validity of our scaling Ansatz.  Confirming that
$m_l/m_s=1/20$ lies within the scaling window is important for our
analysis because simulations on the $N_\tau=12$ lattices have been done
only at this LCP for both the HISQ/tree and asqtad actions.

\begin{figure}[t]
\begin{center}
\includegraphics[width=8cm]{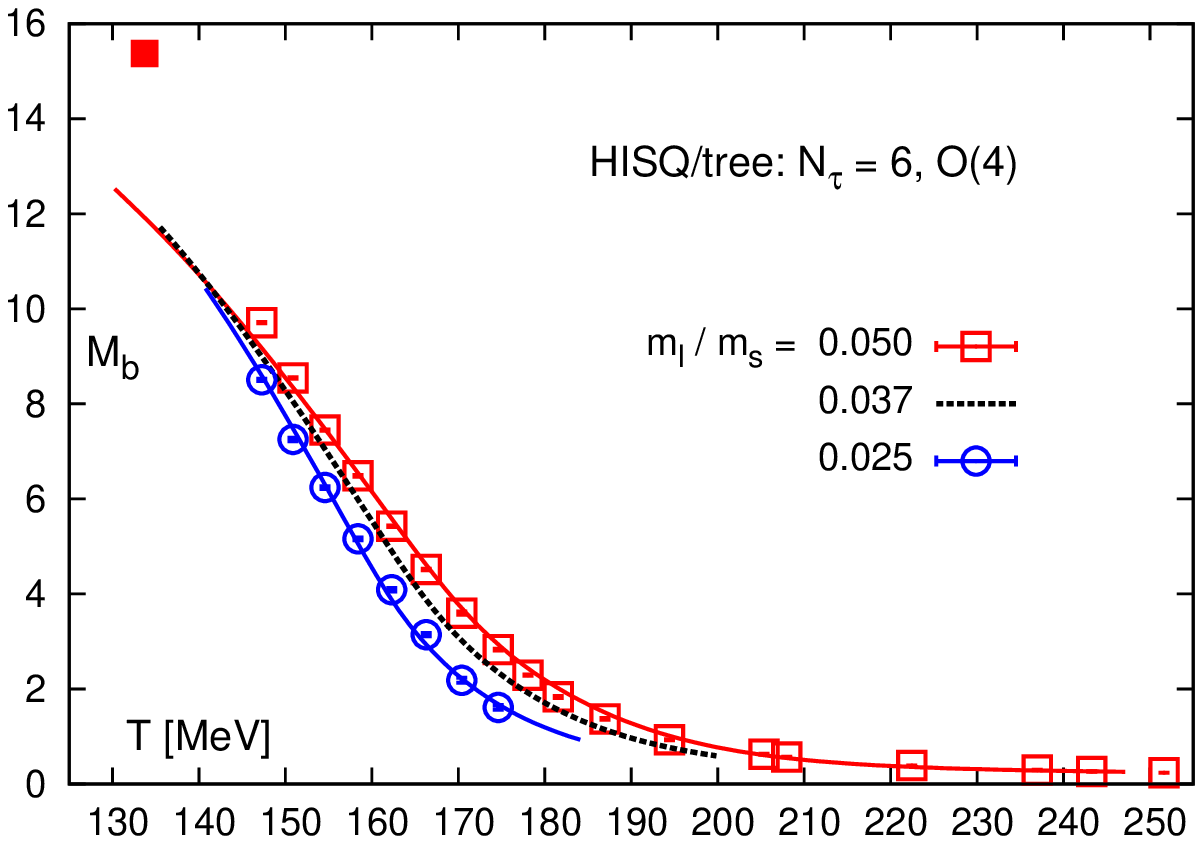}
\includegraphics[width=8cm]{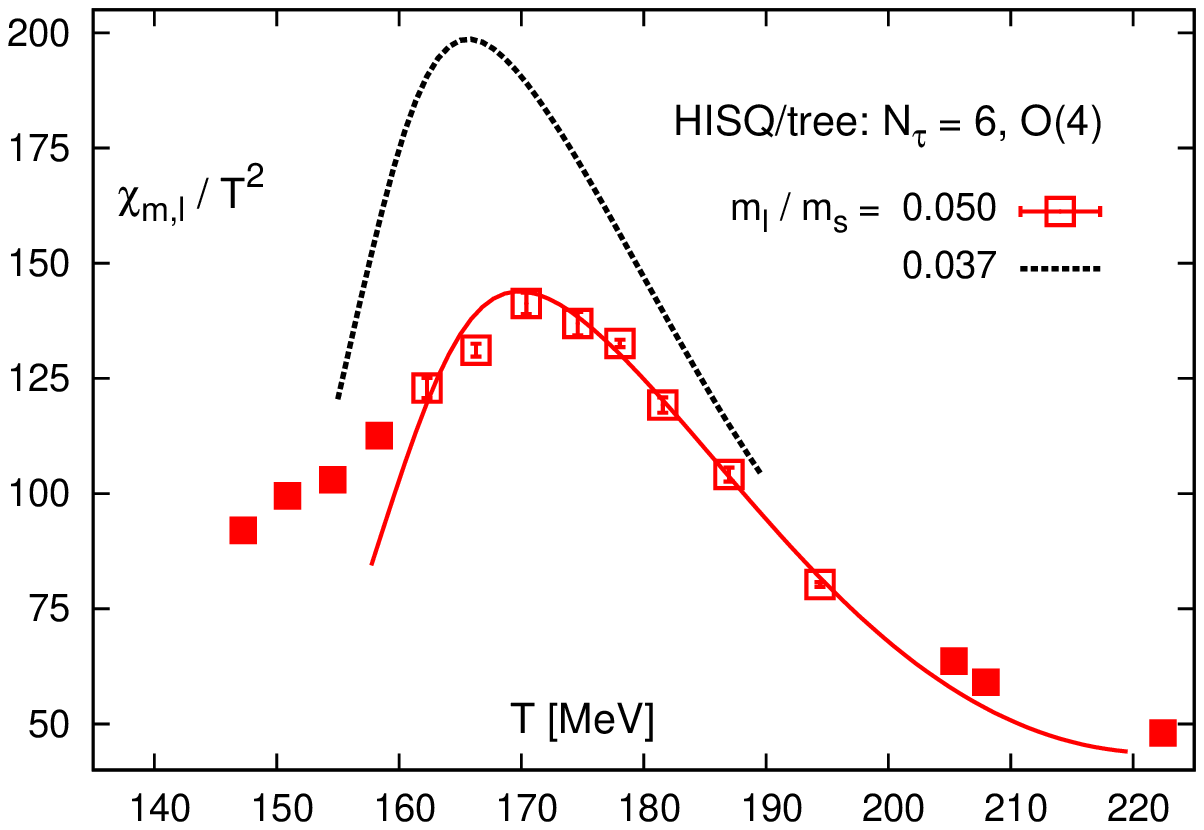}
\end{center}
\caption{
\label{fig:sfit_hisq6_O2} 
  Scaling fits and data for the chiral condensate $M_b$ calculated
  with the HISQ/tree action on lattices with temporal extent
  $N_\tau=6$ (left) and the chiral susceptibility
  $\chi_{m,l}$(right). The data for $M_b$ at $m_l/m_s=0.025$ and for
  $M_b$ and $\chi_{m,l}$ at $m_l/m_s=0.05$ are fit simultaneously
  using the $O(4)$ scaling Ansatz. The fits using the $O(2)$ Ansatz
  are similar. The points used in the scaling fits are plotted using 
  open symbols.  The dotted lines give the data scaled
  to the physical quark masses.}
\end{figure}

\begin{figure}[b]
\begin{center}
\includegraphics[width=8cm]{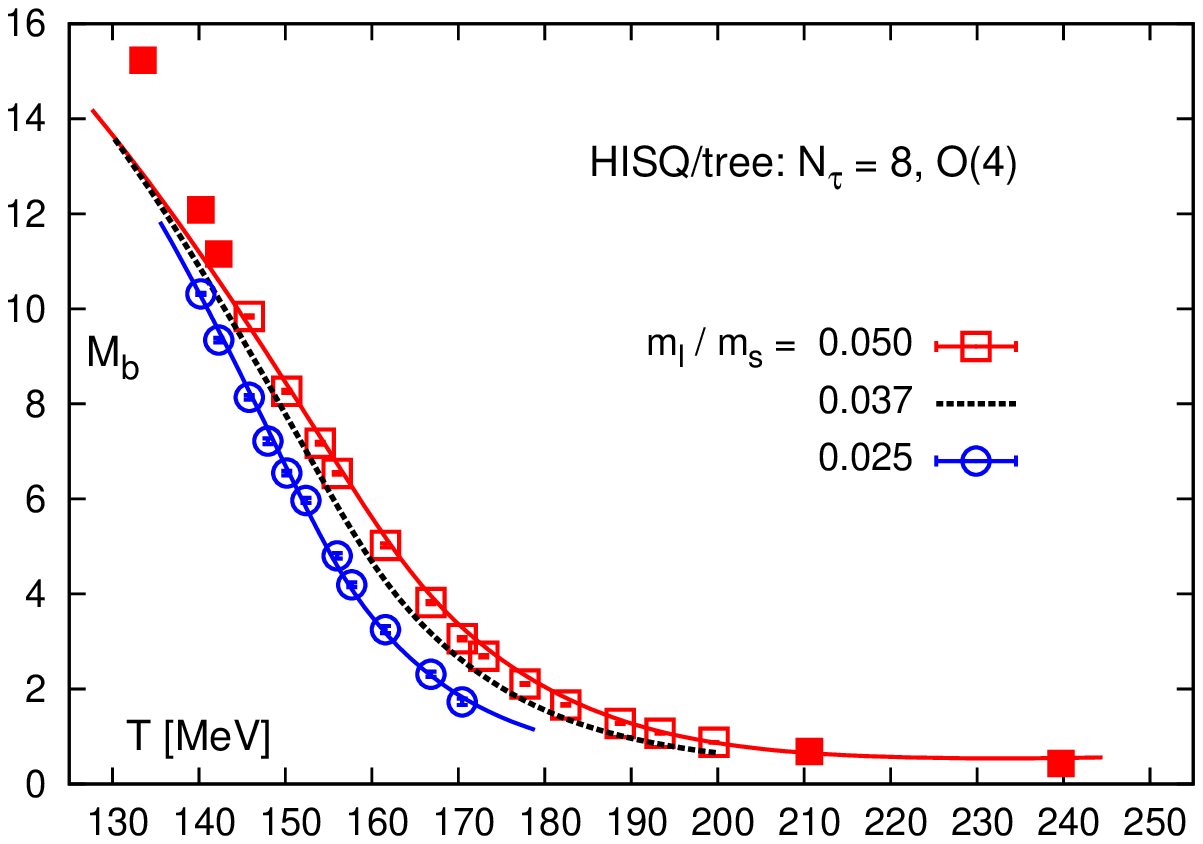}
\includegraphics[width=8cm]{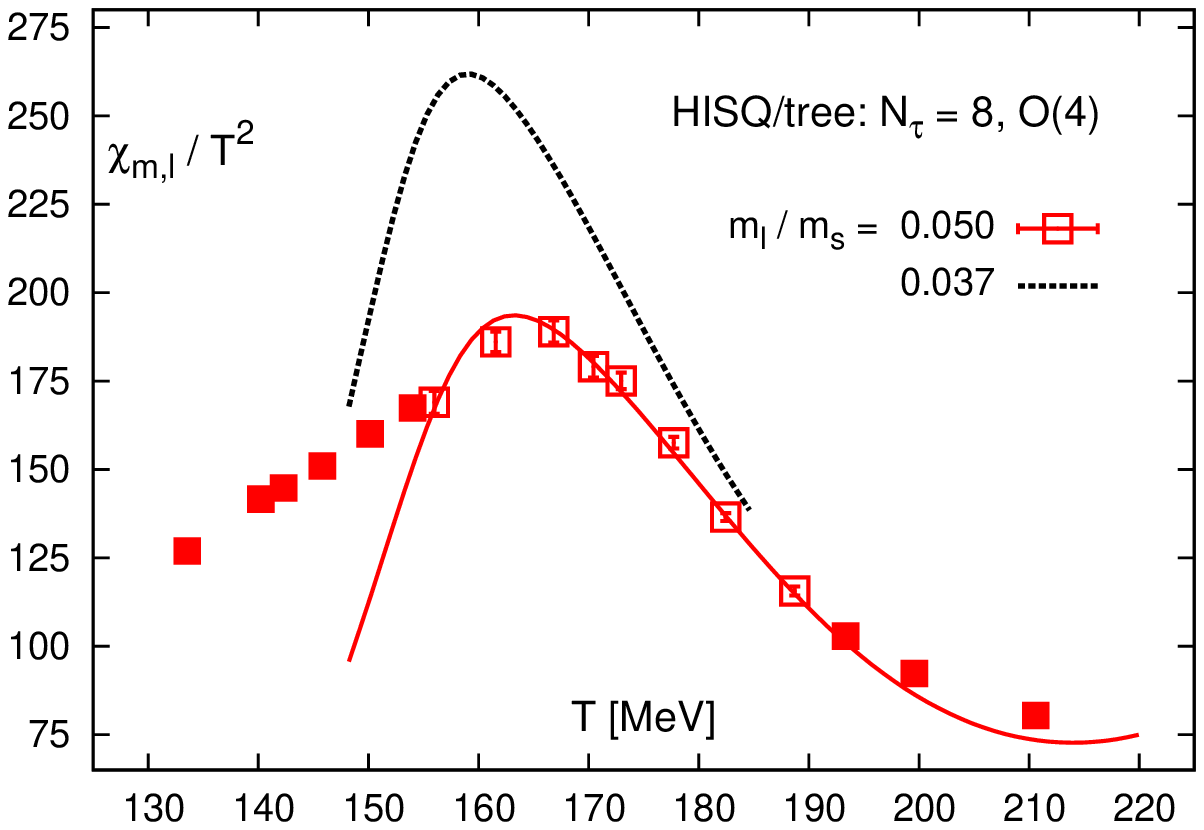}
\end{center}
\caption{
\label{fig:sfit_hisq8_O4} 
  Scaling fits and data for the chiral condensate $M_b$ calculated
  with the HISQ/tree action on lattices with temporal extent
  $N_\tau=8$ (left) and the chiral susceptibility
  $\chi_{m,l}$(right). The data for $M_b$ at $m_l/m_s=0.025$ and for
  $M_b$ and $\chi_{m,l}$ at $m_l/m_s=0.05$ are fit simultaneously
  using the $O(2)$ scaling Ansatz. The fits using the $O(4)$ Ansatz
  are similar.  The points used in the scaling fits are plotted using 
  open symbols.  The dotted lines give the data scaled to the physical
  quark masses.}
\end{figure}

\begin{figure}[t]
\begin{center}
\includegraphics[width=8cm]{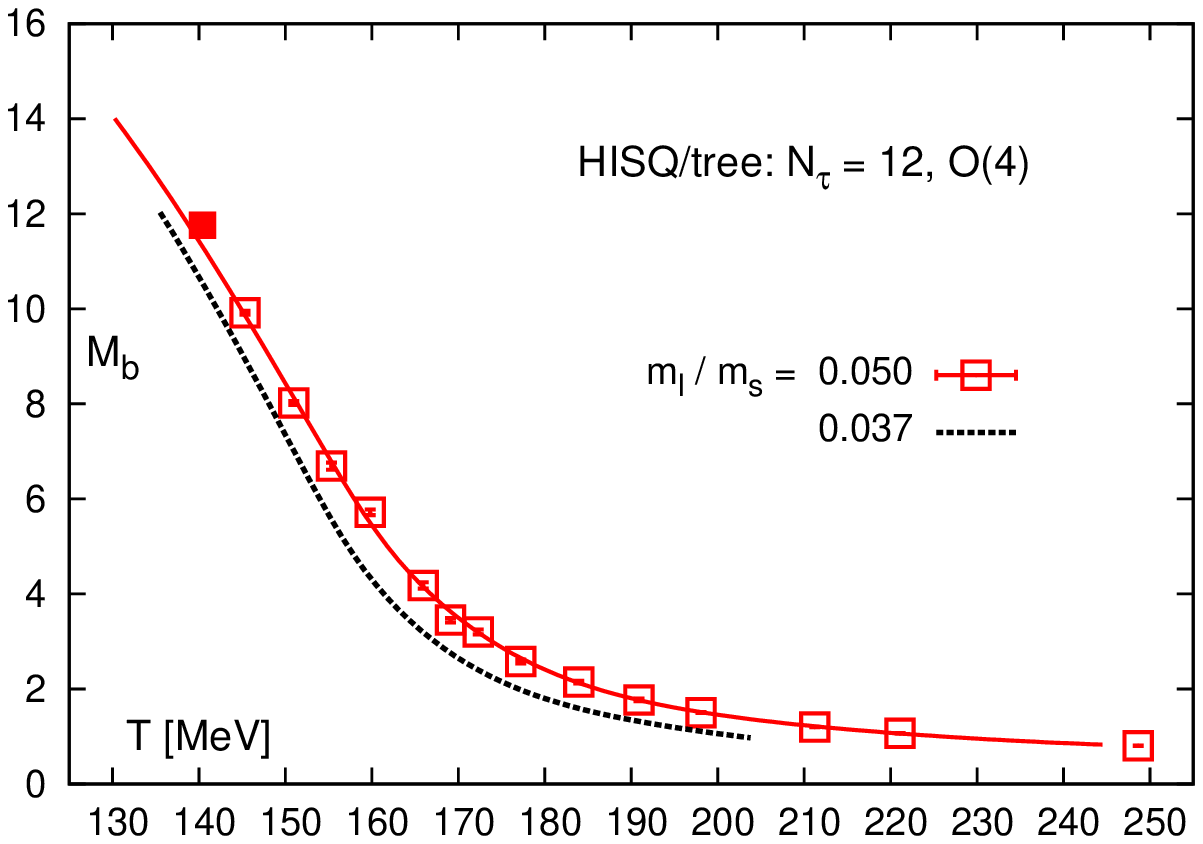}
\includegraphics[width=8cm]{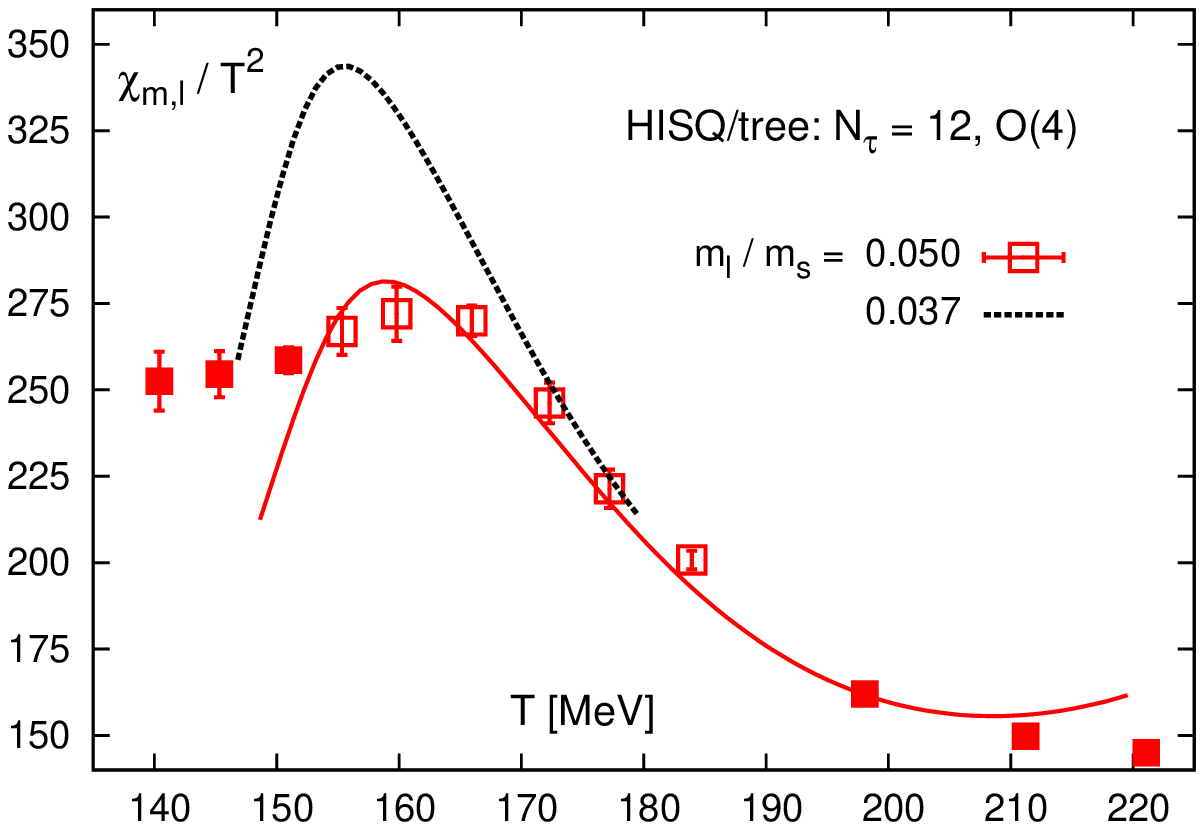}
\end{center}
\caption{
\label{fig:sfit_hisq12_O4} 
  Scaling fits and data for the chiral condensate $M_b$ calculated
  with the HISQ/tree action on lattices with temporal extent
  $N_\tau=12$ (left) and the chiral susceptibility
  $\chi_{m,l}$(right). The data for $M_b$ and $\chi_{m,l}$ at
  $m_l/m_s=0.05$ are fit simultaneously using the $O(4)$ scaling
  Ansatz. The fits using the $O(2)$ Ansatz are similar. The points
  used in the scaling fits are plotted using open symbols.  The dotted
  lines give the data scaled to the physical quark masses.}
\end{figure}

\begin{figure}[t]
\begin{center}
\includegraphics[width=8cm]{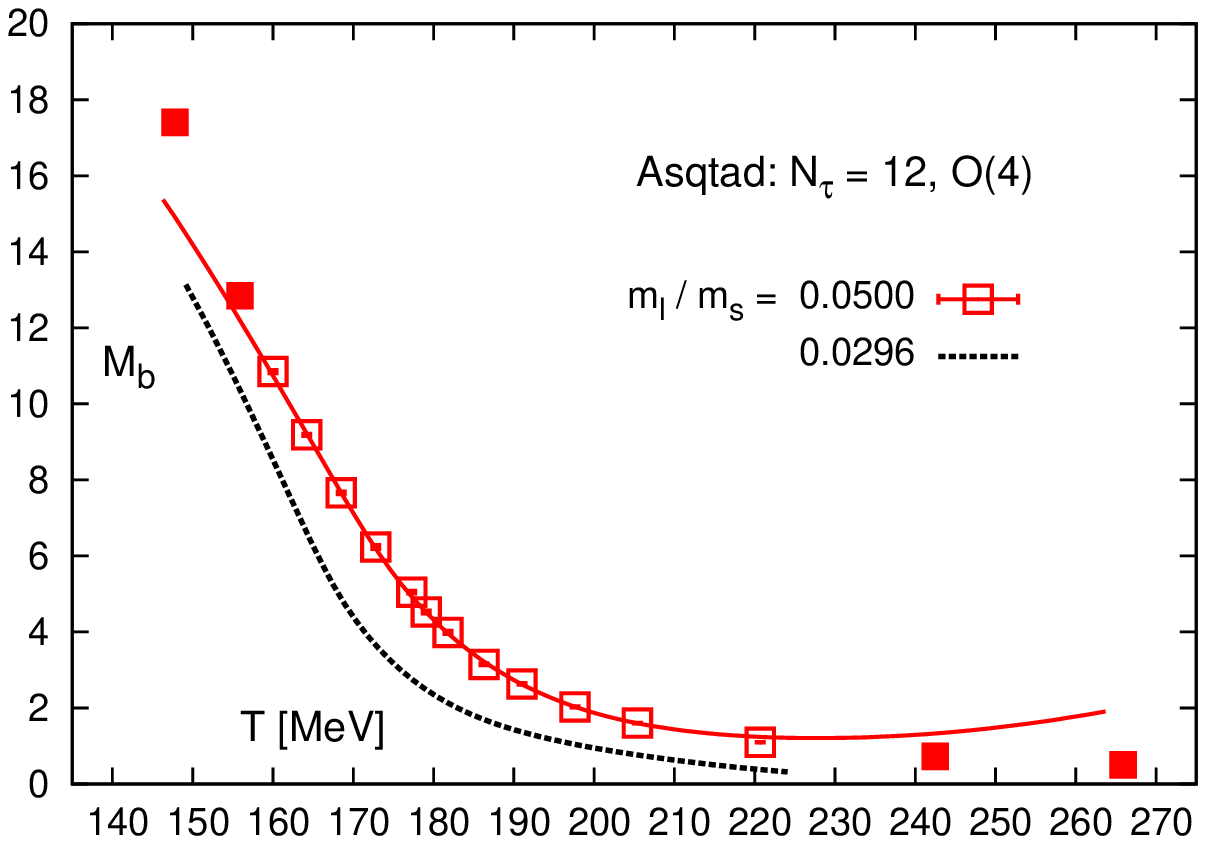}
\includegraphics[width=8cm]{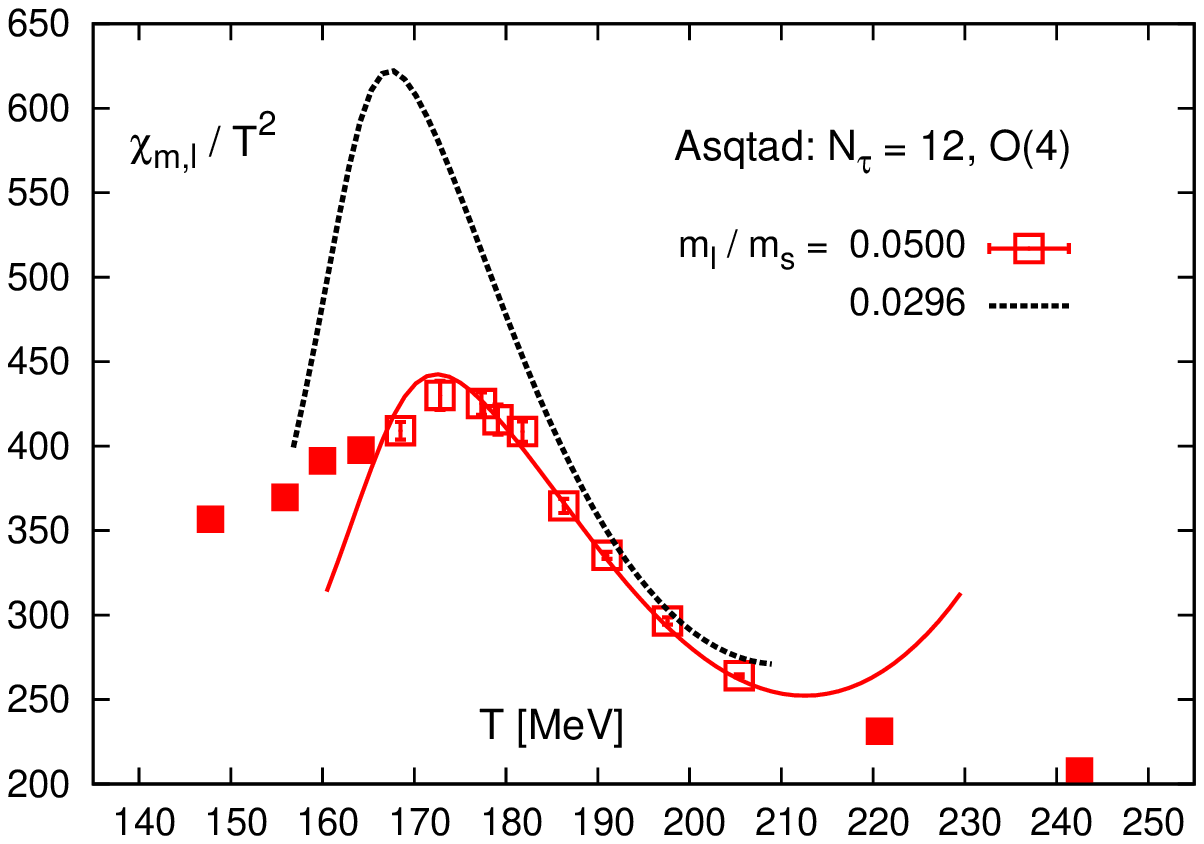}
\end{center}
\caption{
\label{fig:sfit_asqtad12_O4} 
  Scaling fits and data for the chiral condensate $M_b$ calculated
  with the asqtad action on lattices with temporal extent $N_\tau=12$
  (left) and the chiral susceptibility $\chi_{m,l}$(right). The data
  for $M_b$ and $\chi_{m,l}$ at $m_l/m_s=0.05$ are fit simultaneously
  using the $O(4)$ scaling Ansatz. The fits using the $O(2)$ Ansatz
  are similar. The points used in the scaling fits are plotted using 
  open symbols.  The dotted lines give the data scaled
  to the physical quark masses.}
\end{figure}

The most challenging part of determining the best fit was including a
sufficient number of points below the location of the peak in
$\chi_{m,l}$. This is evident from
Figs.~\ref{fig:sfit_hisq6_O2}--\ref{fig:sfit_asqtad12_O4} which show
that the fits rapidly deviate from the $\chi_{m,l}$ data below
the location of the peak.  In most cases, only two points below the
peak could be included in the fits.

The $\chi^2/{\rm dof}$ of these fits are not good. For the $O(4)$ fits
they are $96/12$ and $30/12$ for the asqtad data on $N_\tau=8$ and
$12$ lattices; and $430/28,\ 125/26$ and $70/14$ for the $N_\tau=6,
\ 8$ and $12$ HISQ/tree data, respectively. (The $\chi^2/{\rm dof}$
for the $O(2)$ fits is about 20\% larger for the $N_\tau=6$ and $12$
HISQ/tree data and comparable for the others.)  These large
$\chi^2/{\rm dof}$ reflect the fact that the statistical errors in
$M_b$ are small and do not include the systematic errors due to
fluctuations in points along the LCP coming from less than perfect
tuning of $m_s$ (or $T$), and partly because we have used a truncated
form for the regular part of the free energy. We attempted various
truncations of the regular part and found that the variation in the
location of the peak is negligible and insensitive to the number of
parameters included in the fits.  Note that the location of the peak
and its scaling with $m_l/m_s$ are the only quantities we extract from
the fits.  Overall, we find that the fits using the $O(4)$ scaling
Ansatz are more stable with respect to variation of the fit range.
We, therefore, use estimates from the $O(4)$ fits to obtain our final
results for $T_c$ which are about $2$ MeV lower than those with the
$O(2)$ Ansatz. Fits to the HISQ/tree data are shown in
Figs.~\ref{fig:sfit_hisq6_O2}, \ref{fig:sfit_hisq8_O4}
and~\ref{fig:sfit_hisq12_O4} for $N_\tau=6$, $8$ and $12$,
respectively.

The above analysis using simultaneous fits to data for the order
parameter $M_b$ and the total chiral susceptibility $\chi$ was
repeated for the asqtad action on $N_\tau=8$ lattices with the two LCP
at $m_l/m_s=1/20$ and $1/10$. We failed to find a fit that reproduced
the data for the chiral susceptibility at the heavier mass.  We
conclude that $m_l/m_s=1/10$ is not within the scaling region and
restrict the simultaneous fit to just data with $m_l/m_s=1/20$.  Fits
to the $N_\tau=12$ lattices for both the asqtad and HISQ/tree lattices
are made at the single value of $m_l/m_s=1/20$ where simulations were
carried out.  Our best fits to the $N_\tau=12$ asqtad data are
shown in Fig.~\ref{fig:sfit_asqtad12_O4} for the $O(4)$ Ansatz.

The final value of $T_c$ is calculated by finding the location of the
peak in the susceptibility. Errors on $T_c$ for each action and
$N_\tau$ value are estimated by carrying out the entire analysis for
400 synthetic samples. Each point in these samples is taken from a
Gaussian distribution with mean and standard deviation given by the
central value and the quoted errors on the data points. Once the
scaling function is determined, we can extract $T_c$ at any desired
value of $m_l/m_s$. Our final results at $m_l/m_s=1/20$ and at the
physical quark masses (defined as $m_l/m_s = 0.037$ with $m_s$ tuned
to its physical value) are summarized in Table~\ref{tab:tcall}. For 
the asqtad data, we extrapolate to $m_l/m_s = 0.0296$ to
partially correct for the fact that $m_s$ on the LCP is about $20\%$
heavier than the physical value. Changing the extrapolation point provides
estimates at the correct physical light quark mass but does not
correct for the heavier strange quark mass, so we expect the asqtad
results to overestimate $T_c$. A final point, these estimates are also
consistent with the location of the peak in $\chi_{l,disc}$ as shown
in Figs.~\ref{fig:chi_disc} and~\ref{fig:chi_disc_asqtad_hisq} and
discussed in Sec.~\ref{ssec:chidisc}.

\begin{figure}
\includegraphics[width=8cm]{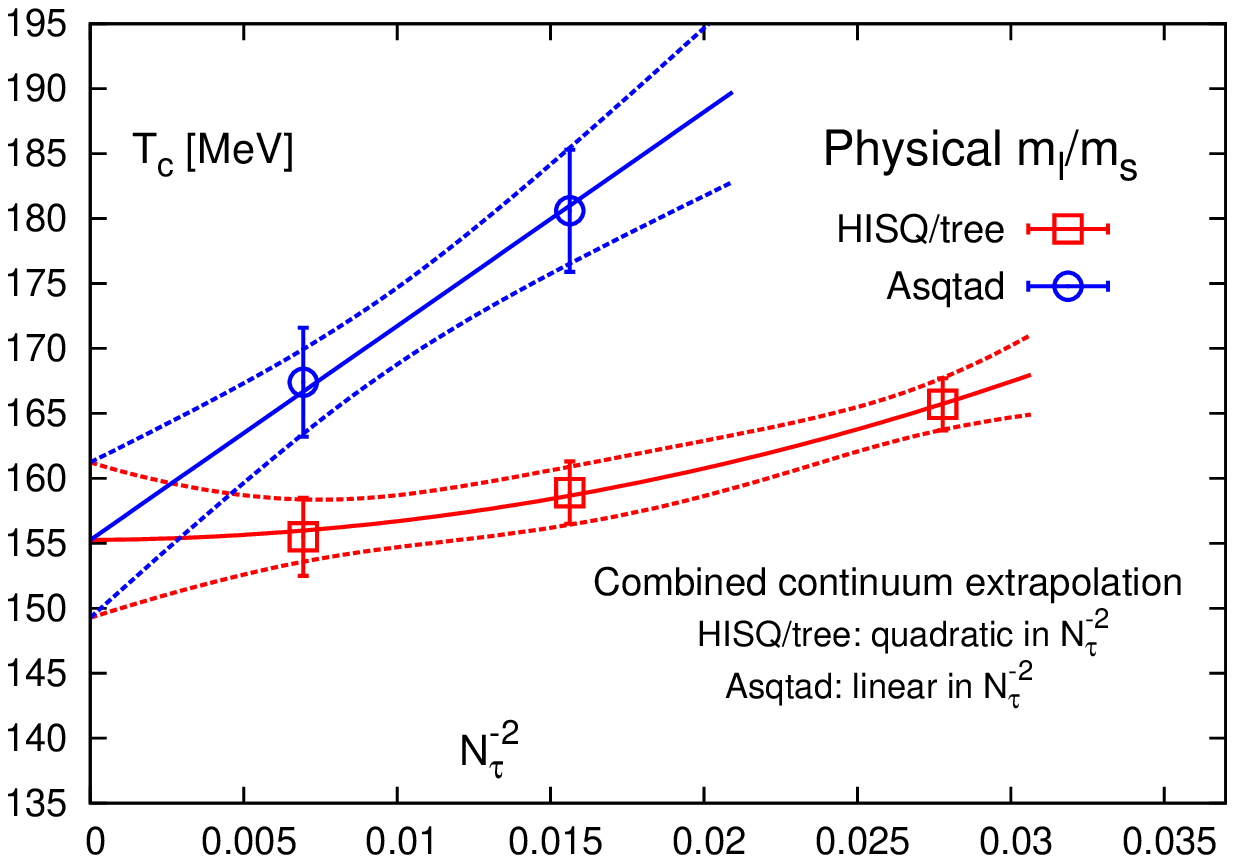}
\includegraphics[width=8cm]{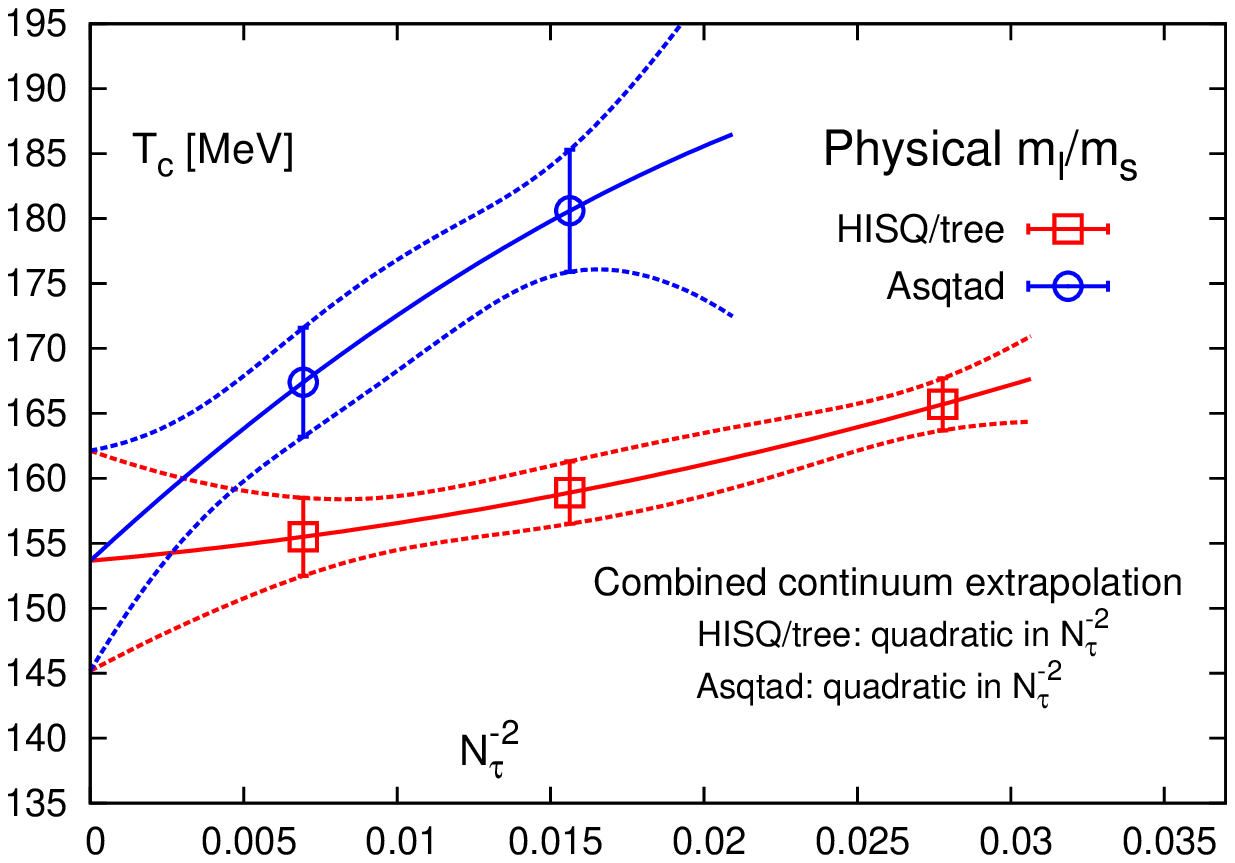}
\caption{ The data and the continuum extrapolation of the transition
  temperature obtained from the $O(4)$ scaling analysis for $m_l/m_s$
  extrapolated to the physical value. The left figure shows a combined 
  fit using a linear extrapolation to the asqtad data and a quadratic to the HISQ/tree data. 
  The right figure shows a combined quadratic fit to both data sets. }
\label{fig:tc_extra} 
\end{figure}

The last step in the determination of $T_c$ is to extraplolate estimates at the
physical value of $m_l/m_s$ obtained at different $N_\tau$ to the
continuum limit. With two asqtad data points and three HISQ/tree
points, we explored all possible combinations of linear and quadratic fits to
the data for each action and combined fits to the asqtad and
HISQ/tree data in which the intercept (continuum value) is constrained
to be the same for the two actions.  The results of individual 
extrapolations linear in $1/N_{\tau}^2$ for the asqtad data and quadratic for 
the HISQ/tree data are given in rows marked $\infty$ in~Table.~\ref{tab:tcall}. 
The results of a combined quadratic fit are given in the final row marked 
``$\infty$ (asqtad+HISQ/tree)''. 
Two examples of combined fits are shown in Fig.~\ref{fig:tc_extra}.
In Fig.~\ref{fig:tc_extra}(left) we show the result of a combined fit
linear in the asqtad data and quadratic in the HISQ/tree data, while the 
combined quadratic fit to both data sets is shown in
Fig.~\ref{fig:tc_extra}(right). 

We take the result, $T=154(8)$ MeV, of the quadratic fit with the
$O(4)$ Ansatz to the HISQ/tree data as our best estimate. These fits
are less sensitive to variations in the range of temperature selected.
The estimate from the $O(2)$ fits is about $2$ MeV higher as shown in
Table~\ref{tab:tcall}.  Since the quadratic fit to just the HISQ/tree
data is identical to the fit shown in Fig.~\ref{fig:tc_extra}(right)
we do not show it separately.  For both $O(2)$ and $O(4)$ Ans\"atze, we
find that the range of variation of the central value of the various
fits is about $5$ MeV, so $8$ MeV is a conservative estimate of the
combined statistical and systematic error and includes the full range
of variation.  Note
that the error estimates, as expected, increase significantly for
quadratic fits compared to linear fits, because in this case there are
as many parameters as data points. In all these fits $\chi^2$ is much
less than one. Furthermore, when the error bars on individual points
do not represent normally distributed statistical errors and there are
one or zero degrees of freedom, chi-square is not a useful guide for
selecting the best fit.  

In determining this final estimate, we have mostly used the asqtad
data as confirmatory for two reasons. First, the slope of the fits to
the asqtad data is 2--2.5 times that in the HISQ/tree fits and the
undetermined quadratic term may be large.  Note that the large slope confirms
our discussion in Sec.~\ref{sec:parameters} that the discretization
errors in the asqtad formulation are larger. Second, the LCP defined
by the strange quark mass is $\approx 20\%$ heavier than the physical
value, so the asqtad action data overestimate $T_c$.

In addition to the above error estimate, there is a $1$ MeV
uncertainty in the setting of the temperature scale as discussed in
Sec.~\ref{sec:parameters} which will shift all estimates. We therefore
quote it as an independent error estimate. Thus, our result
after continuum extrapolation, and at the physical light quark masses,
$m_l/m_s=1/27$, is $T_c=(154  \pm 8 \pm 1){\rm MeV}$.
To obtain an overall single error estimate we 
add the uncertainty in the scale coming from $r_1$ 
to the statistical and systematic errors. Thus, our final 
estimate is 
\begin{equation}
T_c=(154  \pm 9){\rm MeV}  \, .
\label{tc_final}
\end{equation}

\begin{table}[t]
\begin{center}
\vspace{0.3cm}
\begin{tabular}{|l|c|c|c|c|c|}
\hline
$N_{\tau}$ & $T_c,~O(4)$ & $T_c ,~O(2)$  & $T_c,~O(4)$ & $T_c,~O(2)$  \\
           & $m_l/m_s=1/27$ & $m_l/m_s=1/27$ & $m_l/m_s=1/20$   & $m_l/m_s=1/20$  \\
\hline
8 (asqtad)                    & 180.6(4.7) & 181.9(3.2) & 184.5(4.4) & 185.0(3.4) \\
12                            & 167.4(4.2) & 170.1(3.1) & 172.5(3.9) & 174.0(2.9) \\
$\infty$                      & 156.8(8.4) & 160.7(6.1) & 162.9(7.9) & 165.2(5.9) \\
\hline
6 (HISQ/tree)                 & 165.7(2.0) & 167.7(2.3) & 169.8(2.0) & 171.3(2.1) \\
8                             & 158.9(2.4) & 160.6(1.6) & 163.1(2.7) & 164.1(1.8) \\
12                            & 155.5(3.0) & 157.4(3.0) & 158.9(2.7) & 160.9(2.1) \\
$\infty$                      & 153.7(8.4) & 156.0(7.8) & 155.9(8.2) & 159.5(6.1) \\
\hline
$\infty$ (asqtad + HISQ/tree) & 153.7(8)   & 156.0(8)   & 155.9(8)   & 159.5(6) \\
\hline
\end{tabular}
\end{center}
\caption{Pseudocritical temperature $T_c$ determined from $O(2)$ and
  $O(4)$ scaling fits to the chiral condensate and the chiral
  susceptibility. The results are shown at $m_l/m_s=0.05$ at which
  simulations have been performed for all three values of $N_\tau$ and
  at the physical value of $m_l/m_s$.  The temperature scale used in
  the fits is set using $r_1$.  Rows labeled $\infty$ give results
  after a linear extrapolation to the continuum for the asqtad data
  and quadratic for the HISQ data. The last row gives the results of a
  combined fit using an extrapolation quadratic in $1/N_\tau^2$, which
  coincides with the extrapolation of just the HISQ/tree data.}
\label{tab:tcall}
\end{table}

\subsection{Comparison with previous results}

We compare the result in Eq.~(\ref{tc_final}) with three previous $2+1$
flavor studies that also extrapolated $T_c$ data to the continuum
limit and to the physical light quark mass.  It should be emphasized
that in our previous work~\cite{Bazavov:2009zn} such an extrapolation was not
carried out.

The 2004 study by the MILC collaboration used the asqtad
action~\cite{Bernard:2004je}. They extrapolated $T_c$ defined as the
peak position in the total chiral susceptibility using an expression
that incorporated the $O(4)$ critical exponent. They found $T_c =
169(12)(4)$ in the chiral limit, which is just consistent, within
errors, with our current result.  Note, however, that the present data
are more extensive and the scaling analysis is more comprehensive.

\begin{figure}
\includegraphics[width=8cm]{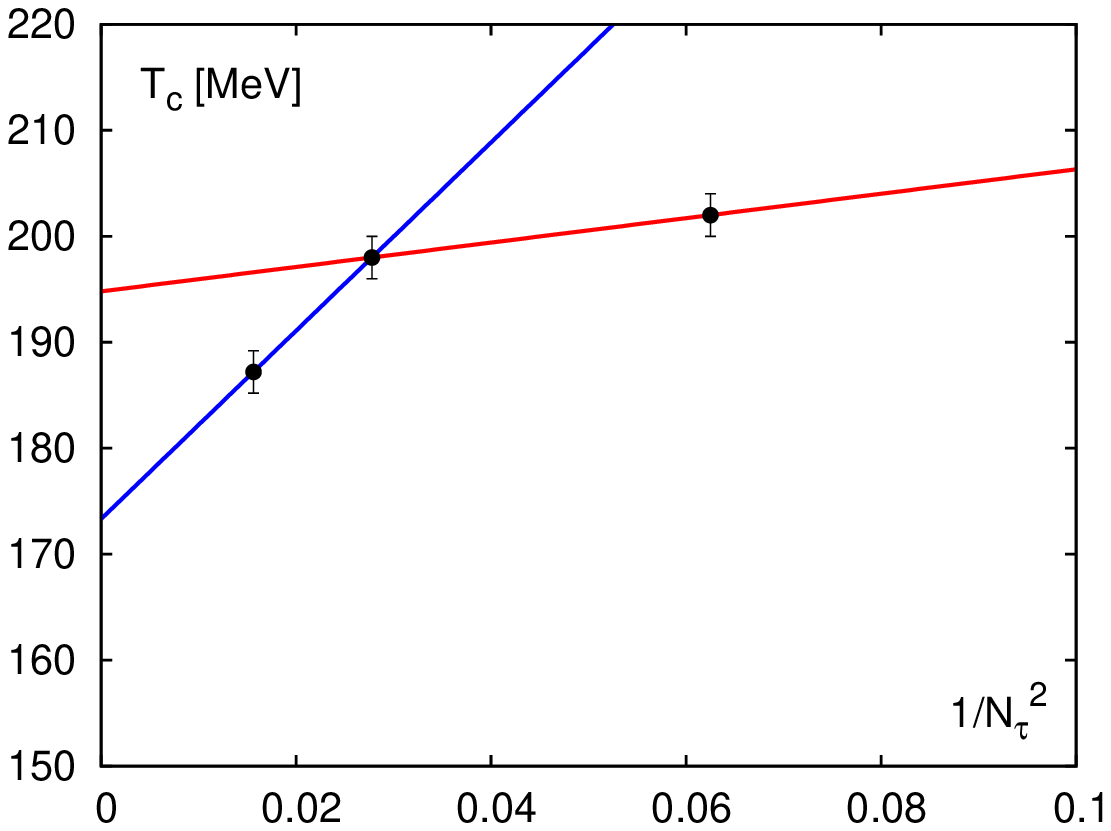}
\caption{Linear extrapolations to the continuum limit of $T_c$
  data for the p4 action. The fit to
  the $N_\tau=4$ and $6$ data is shown in red, while that to the
  $N_\tau=6$ and $8$ points is in blue. The two fits illustrate 
  the change in the slope as $N_\tau$ is increased.
}
\label{fig:p4curvature} 
\end{figure}

The RBC-Bielefeld collaboration studied the p4 action on $N_\tau = 4$
and 6 lattices with several values of the light quark
masses~\cite{Cheng:2006qk}.  The result of a combined extrapolation in
the quark mass and $1/N_\tau^2$ gave $T_c = 192(7)(4)$ MeV.  This
value is significantly higher than the one given in
Eq.~(\ref{tc_final}).  Based on analyses done subsequently, and the
new data on $N_\tau=8$ lattices presented here, we find that this
discrepancy is due to an underestimate of the slope of the linear fit,
$T_c$ $vs.$ $1/N_\tau^2$. The change in the slope between $N_\tau = 4$
and 6 data and $N_\tau = 6$ and 8 data is shown in
Fig.~\ref{fig:p4curvature}. 
To perform a continuum extrapolation of $T_c$ obtained with the p4 action
requires new calculations on lattices with $N_{\tau} \ge 12$. Having
demonstrated that the discretization errors are similar for the p4 and
asqtad actions and given the consistency between results obtained with
the HISQ/tree, asqtad and stout actions, we do not intend to pursue
further calculations with the p4 action.

Lastly, the Wuppertal-Budapest collaboration has carried out a
continuum extrapolation using the stout action on $N_\tau = 6$, 8, and
10 lattices, and more recently including $N_\tau=12$ and 16
lattices~\cite{Aoki:2006br,Aoki:2009sc,Borsanyi:2010bp}. They work
directly at $m_l/m_s = 0.037$ and perform a linear extrapolation in
$1/N_\tau^2$ of their data to obtain continuum estimates. From the
position of the peak in the renormalized chiral susceptibility they
extract $T_c = 147(2)(3)$ MeV.  This value is approximately $ 1 \ \sigma$ lower
than our result.
They also report
higher transition temperatures of $T_c=157(3)(3)$ and $155(3)(3)$
derived from inflection points in $\Delta_{l,s}$ and $\langle \bar
\psi \psi \rangle_R$ respectively.  As discussed in
Sec.~\ref{sec:observables} and confirmed by the significant difference
in their three estimates, using inflection points is a less sensitive
probe of the critical behavior, and thus less reliable for extracting
$T_c$. By making simultaneous fits to the chiral condensate and the
susceptibility, and defining the transition temperature as the peak in
the susceptibility, our analysis overcomes this ambiguity.

\section{Quark number susceptibilities and the Polyakov loop}
\label{sec:deconf}

In this section, we present our results for light and strange quark
number susceptibilities as well as the Polyakov loop expectation
value.  In Sec.~\ref{sec:observables}, we stated that we do not use
these observables for the determination of the QCD transition
temperature because the singular part of the free energy, which drives
the chiral transition, is subleading in the quark number
susceptibilities and thus difficult to isolate. Similarly, in the case
of the Polyakov loop expectation value, a direct relation to the
critical behavior in the light quark mass regime has not been
established.  Nonetheless, we study these observables as they provide
important insight into the deconfining aspects of the QCD transition.

\begin{figure}
\includegraphics[width=0.49\textwidth]{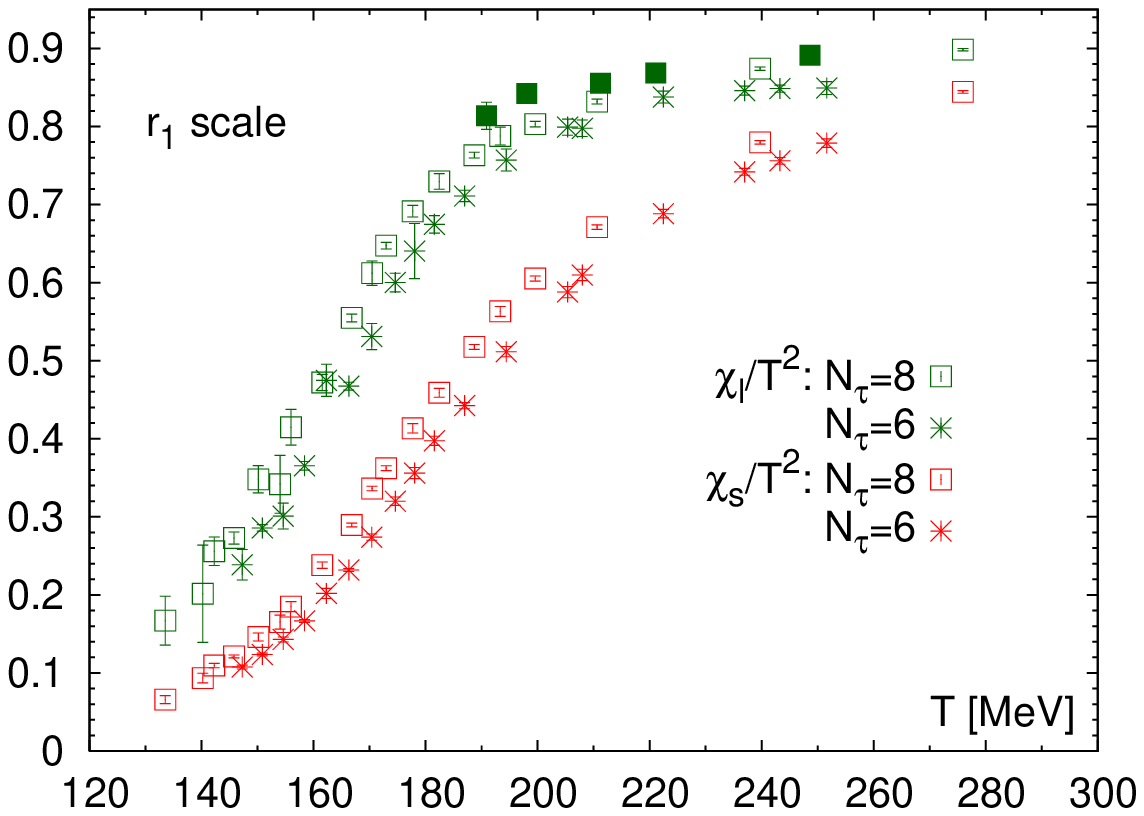}
\includegraphics[width=0.49\textwidth]{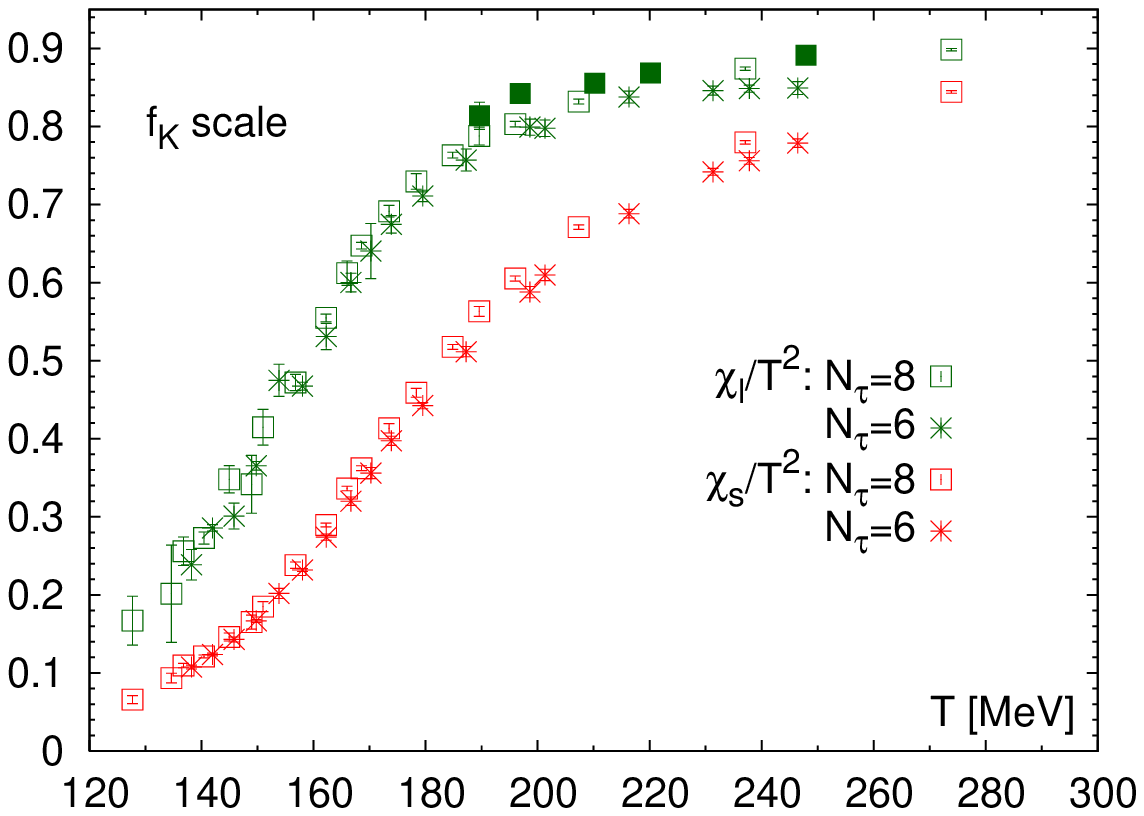}
\caption{Light quark number susceptibility for the HISQ/tree action
  with $m_l/m_s=1/20$ are compared with the strange quark number
  susceptibility. In the left panel $r_1$ is used to set the lattice
  scale, while in the right panel $f_K$ is used. The filled squares correspond to $N_{\tau}=12$. }
\label{fig:chil}
\end{figure}

The discussion in Sec.~\ref{sec:observables} shows that quark number
susceptibilities probe whether the relevant degrees of freedom of the
system at a given temperature are hadronic or partonic; a rapid rise
in the quark number susceptibilities signals the increasing
contribution of light (partonic) degrees of freedom.  Theoretically,
the reduced temperature $t$ defined in Eq.~(\ref{reduced2}) is
symmetric in the light and strange flavors at leading order,
therefore, the structure of the singular contribution to both
observables is similar and should differ only in magnitude.  We
compare the temperature dependence of $\chi_l$ and $\chi_s$ on
$N_\tau=6$ and $8$ lattices in Fig.~\ref{fig:chil} and find a similar
behavior with $\chi_l$ exhibiting a more rapid rise at temperatures
closer to the chiral transition temperature. From these data, one would
be led to conclude that $\chi_l$ and $\chi_s$ are probing the
singular part of the free energy and deduce a higher transition
temperature from $\chi_s$ than from $\chi_l$. However, in
Sec.~\ref{sec:observables} we pointed out that the temperature
dependence of these observables, even in the chiral limit, is
dominated by the regular part of the free energy.  This feature is
highlighted by the difference in the temperature dependence of
$\chi_l$ and $\chi_s$ versus that of the chiral condensate, shown in
Fig.~\ref{fig:pbpR}, which is dominated by the singular part in the
chiral limit.  Thus, even though the data in Fig.~\ref{fig:chil} show
a rapid crossover, extracting information about the singular part from
them is nontrivial, and consequently the determination of $T_c$ from these
observables is less reliable.

In Fig.~\ref{fig:chil}, we also compare the temperature dependence of
$\chi_l$ and $\chi_s$ using $r_1$ and $f_K$ to set the temperature
scale.  Both light and strange quark number susceptibilities show a
much smaller cutoff dependence when using the $f_K$ scale.  A similar
behavior was observed for the chiral susceptibility as discussed in
Sec.~\ref{ssec:chiralsus}.

\begin{figure}
\includegraphics[width=0.44\textwidth]{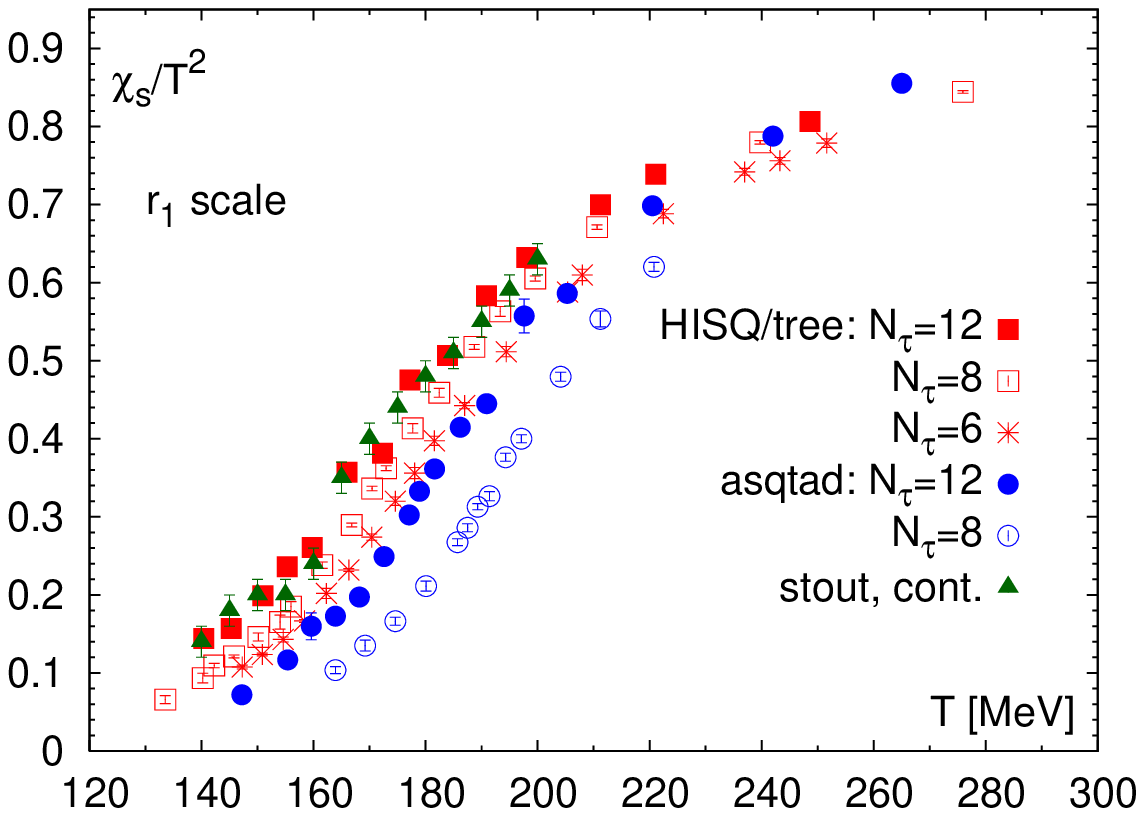}
\includegraphics[width=0.44\textwidth]{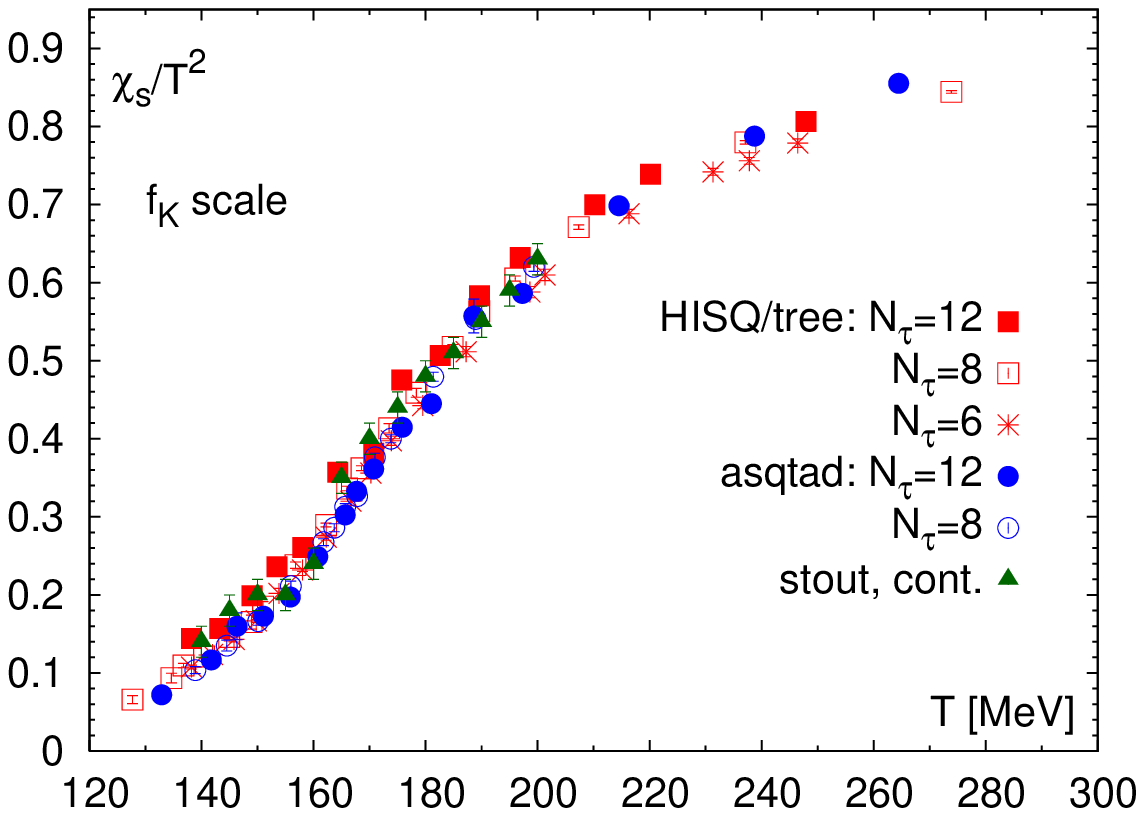}
\caption{Data for the strange quark number susceptibility for $m_l=0.05m_s$
with the asqtad and HISQ/tree actions are compared to 
the continuum extrapolated stout results \cite{Borsanyi:2010bp}. 
In the right panel, we show that all the data collapse to a single curve when 
$f_K$ is used to set the temperature scale. }
\label{fig:chis}
\end{figure}

In order to analyze further the cutoff dependence and to compare our
data with the continuum extrapolated stout results \cite{Borsanyi:2010bp}, we
examine the strange quark number susceptibility, which is
statistically under better control. In Fig.~\ref{fig:chis}, we show
$\chi_s/T^2$ for $m_l=0.05m_s$ calculated with the HISQ/tree and 
asqtad actions. The larger cutoff effects are in the asqtad data.
We also plot the data using
$f_K$ to set the temperature scale and find that all data collapse
into one curve as shown in Fig.~\ref{fig:chis}(right), suggesting that
the discretization errors in $\chi_s$ and $f_K$ are very similar. On
the basis of this agreement, we conclude that the differences in
results for $\chi_s$ seen in Fig.~\ref{fig:chis}(left) are accounted
for by discretization errors.

The Polyakov loop is a sensitive probe of the thermal properties of
the medium even though it is not a good measure of the critical
behavior as discussed in Sec.~\ref{sec:observables}. After proper
renormalization, the square of the Polyakov loop characterizes the
long distance behavior of the static quark anti-quark free energy; it
gives the excess free energy needed to screen two well separated color
charges. The renormalized Polyakov loop $L_{ren}(T)$ has been studied
in the past in the pure gauge theory \cite{Kaczmarek:2002mc,Digal:2003jc} as well as
in QCD with two \cite{Kaczmarek:2005ui}, three \cite{Petreczky:2004pz} and two plus one
flavors of quarks \cite{Cheng:2007jq,Bazavov:2009zn}. Following Ref. \cite{Aoki:2006br} 
$L_{ren}(T)$, for a given
$N_\tau$, is obtained from the bare Polyakov loop defined in
Eq.~(\ref{Ldef}) as
\begin{equation}
L_{ren}(T)=\exp(-N_\tau c(\beta)/2) L_{bare} \; ,
\label{eq:Lrenorm}
\end{equation}
where the renormalization constant $c(\beta)$ is deduced from the heavy quark
potential discussed in Sec.~\ref{sec:parameters}.

\begin{figure}
\centering
\includegraphics[width=7.5cm]{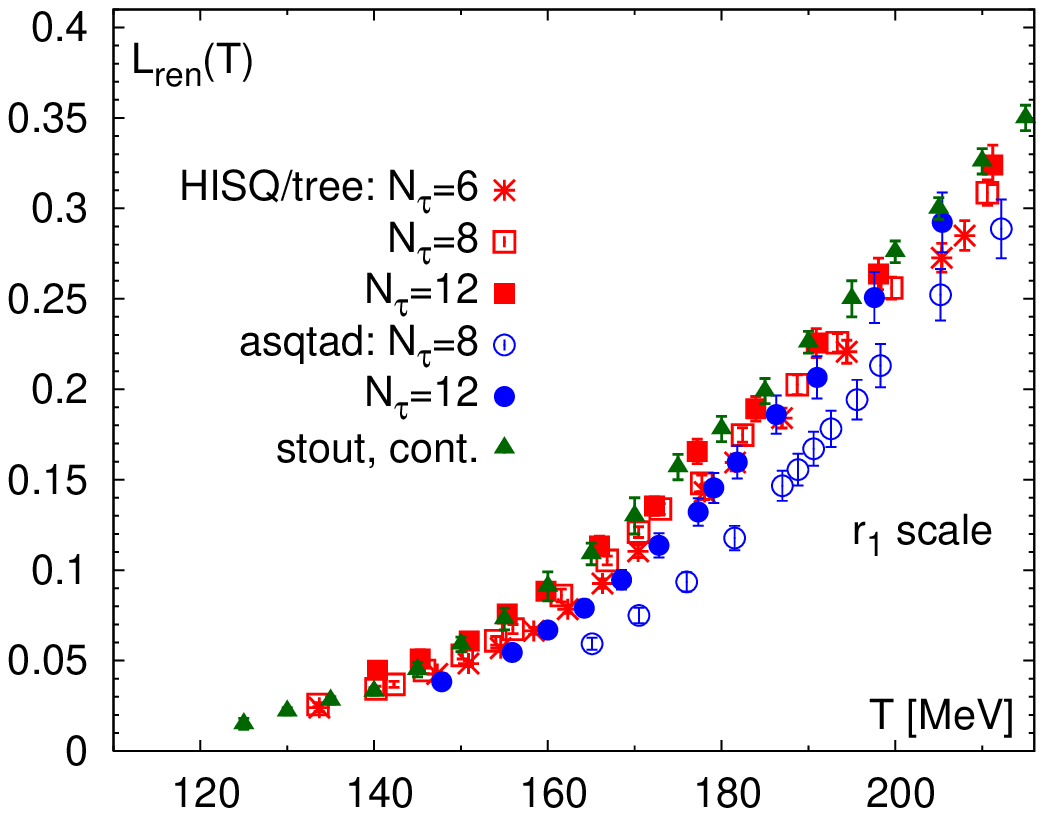}
\includegraphics[width=7.5cm]{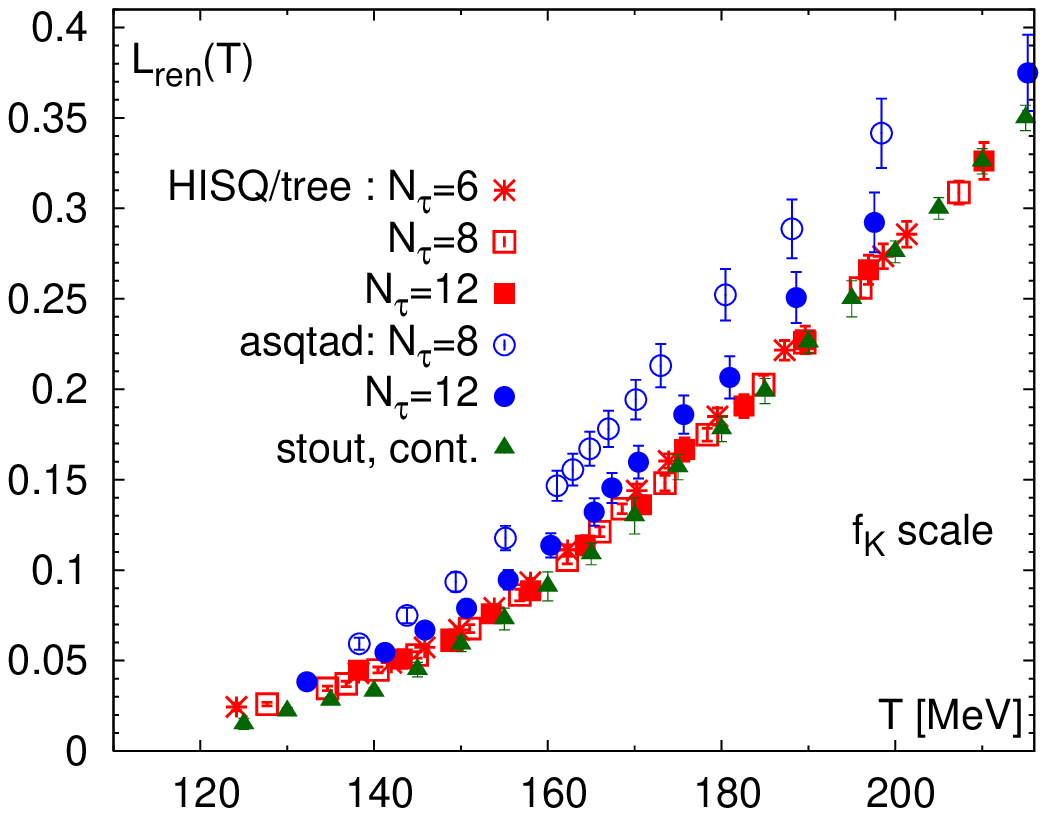}
\caption{The renormalized Polyakov loop calculated for the asqtad and
  HISQ/tree actions and compared with the continuum extrapolated stout
  result \cite{Borsanyi:2010bp}.  In the right panel, we show the same data
  using $f_K$ to set the scale.  }
\label{fig:lren}
\end{figure}

We compare results for $L_{ren}(T)$ from the asqtad, HISQ/tree and stout actions 
in Fig.~\ref{fig:lren} to determine the cutoff effects in the
description of basic thermal properties, {\it e.g.}, in the asymptotic
long distance structure of a heavy quark free energy and the screening
of the quark-antiquark force.  We find significant cutoff effects in
$L_{ren}$ with the asqtad action; however, these are considerably
smaller than seen in the other quantities discussed so far. This is
partially expected as the Polyakov loop is a purely gluonic observable
and is thus less affected by the taste symmetry breaking in the
staggered formulations.  In Fig.~\ref{fig:lren}(right), we also show
the temperature dependence of $L_{ren}$ when $f_K$ is used to set the
scale. For the HISQ/tree action, the already small
cutoff effects are further reduced and the data move even closer to the continuum
extrapolated stout result. For the asqtad data, on the other hand, the
$N_{\tau}$ dependence is slightly more pronounced. The more remarkable
feature is the switch in the approach to the continuum limit; it is
from above when $f_K$ is used to set the temperature scale and from
below when using the $r_1$ scale. This is similar to what has been
observed in calculations with the stout action on coarser lattices
\cite{Aoki:2006br}. Such behavior is not surprising, as there is no
obvious reason for cutoff effects in a hadronic observable like $f_K$
and a gluonic observable like $L_{ren}$ to be related so that there is
a cancellation of lattice artifacts in their ratio. Based on these
observed changes with $N_\tau$ and dependence on whether the scale is
set by $r_1$ or $f_K$, we expect that the asqtad action will need
$L_{ren}$ data on $N_{\tau}> 12$ lattices to agree with the continuum
extrapolated HISQ/tree and stout data.  This confirms the statement
made in Sec.~\ref{sec:parameters} that while asqtad and the HISQ
actions are equivalent at order ${\cal O}(\alpha_s a^2)$ the
difference at higher orders makes sizeable contributions below
$N_{\tau} \sim 12$.  Overall, we conclude that once discretization effects
are accounted for, the Polyakov loop measured using different
staggered discretization schemes yields consistent estimates of the
thermal properties of the theory.

\begin{figure}
\centering
\includegraphics[width=7.5cm]{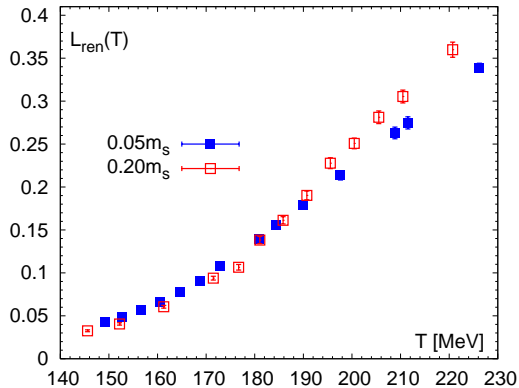}
\caption{The renormalized Polyakov loop for the HISQ/tree
action on $N_{\tau}=6$ lattices for two values of the light
quark mass $m_l/m_s=0.05$ and $0.20$.
}
\label{fig:lren02}
\end{figure}

Lastly, we briefly discuss the quark mass dependence of the Polyakov
loop.  In Fig.~\ref{fig:lren02}, we show the renormalized Polyakov loop
for two quark masses $m_l/m_s=0.05$ and $0.20$ on $N_{\tau}=6$
HISQ/tree ensembles. The quark mass dependence is small in the
transition region, and at higher temperatures the renormalized
Polyakov loop rises faster for the heavier quark mass. This
dependence on the quark mass is consistent with the results obtained
using the p4 action \cite{Cheng:2009zi}. Previous work suggests that for
larger values of the light quark masses, $m_l>0.2m_s$, the quark mass
dependence becomes more significant as $L$ is influenced by the phase
transition in the pure gauge theory~\cite{Petreczky:2004pz}.

\section{Conclusions}
\label{sec:conclusions}

In this paper, we present results for the chiral and deconfinement
aspects of the QCD transition using three improved staggered quark
actions called p4, asqtad and HISQ/tree.  The chiral transition is
studied using the chiral condensate and the chiral susceptibility,
while the deconfining aspects of the transition have been addressed
using the quark number susceptibilities and the renormalized Polyakov
loop. By comparing the results obtained at different lattice spacings,
we analyze the cutoff dependence of these quantities, in particular, 
the effects due to taste symmetry breaking. We find that the effects
of taste symmetry breaking are significantly reduced in the HISQ/tree
action. After demonstrating control over systematic errors, we compare
our data with the continuum extrapolated stout
results and find agreement for all quantities. 

We analyze the chiral transition in terms of universal $O(N)$ scaling
functions using simulations performed at several values of the light
quark mass and $N_\tau$. We show that for $m_l/m_s < 0.1$ the
corrections to scaling are small and determine the critical
temperature in the 2-flavor (degenerate $u$ and $d$ quarks) chiral
limit. We define the pseudocritical temperature $T_c$ at non-zero
values of the light quark mass, in particular the physical value
$m_l/m_s=1/27$, as the peak in the total chiral susceptibility. We
find that this definition of the pseudocritical temperature gives
estimates consistent with those obtained from the location of the peak
in the disconnected chiral susceptibility used in previous studies.
Our final result, after extrapolation to the continuum limit and for
the physical value of light quarks, $m_l/m_s=1/27$, is $T_c=(154 \pm
9){\rm MeV}$ based on the analysis summarized in
Table~\ref{tab:tcall}.

The Wuppertal-Budapest collaboration has extracted $T_c$ using the
inflection point in the data for the chiral condensate and the peak in
the chiral susceptibility.  In their most recent
publication~\cite{Borsanyi:2010bp}, they quote values ranging from $147$ to
$157$ MeV depending on the chiral observable.  
While their methods for extracting $T_c$ differ, their values are, within errors, 
in agreement with our determination. The consistency of the results obtained with the HISQ/tree, 
asqtad and stout actions 
shows that these results represent the continuum limit value for the staggered formulation.

Finally, we also consider it interesting to determine the
pseudocritical temperature in terms of other scaling functions, namely
the scaling functions related to the mixed susceptibility, Eq.~(\ref{chit}),
and the specific heat. This would allow estimation of the intrinsic
uncertainty in defining the pseudocritical temperature at nonzero
values of the light quark mass where, {\it a priori}, it can depend on
the probe used to estimate it. We will address this issue in future work. 


\section*{Acknowledgments}
\label{ackn}
This work has been supported in part by contracts DE-AC02-98CH10886,
DE-AC52-07NA27344, DE-FG02-92ER40699, DE-FG02-91ER-40628,
DE-FG02-91ER-40661, DE-FG02-04ER-41298, DE-KA-14-01-02 with the
U.S. Department of Energy, and NSF grants PHY08-57333, PHY07-57035,
PHY07-57333 and PHY07-03296, the Bundesministerium f\"ur Bildung und
Forschung under grant 06BI9001, the Gesellschaft f\"ur
Schwerionenforschung under grant BILAER and the Deutsche
Forschungsgemeinschaft under grant GRK881, and the EU Integrated
Infrastructure Initiative HadronPhysics2.  The numerical simulations
have been performed on BlueGene/L computers at Lawrence Livermore
National Laboratory (LLNL), the New York Center for Computational
Sciences (NYCCS) at Brookhaven National Laboratory, on BlueGene/P
computers at NIC, Juelich, US Teragrid (Texas Advanced Computing
Center), and on clusters of the USQCD collaboration in JLab and FNAL.
We thank Claude Bernard for help with the determination of the scale
for the asqtad action using $f_K$.

\clearpage
\appendix
\section{Lattice parameters for p4, asqtad and HISQ/tree simulations}
\label{sec:appendix1}

In this appendix, we provide details on the parameters used in the
calculations with the p4, asqtad and HISQ/tree actions. These include
the gauge couplings, quark masses and lattice sizes at which
simulations were carried out and the collected statistics.  In the
following tables, $TU_T$ stands for the molecular dynamics time units
simulated for finite temperature runs and $TU_0$ for the corresponding
zero temperature ones.  In Table \ref{tab:p4_runs}, we give the
parameters of the simulations with p4 action on $32^3 \times 8$
lattices. 
In Table \ref{tab:asqtad8_runs} and Table \ref{tab:asqtad12_runs}, we
give the simulation parameters for the asqtad action corresponding to
$N_{\tau}=8$ and $N_{\tau}=12$ lattices, respectively.  The parameters
for the $N_\tau =8$ ensemble with $m_l/m_s = 0.1$ runs are given in
Ref.~\cite{Bazavov:2009zn}.
The details on the extraction of $r_0$ and $r_1$ are
given in Sec.~\ref{subsec:scalesetting} and in
Appendix~\ref{sec:appendix2}.

\begin{table}
\begin{tabular}{|r||r|r|r||r|r|r||r|r|r|}
\hline
 &  \multicolumn{3}{c||}{ $m_l=0.05m_s$ }  &  \multicolumn{3}{c||}{ $m_l=0.10m_s$ } & \multicolumn{3}{c|}{ $m_l=0.20m_s$ }\\           
\hline
\multicolumn{1}{|c||}{ $\beta$ }&\multicolumn{1}{c|}{  $m_l$   }&\multicolumn{1}{c|}{  $m_s$ }&\multicolumn{1}{c||}{ $2\times TU_T$    }&\multicolumn{1}{c|}{  $m_l$  }&\multicolumn{1}{c|}{  $m_s$ }&\multicolumn{1}{c||}{ $2\times TU_T$    }&\multicolumn{1}{c|}{  $m_l$  }&\multicolumn{1}{c|}{  $m_s$ }&\multicolumn{1}{c||}{ $2\times TU_T$ }\\ \hline
 3.4900  & 0.001450 & 0.0290 & 31680     & ---     & ---    & ---       & ---     & ---    & ---    \\
 3.5000  & 0.001265 & 0.0253 & 32520     & ---     & ---    & ---       & ---     & ---    & ---   \\
 3.5100  & 0.001300 & 0.0260 & 30050     & ---     & ---    & ---       & ---     & ---    & ---   \\
 3.5150  & 0.001200 & 0.0240 & 30480     & ---     & ---    & ---       & 0.00480 & 0.0240 &  3510 \\
 3.5200  & 0.001200 & 0.0240 & 39990     & 0.00240 & 0.0240 & 11800     & 0.00480 & 0.0240 &  6470 \\
 3.5225  & 0.001200 & 0.0240 & 54610     & 0.00240 & 0.0240 & 30620     & ---     & ---    & ---   \\
 3.5250  & 0.001200 & 0.0240 & 64490     & 0.00240 & 0.0240 & 31510     & ---     & ---    & ---   \\
 3.5275  & 0.001200 & 0.0240 & 65100     & 0.00240 & 0.0240 & 28060     & ---     & ---    & ---   \\
 3.5300  & 0.001200 & 0.0240 & 70230     & 0.00240 & 0.0240 & 34420     & 0.00480 & 0.0240 & 24900 \\
 3.5325  & ---      & ---    & ---       & 0.00240 & 0.0240 & 27310     & ---     & ---    & ---   \\
 3.5350  & 0.001200 & 0.0240 & 39510     & 0.00240 & 0.0240 & 29250     & 0.00480 & 0.0240 &  4850 \\
 3.5375  & ---      & ---    & ---       & 0.00240 & 0.0240 & 30320     & 0.00480 & 0.0240 &  4890 \\
 3.5400  & 0.001200 & 0.0240 & 56740     & 0.00240 & 0.0240 & 31590     & 0.00480 & 0.0240 &  5190 \\
 3.5425  & ---      & ---    & ---       & 0.00240 & 0.0240 & 38520     & 0.00480 & 0.0240 &  4380 \\
 3.5450  & 0.001075 & 0.0215 & 20700     & 0.00240 & 0.0240 & 15060     & 0.00480 & 0.0240 &  5300 \\
 3.5475  & ---      & ---    & ---       & ---     & ---    & ---       & 0.00480 & 0.0240 &  4630 \\
 3.5500  & ---      & ---    & ---       & ---     & ---    & ---       & 0.00480 & 0.0240 &  5530 \\
 3.5525  & ---      & ---    & ---       & ---     & ---    & ---       & 0.00480 & 0.0240 & 22540 \\
 3.5550  & ---      & ---    & ---       & ---     & ---    & ---       & 0.00480 & 0.0240 & 23420 \\
 3.5600  & 0.001025 & 0.0205 & 11660     & ---     & ---    & ---       & 0.00480 & 0.0240 & 21230 \\
 3.5700  & ---      & ---    & ---       & ---     & ---    & ---       & 0.00480 & 0.0240 &  9250 \\
\hline
\end{tabular}
\caption{Parameters used in simulations with the p4 action on
  $N_\tau=8$ lattices and three lines of constant physics defined by
  $m_l/m_s = 0.2$, $0.1$ and $0.05$.  The quark masses are given in
  units of the lattice spacing $a$.  The finite temperature runs were
  carried out on $32^3 \times 8$ lattices. The statistics are given in
  units of trajectories ($2 \times TU_T$) which are of length $\tau_{MD}=0.5$.}
\label{tab:p4_runs}
\end{table}

\begin{table}
\begin{tabular}{|r||r|r|r|r|r||r|r|r|r|}
\hline
        & \multicolumn{5}{c||}{$m_l=0.05m_s$} & \multicolumn{4}{c|}{$m_l=0.20m_s$} \\\hline
$\beta$ & $m_l$   & $m_s$  & $u_0$  & $TU_0$    & $TU_T$    & $m_l$   & $m_s$  & $u_0$  & $TU_T$        \\
\hline
6.5500  & 0.00353 & 0.0705 & 0.8594 & ---       & 29541     & ---     & ---    & ---    & ---           \\
6.5750  & 0.00338 & 0.0677 & 0.8605 & ---       & 13795     & 0.01390 & 0.0684 & 0.8603 & 8515          \\
6.6000  & 0.00325 & 0.0650 & 0.8616 & 3251      & 25979     & 0.01330 & 0.0655 & 0.8614 & 12695         \\
6.6250  & 0.00312 & 0.0624 & 0.8626 & 3815      & 24442     & 0.01270 & 0.0628 & 0.8624 & 12695         \\
6.6500  & 0.00299 & 0.0599 & 0.8636 & 3810      & 22731     & 0.01210 & 0.0602 & 0.8634 & 12695         \\
6.6580  & 0.00295 & 0.0590 & 0.8640 & 3789      & 32725     & ---     & ---    & ---    & ---           \\
6.6660  & 0.00292 & 0.0583 & 0.8643 & 3445      & 32020     & ---     & ---    & ---    & ---           \\
6.6750  & 0.00288 & 0.0575 & 0.8646 & 4840      & 31067     & 0.01160 & 0.0577 & 0.8645 & 13695         \\
6.6880  & 0.00282 & 0.0563 & 0.8652 & 3895      & 33240     & ---     & ---    & ---    & ---           \\
6.7000  & 0.00276 & 0.0552 & 0.8656 & 5475      & 32324     & 0.01110 & 0.0552 & 0.8655 & 12695         \\
6.7300  & 0.00262 & 0.0525 & 0.8667 & 3827      & 26634     & 0.01050 & 0.0526 & 0.8666 & 12655         \\
6.7600  & 0.00250 & 0.0500 & 0.8678 & 4725      & 30433     & 0.01000 & 0.0500 & 0.8677 & 10695         \\
6.8000  & 0.00235 & 0.0471 & 0.8692 & 5710      & 35690     & 0.00953 & 0.0467 & 0.8692 &  7905         \\
6.8300  & ---     & ---    & ---    & ---       & ---       & 0.00920 & 0.0452 & 0.8702 &  6910         \\
\hline
\end{tabular}
\caption{Parameters used in simulations with the asqtad action on
  $N_\tau=8$ lattices and two lines of constant physics defined by $m_l/m_s = 0.05$ and
  $0.2$.  The quark masses are given in units of the lattice spacing
  $a$. The zero and finite temperature runs were carried out on $32^4$
  and $32^3 \times 8$ lattices, respectively. The number of 
  molecular dynamics time units for zero and finite temperature runs are 
  given in columns labeled by $TU_0$ and $TU_T$, respectively. 
}
\label{tab:asqtad8_runs}
\end{table}

\begin{table}
\begin{tabular}{|r|r|r|r|r|r|r|}
\hline
$\beta$ &  $m_l$  &  $m_s$ & $u_0$  & $TU_0$    &  $TU_T$\\
\hline
 6.800 & 0.00236  & 0.0471 & 0.8692 & 2040      & 29687 \\
 6.850 & 0.00218  & 0.0436 & 0.8709 & 3745      & 30452 \\
 6.875 & 0.00210  & 0.0420 & 0.8718 & 2025      & 18814 \\
 6.900 & 0.00202  & 0.0404 & 0.8726 & 3340      & 33427 \\
 6.925 & 0.00195  & 0.0389 & 0.8733 & 2925      & 27966 \\
 6.950 & 0.00187  & 0.0375 & 0.8741 & 2890      & 27734 \\
 6.975 & 0.00180  & 0.0361 & 0.8749 & ---       & 28464 \\
 6.985 & 0.00178  & 0.0355 & 0.8752 & ---       & 25191 \\
 7.000 & 0.00174  & 0.0347 & 0.8756 & 3055      & 25580 \\
 7.025 & 0.00167  & 0.0334 & 0.8764 & ---       & 21317 \\
 7.050 & 0.00161  & 0.0322 & 0.8771 & 3031      & 25605 \\
 7.085 & 0.00153  & 0.0305 & 0.8781 & ---       & 19070 \\
 7.125 & 0.00144  & 0.0289 & 0.8793 & 3100      & 32640 \\
 7.200 & 0.00128  & 0.0256 & 0.8813 & 4615      & 31175 \\
 7.300 & 0.00110  & 0.0220 & 0.8839 & 2645      & 34240 \\
 7.400 & 0.000946 & 0.0189 & 0.8863 & 1611      & 22805 \\
 7.550 & 0.000754 & 0.0151 & 0.8863 & 1611      & 26415 \\
\hline
\end{tabular}
\caption{Parameters used in simulations with the asqtad action on
  $N_\tau=12$ lattices and the LCP defined by $m_l/m_s = 0.05$.  
  The quark masses are given in units of the lattice spacing
  $a$. The zero and finite temperature runs were carried out on $48^4$
  and $48^3 \times 12$ lattices respectively. }
\label{tab:asqtad12_runs}
\end{table}

For the HISQ/tree action, the $N_\tau=6$ calculations
with $m_l=0.2m_s$ were done on $16^3 \times 32$ and $16^3 \times 6$ lattices
for zero and finite temperature studies, respectively, and the run parameters
are summarized in Table \ref{tab:hisq_0.2ms_runs}. 

The parameters of finite temperature simulations at $m_l=0.05m_s$ for
ensembles with $N_{\tau}=6,~8$ and $12$ and the corresponding zero
temperature calculations are given in Table
\ref{tab:hisq_0.05ms_runs}.  For the $O(N)$ scaling analysis, we have
also used the HISQ/tree data on $N_\tau=6$ and $8$ lattices at $m_l=0.025m_s$ as 
summarized in Table~\ref{tab:hisq_0.025ms_runs}. These data are part of 
the ongoing RBC-Bielefeld study of $O(N)$ scaling~\cite{rbcbi_future}. 

\begin{table}
\begin{tabular}{|r|r|r|r|r|}
\hline

$\beta$ & $m_l$  & $m_s$ & $TU_0$      & $TU_T$ \\
\hline
6.000   & 0.0230 & 0.115 & 3,000       & 6,000 \\
6.038   & 0.0216 & 0.108 & 3,000       & 6,000 \\
6.100   & 0.0200 & 0.100 & 3,000       & 6,000 \\
6.167   & 0.0182 & 0.091 & 3,000       & 6,000 \\
6.200   & 0.0174 & 0.087 & 3,000       & 6,000 \\
6.227   & 0.0168 & 0.084 & 3,000       & 6,000 \\
6.256   & 0.0162 & 0.081 & 3,000       & 6,000 \\
6.285   & 0.0158 & 0.079 & 3,000       & 6,000 \\
6.313   & 0.0152 & 0.076 & 3,000       & 6,000 \\
6.341   & 0.0148 & 0.074 & 3,000       & 6,000 \\
6.369   & 0.0144 & 0.072 & 3,000       & 6,000 \\
6.396   & 0.0140 & 0.070 & 3,000       & 6,000 \\
6.450   & 0.0136 & 0.068 & 3,000       & 6,000 \\
6.800   & 0.0100 & 0.050 & 3,000       & ---   \\
\hline
\end{tabular}
\caption{Parameters used in simulations with the HISQ/tree action on
  $N_\tau=6$ lattices and the LCP defined by $m_l/m_s = 0.2$.  
  The quark masses are given in units of the lattice spacing
  $a$. The zero and finite temperature runs were carried out on $16^3 \times 32$
  and $16^3 \times 6$ lattices respectively. }
\label{tab:hisq_0.2ms_runs}
\end{table}

\begin{table}[p]
\begin{tabular}{|r|r|r||l|r||r|r|r|}
\hline

        &         &         & \multicolumn{2}{c||}{$T=0$} & $N_\tau=6$ & $N_\tau=8$ & $N_\tau=12$ \\\hline
$\beta$ &  $m_l$  &  $m_s$  & $N_s^3\times N_\tau$ & $TU_0$   & $TU_T$    & $TU_T$    & $TU_T$    \\
\hline
5.900   & 0.00660 & 0.1320  & $24^3 \times 32$     & 3700 & 22280 & ---   & ---   \\       
6.000   & 0.00569 & 0.1138  & $24^3 \times 32$     & 5185 & 12030 & ---   & ---   \\
6.025   & 0.00550 & 0.1100  & $24^3 \times 32$     & 1345 & 16420 & ---   & ---   \\
6.050   & 0.00532 & 0.1064  & $24^3 \times 32$     & 4850 & 19990 & ---   & ---   \\
6.075   & 0.00518 & 0.1036  & ---                  & ---  & 20470 & ---   & ---   \\         
6.100   & 0.00499 & 0.0998  & $28^3 \times 32$     & 4190 & 29380 & ---   & ---   \\
6.125   & 0.00483 & 0.0966  & ---                  & ---  & 20320 & ---   & ---   \\         
6.150   & 0.00468 & 0.0936  & ---                  & ---  & 11220 & ---   & ---   \\
6.175   & 0.00453 & 0.0906  & ---                  & ---  & 10860 & ---   & ---   \\
6.195   & 0.00440 & 0.0880  & $32^4$               & 3175 & 22330 & 16520 & ---   \\
6.215   & 0.00431 & 0.0862  & ---                  & ---  &  6390 & ---   & ---   \\  
6.245   & 0.00415 & 0.0830  & ---                  & ---  &  6400 &  8560 & ---   \\   
6.260   & 0.00405 & 0.0810  & ---                  & ---  & ---   & 10340 & ---   \\
6.285   & 0.00395 & 0.0790  & $32^4$               & 3560 &  6750 & 16900 & ---   \\
6.315   & 0.00380 & 0.0760  & ---                  & ---  & ---   &  7950 & ---   \\ 
6.341   & 0.00370 & 0.0740  & $32^4$               & 3160 &  6590 & 11990 & ---   \\
6.354   & 0.00364 & 0.0728  & $32^4$               & 2295 &  5990 & 11990 & ---   \\
6.390   & 0.00347 & 0.0694  & $32^4$               & 4435 & ---   & 16120 & ---   \\
6.423   & 0.00335 & 0.0670  & $32^4$               & 2295 &  5990 & 11990 & ---   \\
6.445   & 0.00326 & 0.0652  & ---                  & ---  & ---   &  9000 & ---   \\     
6.460   & 0.00320 & 0.0640  & $32^3 \times 64$     & 2610 & ---   & 10990 & ---   \\
6.488   & 0.00310 & 0.0620  & $32^4$               & 2295 &  8790 & 11990 & ---   \\
6.515   & 0.00302 & 0.0604  & $32^4$               & 2520 & 10430 & 10100 & ---   \\
6.550   & 0.00291 & 0.0582  & $32^4$               & 2295 &  7270 & 11900 & ---   \\
6.575   & 0.00282 & 0.0564  & $32^4$               & 2650 &  7330 & 14500 & ---   \\
6.608   & 0.00271 & 0.0542  & $32^4$               & 2295 &  6560 & 11990 & ---   \\
6.664   & 0.00257 & 0.0514  & $32^4$               & 2295 &  8230 & 11990 &  4240 \\
6.700   & 0.00248 & 0.0496  & ---                  & ---  & ---   & ---   &  7000 \\
6.740   & 0.00238 & 0.0476  & $48^4$               & 1350 & ---   & ---   &  6670 \\
6.770   & 0.00230 & 0.0460  & ---                  & ---  & ---   & ---   &  6820 \\  
6.800   & 0.00224 & 0.0448  & $32^4$               & 5650 &  7000 & 11990 &  7090 \\
6.840   & 0.00215 & 0.0430  & ---                  & ---  & ---   & ---   &  8410 \\ 
6.860   & 0.00210 & 0.0420  & ---                  & ---  & ---   & ---   &  2740 \\ 
6.880   & 0.00206 & 0.0412  & $48^4$               & 1400 & ---   & ---   & 10120 \\
6.910   & 0.00200 & 0.0400  & ---                  & ---  & ---   & ---   &  4630 \\
6.950   & 0.00193 & 0.0386  & $32^4$               &10830 &  7480 & 11990 &  6700 \\
6.990   & 0.00185 & 0.0370  & ---                  & ---  & ---   & ---   &  5470 \\ 
7.030   & 0.00178 & 0.0356  & $48^4$               & 1355 & ---   & ---   &  7290 \\ 
7.100   & 0.00166 & 0.0332  & ---                  & ---  & ---   & ---   & 10300 \\  
7.150   & 0.00160 & 0.0320  & $32^4$               & 2295 &  4770 & 11990 & 10390 \\
7.150   & 0.00160 & 0.0320  & $48^3 \times 64$     & 1458 & ---   & ---   & ---   \\
7.280   & 0.00142 & 0.0284  & $48^3 \times 64$     & 1734 & ---   & ---   & 11620 \\
\hline
\end{tabular}
\caption{Parameters used in simulations with the HISQ/tree action on
  $N_\tau=6$, $8$ and $12$ lattices and the LCP defined by $m_l/m_s =
  0.05$.  The quark masses are given in units of the lattice spacing
  $a$. The statistics in molecular dynamics time units $TU$ are given
  for both the zero and finite temperature runs. The lattice sizes
  used for the $N_\tau=6$, $8$ and $12$ finite temperature simulations
  were $24^3 \times 6$, $32^3 \times 8$ and $48^3 \times 12$, 
  respectively. }
\label{tab:hisq_0.05ms_runs}
\end{table}

\begin{table}
\begin{tabular}{|r|r|r|r|}
\hline

$\beta$ &  $m_l$    &  $m_s$ & $TU_T$ \\
\hline
\multicolumn{4}{|c|}{$N_\tau=6$} \\ \hline
6.000   & 0.0028450 & 0.1138 & 3510 \\
6.025   & 0.0027500 & 0.1100 & 3460 \\
6.050   & 0.0026600 & 0.1064 & 3710 \\
6.075   & 0.0025900 & 0.1036 & 3930 \\
6.100   & 0.0024950 & 0.0998 & 3200 \\
6.125   & 0.0024150 & 0.0966 & 4020 \\ \hline
\multicolumn{4}{|c|}{$N_\tau=8$} \\ \hline
6.215   & 0.0021425 & 0.0857 &  860 \\
6.230   & 0.0021025 & 0.0841 &  950 \\
6.245   & 0.0020750 & 0.0830 & 3730 \\
6.260   & 0.0020250 & 0.0810 & 4090 \\
6.285   & 0.0019750 & 0.0790 & 4050 \\
6.300   & 0.0019300 & 0.0772 & 4040 \\
6.315   & 0.0019000 & 0.0760 & 4170 \\
6.330   & 0.0018650 & 0.0746 & 4040 \\
6.341   & 0.0018500 & 0.0740 & 1340 \\
6.354   & 0.0018200 & 0.0728 & 4070 \\
6.365   & 0.0017900 & 0.0716 & 4420 \\
6.390   & 0.0017350 & 0.0694 & 4490 \\
6.423   & 0.0016750 & 0.0670 & 1710 \\
\hline
\end{tabular}
\caption{Parameters used in simulations with the HISQ/tree action on
  $N_\tau=6$ and $8$ lattices and the LCP defined by $m_l/m_s = 0.025$.  
  The quark masses are given in units of the lattice spacing
  $a$. The finite temperature runs were carried out on $32^3 \times 6$
  and $32^3\times8$ lattices as part of the RBC-Bielefeld study of 
  $O(N)$ scaling~\cite{rbcbi_future}.
}
\label{tab:hisq_0.025ms_runs}
\end{table}

Data for the light and strange quark chiral condensate, disconnected and connected
chiral susceptibility, the light and strange quark number
susceptibility and the bare Polyakov loop on finite temperature ensembles for the asqtad action on
$N_\tau=8$ and $12$ lattices are given in
Tables~\ref{tab:asqtad_NT8_obs} and~\ref{tab:asqtad_0.05ms_NT12_obs}.
The corresponding data for the HISQ/tree action are given in
Tables~\ref{tab:hisq_NT6_obs}, \ref{tab:hisq_NT8_obs}
and~\ref{tab:hisq_0.05ms_NT12_obs} for $N_\tau=6, \ 8$ and $12$
lattices, respectively.

\begin{table}[thbp]
\begin{tabular}{|c|c|c||c|c||c|c||c|c|}
\hline
$\beta$ &  $T$ MeV &  $m_s$  & $2 \times \langle \bar \psi \psi \rangle_l$ & $2 \times \langle \bar \psi \psi \rangle_s$ & $\chi_l^{\rm dis}$   & $\chi_l^{\rm con}$  & $\chi_s$ & $3 \times L_{\rm bare}$ \\
\hline
\multicolumn{9}{|c|}{$m_l/m_s = 0.05$} \\
\hline
6.550 & 165.1 & 0.0705 & 0.04306(06) & 0.15461(3) & 0.915(39) & 1.2570(15) & 0.1037(45) & 0.00578(06) \\
6.575 & 170.5 & 0.0677 & 0.03698(11) & 0.14547(5) & 1.035(50) & 1.3068(26) & 0.1350(70) & 0.00730(08) \\
6.600 & 176.0 & 0.0650 & 0.03111(13) & 0.13672(5) & 1.179(56) & 1.3682(51) & 0.1664(51) & 0.00910(09) \\
6.625 & 181.5 & 0.0624 & 0.02531(20) & 0.12832(7) & 1.536(73) & 1.4496(80) & 0.2112(64) & 0.01146(14) \\
6.650 & 187.0 & 0.0599 & 0.01979(16) & 0.12027(5) & 1.490(65) & 1.5282(51) & 0.2675(45) & 0.01428(15) \\
6.658 & 188.8 & 0.0590 & 0.01837(13) & 0.11765(6) & 1.381(50) & 1.5381(33) & 0.2861(51) & 0.01515(09) \\
6.666 & 190.6 & 0.0583 & 0.01681(08) & 0.11528(4) & 1.183(88) & 1.5506(20) & 0.3130(38) & 0.01627(09) \\
6.675 & 192.6 & 0.0575 & 0.01490(12) & 0.11264(4) & 1.024(57) & 1.5609(26) & 0.3341(51) & 0.01735(09) \\
6.688 & 195.6 & 0.0563 & 0.01313(14) & 0.10890(5) & 0.912(45) & 1.5274(30) & 0.3725(38) & 0.01893(11) \\
6.700 & 198.3 & 0.0552 & 0.01123(08) & 0.10539(3) & 0.687(34) & 1.4977(29) & 0.4006(38) & 0.02078(09) \\
6.730 & 205.2 & 0.0525 & 0.00843(08) & 0.09741(4) & 0.344(34) & 1.3621(66) & 0.4781(38) & 0.02465(12) \\
6.760 & 212.2 & 0.0500 & 0.00671(05) & 0.09023(4) & 0.163(12) & 1.2247(45) & 0.5478(32) & 0.02830(14) \\
6.800 & 221.7 & 0.0471 & 0.00526(03) & 0.08185(4) & 0.069(05) & 1.0719(49) & 0.6227(38) & 0.03365(13) \\
\hline             
\multicolumn{9}{|c|}{$m_l/m_s = 0.1$} \\
\hline
6.4580 & 144.0 & 0.0820 & 0.07942(07) &  & 0.531(20) &  1.0261(05) &   &  \\
6.5000 & 152.4 & 0.0765 & 0.06684(06) &  & 0.564(40) &  1.0416(06) &   &  \\
6.5500 & 162.8 & 0.0705 & 0.05359(07) &  & 0.554(25) &  1.0623(08) &   &  \\
6.6000 & 173.5 & 0.0650 & 0.04183(10) &  & 0.668(31) &  1.0944(19) &   &  \\
6.6250 & 179.0 & 0.0624 & 0.03657(09) &  & 0.724(64) &  1.1150(19) &   &  \\
6.6500 & 184.5 & 0.0599 & 0.03116(15) &  & 0.925(78) &  1.1433(28) &   &  \\
6.6580 & 186.4 & 0.0590 & 0.02900(11) &  & 0.891(48) &  1.1614(23) &   &  \\
6.6660 & 188.2 & 0.0583 & 0.02770(14) &  & 0.916(64) &  1.1661(31) &   &  \\
6.6750 & 190.2 & 0.0575 & 0.02579(13) &  & 0.946(88) &  1.1763(28) &   &  \\
6.6830 & 192.1 & 0.0567 & 0.02394(17) &  & 0.996(69) &  1.1873(23) &   &  \\
6.6910 & 193.8 & 0.0560 & 0.02278(15) &  & 0.803(56) &  1.1877(22) &   &  \\
6.7000 & 196.0 & 0.0552 & 0.02108(12) &  & 0.799(51) &  1.1881(22) &   &  \\
6.7080 & 197.8 & 0.0544 & 0.01976(16) &  & 0.759(89) &  1.1880(19) &   &  \\
6.7150 & 199.4 & 0.0538 & 0.01872(13) &  & 0.664(50) &  1.1833(16) &   &  \\
6.7300 & 202.9 & 0.0525 & 0.01648(10) &  & 0.501(33) &  1.1675(17) &   &  \\
6.7450 & 206.5 & 0.0512 & 0.01474(10) &  & 0.414(33) &  1.1425(20) &   &  \\
6.7600 & 210.0 & 0.0500 & 0.01343(07) &  & 0.349(20) &  1.1131(13) &   &  \\
6.8000 & 219.6 & 0.0471 & 0.01054(05) &  & 0.170(11) &  1.0200(20) &   &  \\
6.8500 & 232.0 & 0.0437 & 0.00840(03) &  & 0.058(03) &  0.9231(20) &   &  \\
\hline
\end{tabular}
\caption{Data for the light and strange quark chiral condensate, disconnected and connected
  chiral susceptibility, the strange quark number suceptibility and the bare Polyakov loop
  for the asqtad action on $N_\tau=8$ lattices and the two lines of constant physics defined by
  $m_l/m_s = 0.05$ and $m_l/m_s = 0.1$. 
  The data for the light and strange quark condensate are presented using the 2-flavor normalization.}
\label{tab:asqtad_NT8_obs}
\end{table}

\begin{table}[thbp]
\begin{tabular}{|c|c|c||c|c||c|c||c|c|}
\hline
$\beta$ &  $T$ MeV &  $m_s$  & $2 \times \langle \bar \psi \psi \rangle_l$ & $2 \times \langle \bar \psi \psi \rangle_s$ & $\chi_l^{\rm dis}$   & $\chi_l^{\rm con}$  & $\chi_s$ & $3 \times L_{\rm bare}$ \\
\hline
6.800 & 147.8 & 0.0471 & 0.017823(21) & 0.088585(10) & 0.4343(155) & 1.0215(07) & 0.0720(72) & 0.000683(21) \\
6.850 & 155.9 & 0.0436 & 0.014203(27) & 0.079482(11) & 0.4999(142) & 1.0336(11) & 0.1152(72) & 0.000985(18) \\
6.875 & 160.0 & 0.0420 & 0.012458(46) & 0.075317(17) & 0.6227(384) & 1.0471(23) & 0.1598(86) & 0.001218(25) \\
6.900 & 164.2 & 0.0404 & 0.010962(33) & 0.071361(14) & 0.6410(265) & 1.0602(15) & 0.1728(72) & 0.001450(20) \\
6.925 & 168.5 & 0.0389 & 0.009495(45) & 0.067657(14) & 0.6843(356) & 1.0783(28) & 0.1973(72) & 0.001752(23) \\
6.950 & 172.8 & 0.0375 & 0.008019(64) & 0.064170(26) & 0.7850(596) & 1.1005(29) & 0.2491(86) & 0.002124(38) \\
6.975 & 177.3 & 0.0361 & 0.006741(45) & 0.060787(20) & 0.7348(458) & 1.1088(17) & 0.3024(72) & 0.002494(29) \\
6.985 & 179.1 & 0.0355 & 0.006148(47) & 0.059361(21) & 0.6692(617) & 1.1085(16) & 0.3326(86) & 0.002757(30) \\
7.000 & 181.8 & 0.0347 & 0.005544(53) & 0.057468(22) & 0.6432(421) & 1.0973(13) & 0.3614(72) & 0.003048(31) \\
7.025 & 186.3 & 0.0334 & 0.004544(36) & 0.054397(17) & 0.3990(286) & 1.0665(14) & 0.4147(86) & 0.003591(34) \\
7.050 & 191.0 & 0.0322 & 0.003943(28) & 0.051647(17) & 0.2760(131) & 1.0264(25) & 0.4450(86) & 0.004037(30) \\
7.085 & 197.6 & 0.0305 & 0.003209(22) & 0.047827(21) & 0.1585(137) & 0.9502(28) & 0.5386(86) & 0.004993(39) \\
7.125 & 205.4 & 0.0289 & 0.002672(11) & 0.044212(13) & 0.0756(047) & 0.8793(19) & 0.5861(58) & 0.005955(37) \\
7.200 & 220.7 & 0.0256 & 0.002068(08) & 0.037549(10) & 0.0266(022) & 0.7898(22) & 0.6984(58) & 0.008019(38) \\
7.300 & 242.5 & 0.0220 & 0.001588(03) & 0.030717(07) & 0.0057(009) & 0.7180(12) & 0.7862(58) & 0.010999(39) \\
7.400 & 265.9 & 0.0189 & 0.001286(02) & 0.025334(06) & 0.0017(003) & 0.6785(07) & 0.8510(58) & 0.014392(84) \\
7.550 & 304.7 & 0.0151 & 0.000970(03) & 0.019278(04) & 0.0012(011) & 0.6419(12) &            &              \\
\hline
\end{tabular}
\caption{Data for the light and strange quark chiral condensate, disconnected and connected
  chiral susceptibility, the strange quark number susceptibility and the bare Polyakov loop
  for the asqtad action on $N_\tau=12$ lattices and the LCP defined by
  $m_l/m_s = 0.05$. 
  The data for the light and strange quark condensate are presented using the 2-flavor normalization.
 }
\label{tab:asqtad_0.05ms_NT12_obs}
\end{table}

\begin{table}[thbp]
\begin{tabular}{|c|c|c||c|c||c|c||c|c|c|}
\hline
$\beta$ &  $T$ MeV &  $a m_s$  & $2 \times \langle \bar \psi \psi \rangle_l$ & $2 \times \langle \bar \psi \psi \rangle_s$ & $\chi_l^{\rm dis}$   & $\chi_l^{\rm con}$  & $\chi_l$ & $\chi_s$ & $3 \times L_{\rm bare}$ \\
\hline
\multicolumn{10}{|c|}{$m_l/m_s = 0.025$} \\
\hline
6.0000 & 147.262 & 0.1138 & 0.05766(15) & 0.17718(8)    & 1.32(16) &    &    &    &  \\
6.0250 & 150.876 & 0.1100 & 0.05086(23) & 0.16948(11)   & 1.33(23) &    &    &    &  \\
6.0500 & 154.584 & 0.1064 & 0.04525(20) & 0.16223(6)    & 1.61(25) &    &    &    &  \\
6.0750 & 158.389 & 0.1036 & 0.03842(22) & 0.15533(8)    & 1.72(21) &    &    &    &  \\
6.1000 & 162.292 & 0.0998 & 0.03161(29) & 0.14777(12)   & 2.23(18) &    &    &    &  \\
6.1250 & 166.295 & 0.0966 & 0.02511(28) & 0.14063(10)   & 1.48(12) &    &    &    &  \\
6.1500 & 170.400 & 0.0936 & 0.01799(44) & 0.13339(14)   & 1.68(09) &    &    &    &  \\
6.1750 & 174.610 & 0.0906 & 0.01372(41) & 0.12671(24)   & 1.18(11) &    &    &    &  \\
\hline
\multicolumn{10}{|c|}{$m_l/m_s = 0.05$} \\                           
\hline
5.900 & 133.710 & 0.1320 & 0.089873(076) & 0.210216(046) &  0.5767(237) & 0.7967(025) &           &           &  0.00786(65)  \\
6.000 & 147.262 & 0.1138 & 0.065834(071) & 0.177805(037) &  0.7307(182) & 0.9124(046) & 0.239(20) & 0.1076(25)&  0.01235(14)  \\
6.025 & 150.876 & 0.1100 & 0.059957(105) & 0.170305(035) &  0.8508(277) & 0.9564(048) & 0.286(04) & 0.1236(17)&  0.01367(21)  \\
6.050 & 154.584 & 0.1064 & 0.054023(089) & 0.162915(051) &  0.9711(409) & 0.9448(060) & 0.301(17) & 0.1429(32)&  0.01573(14)  \\
6.075 & 158.389 & 0.1036 & 0.048293(128) & 0.156244(044) &  1.0169(307) & 1.0551(037) & 0.365(05) & 0.1667(18)&  0.01805(16)  \\
6.100 & 162.292 & 0.0998 & 0.041948(105) & 0.148575(040) &  1.1426(278) & 1.1358(273) & 0.475(21) & 0.2020(65)&  0.02098(16)  \\
6.125 & 166.295 & 0.0966 & 0.036062(123) & 0.141563(049) &  1.2281(366) & 1.2068(061) & 0.467(05) & 0.2319(19)&  0.02441(15)  \\
6.150 & 170.400 & 0.0936 & 0.029645(259) & 0.134460(098) &  1.3652(617) & 1.2792(091) & 0.531(17) & 0.2740(40)&  0.02875(36)  \\
6.175 & 174.610 & 0.0906 & 0.024075(159) & 0.127619(079) &  1.1297(692) & 1.3365(081) & 0.600(12) & 0.3200(54)&               \\
6.195 & 178.054 & 0.0880 & 0.020080(104) & 0.122064(051) &  1.0001(172) & 1.3415(068) & 0.640(35) & 0.3560(72)&  0.03664(29)  \\
6.215 & 181.568 & 0.0862 & 0.016410(172) & 0.117038(095) &  0.7007(439) & 1.3053(074) & 0.675(11) & 0.3974(58)&  0.04057(31)  \\
6.245 & 186.969 & 0.0830 & 0.012735(179) & 0.109962(123) &  0.4768(400) & 1.2076(075) & 0.711(07) & 0.4424(40)&  0.04646(49)  \\
6.285 & 194.423 & 0.0790 & 0.009099(091) & 0.100599(080) &  0.1752(094) & 1.0271(050) & 0.757(14) & 0.5116(65)&  0.05534(40)  \\
6.341 & 205.355 & 0.0740 & 0.006545(101) & 0.089489(156) &  0.0812(104) & 0.8432(088) & 0.799(10) & 0.5879(72)&  0.06806(54)  \\
6.354 & 207.977 & 0.0728 & 0.006011(038) & 0.086944(096) &  0.0476(049) & 0.7958(056) & 0.798(10) & 0.6098(72)&  0.07110(54)  \\
6.423 & 222.450 & 0.0670 & 0.004457(023) & 0.075293(079) &  0.0174(036) & 0.6583(028) & 0.838(08) & 0.6883(54)&  0.08810(38)  \\
6.488 & 236.960 & 0.0620 & 0.003643(019) & 0.066299(040) &  0.0065(022) & 0.5854(017) & 0.846(06) & 0.7420(43)&  0.10367(41)  \\
6.515 & 243.246 & 0.0603 & 0.003387(013) & 0.063337(051) &  0.0019(012) & 0.5632(016) & 0.849(05) & 0.7560(43)&  0.11238(52)  \\
6.550 & 251.627 & 0.0581 & 0.003149(019) & 0.059679(031) &  0.0003(008) & 0.5410(012) & 0.849(08) & 0.7787(54)&  0.12180(61)  \\
6.575 & 257.777 & 0.0564 & 0.002977(011) & 0.057089(046) &  0.0019(010) &             &           &           &  0.12714(58)  \\
6.608 & 266.106 & 0.0542 & 0.002798(011) & 0.053897(029) &  0.0021(015) &             &           &           &  0.13663(63)  \\
6.664 & 280.807 & 0.0514 & 0.002549(017) & 0.049855(046) &  0.0019(011) & 0.4966(009) &           &           &  0.15203(70)  \\
6.800 & 319.609 & 0.0448 & 0.002101(012) & 0.041501(018) &  0.0005(007) & 0.4669(003) & 0.861(04) & 0.8248(36)&  0.18747(68)  \\
6.950 & 367.875 & 0.0386 & 0.001746(007) & 0.034571(012) &  0.0004(009) & 0.4495(003) & 0.864(06) & 0.8374(58)&  0.22876(74)  \\
7.150 & 442.160 & 0.0320 & 0.001412(010) & 0.027810(016) & -0.0025(014) & 0.4353(001) & 0.849(05) & 0.8395(43)&  0.28241(73)  \\
\hline
\end{tabular}
\caption{Data for the light and strange quark chiral condensate,
  disconnected and connected chiral susceptibility, the light and
  strange quark number susceptibility and the bare Polyakov loop for
  the HISQ/tree action on $N_\tau=6$ lattices and the two lines of
  constant physics defined by $m_l/m_s = 0.025$ and $m_l/m_s = 0.05$.
  A thousand trajectories were discarded for thermalization.}
\label{tab:hisq_NT6_obs}
\end{table}

\begin{table}[thbp]
\begin{tabular}{|c|c|c||c|c||c|c||c|c|c|}
\hline
$\beta$ &  $T$ MeV &  $a m_s$  & $2 \times \langle \bar \psi \psi \rangle_l$ & $2 \times \langle \bar \psi \psi \rangle_s$ & $\chi_l^{\rm dis}$   & $\chi_l^{\rm con}$  & $\chi_l$ & $\chi_s$ & $3 \times L_{\rm bare}$ \\
\hline
\multicolumn{10}{|c|}{$m_l/m_s = 0.025$} \\
\hline
6.2450 & 140.227 & 0.0830 & 0.03034(07) & 0.11810(4)   & 0.90(07)  &   &   &   &  \\
6.2600 & 142.298 & 0.0811 & 0.02812(14) & 0.11447(7)   & 1.09(10)  &   &   &   &  \\
6.2850 & 145.817 & 0.0790 & 0.02514(15) & 0.10994(8)   & 0.96(09)  &   &   &   &  \\
6.3000 & 147.970 & 0.0773 & 0.02279(19) & 0.10643(11)  & 1.16(11)  &   &   &   &  \\
6.3150 & 150.154 & 0.0759 & 0.02105(15) & 0.10381(7)   & 1.30(07)  &   &   &   &  \\
6.3300 & 152.370 & 0.0746 & 0.01953(17) & 0.10111(8)   & 1.32(11)  &   &   &   &  \\
6.3540 & 155.983 & 0.0728 & 0.01610(20) & 0.09688(8)   & 1.47(07)  &   &   &   &  \\
6.3650 & 157.666 & 0.0717 & 0.01427(17) & 0.09449(6)   & 1.44(07)  &   &   &   &  \\
6.3900 & 161.558 & 0.0695 & 0.01142(26) & 0.09010(9)   & 1.32(09)  &   &   &   &  \\
6.4230 & 166.837 & 0.0670 & 0.00842(20) & 0.08503(9)   & 0.77(10)  &   &   &   &  \\
6.4450 & 170.448 & 0.0653 & 0.00646(25) & 0.08143(15)  & 0.72(10)  &   &   &   &  \\
\hline             
\multicolumn{10}{|c|}{$m_l/m_s = 0.05$} \\
\hline
6.1950 & 133.541 & 0.0880 & 0.042262(031) & 0.12903(2) & 0.3855(121) & 0.7987(25) & 0.1670(314) & 0.0659(51) & 0.00289(06) \\
6.2450 & 140.227 & 0.0830 & 0.035540(065) & 0.11837(4) & 0.5719(183) & 0.8203(52) & 0.2016(623) & 0.0935(61) & 0.00381(07) \\
6.2600 & 142.298 & 0.0811 & 0.033581(085) & 0.11487(4) & 0.5822(301) & 0.8395(42) & 0.2559(183) & 0.1095(28) & 0.00406(10) \\
6.2850 & 145.817 & 0.0790 & 0.030407(073) & 0.11018(3) & 0.6314(262) & 0.8638(44) & 0.2729(078) & 0.1212(19) & 0.00489(09) \\
6.3150 & 150.154 & 0.0759 & 0.026587(105) & 0.10415(4) & 0.7213(276) & 0.8978(56) & 0.3481(175) & 0.1461(50) & 0.00577(06) \\
6.3410 & 154.016 & 0.0740 & 0.023661(103) & 0.09968(4) & 0.7739(205) & 0.9205(58) & 0.3418(370) & 0.1651(90) & 0.00664(08) \\
6.3540 & 155.983 & 0.0728 & 0.021917(088) & 0.09721(5) & 0.7761(500) & 0.9320(53) & 0.4147(230) & 0.1849(64) & 0.00734(06) \\
6.3900 & 161.558 & 0.0695 & 0.017638(160) & 0.09054(5) & 0.9397(436) & 0.9839(55) & 0.4721(104) & 0.2381(41) & 0.00937(18) \\
6.4230 & 166.837 & 0.0670 & 0.013908(119) & 0.08525(5) & 0.8828(466) & 1.0351(67) & 0.5547(051) & 0.2895(24) & 0.01151(10) \\
6.4450 & 170.448 & 0.0653 & 0.011433(130) & 0.08153(7) & 0.7199(464) & 1.0390(49) & 0.6121(157) & 0.3362(29) & 0.01325(17) \\
6.4600 & 172.952 & 0.0642 & 0.010222(097) & 0.07919(5) & 0.6721(343) & 1.0315(65) & 0.6473(043) & 0.3623(30) & 0.01468(11) \\
6.4880 & 177.720 & 0.0620 & 0.008284(082) & 0.07519(5) & 0.4976(242) & 0.9822(39) & 0.6914(074) & 0.4134(59) & 0.01631(36) \\
6.5150 & 182.435 & 0.0603 & 0.006738(062) & 0.07165(5) & 0.3083(153) & 0.9126(44) & 0.7296(100) & 0.4589(58) & 0.01937(16) \\
6.5500 & 188.720 & 0.0582 & 0.005356(051) & 0.06722(5) & 0.1661(165) & 0.8202(54) & 0.7632(037) & 0.5177(31) & 0.02264(38) \\
6.5750 & 193.333 & 0.0564 & 0.004616(022) & 0.06400(3) & 0.0994(078) & 0.7540(65) & 0.7878(114) & 0.5632(64) & 0.02539(20) \\
6.6080 & 199.580 & 0.0542 & 0.003944(024) & 0.06007(4) & 0.0636(043) & 0.6891(47) & 0.8030(036) & 0.6052(32) & 0.02907(30) \\
6.6640 & 210.605 & 0.0514 & 0.003242(020) & 0.05494(5) & 0.0317(027) & 0.6125(26) & 0.8318(031) & 0.6711(29) & 0.03566(28) \\
6.8000 & 239.706 & 0.0448 & 0.002343(009) & 0.04462(2) & 0.0068(011) & 0.5192(13) & 0.8741(021) & 0.7796(21) & 0.05368(24) \\
6.9500 & 275.906 & 0.0386 & 0.001855(005) & 0.03651(1) &-0.0013(007) & 0.4786(06) & 0.8984(016) & 0.8443(17) & 0.07535(24) \\
7.1500 & 331.620 & 0.0320 & 0.001454(003) & 0.02901(1) & 0.0001(009) & 0.4554(04) & 0.9123(014) & 0.8829(14) & 0.10485(30) \\
\hline
\end{tabular}
\caption{Data for the light and strange quark chiral condensate,
  disconnected and connected chiral susceptibility, the light and
  strange quark number susceptibility and the bare Polyakov loop for
  the HISQ/tree action on $N_\tau=8$ lattices and the two lines of
  constant physics defined by $m_l/m_s = 0.025$ and $m_l/m_s = 0.05$.
  Fifteen hundred trajectories were discarded for thermalization.}
\label{tab:hisq_NT8_obs}
\end{table}

\begin{table}[thbp]
\begin{tabular}{|c|c|c||c|c||c|c||c|c|c|}
\hline
$\beta$ &  $T$ MeV &  $a m_s$  & $2 \times \langle \bar \psi \psi \rangle_l$ & $2 \times \langle \bar \psi \psi \rangle_s$ & $\chi_l^{\rm dis}$   & $\chi_l^{\rm con}$  & $\chi_l$ & $\chi_s$ & $3 \times L_{\rm bare}$ \\
\hline
6.6640  & 140.403 &  0.0514 & 0.011032(61) & 0.060172(32) & 0.414(57) & 0.6697(075) & 0.251(57) & 0.1440(072) & 0.001011(095) \\
6.7000  & 145.321 &  0.0496 & 0.009642(38) & 0.057132(18) & 0.437(34) & 0.6652(155) & 0.310(52) & 0.1570(086) & 0.001173(110) \\
6.7400  & 150.967 &  0.0476 & 0.008127(41) & 0.053835(28) & 0.449(23) & 0.6735(060) & 0.403(41) & 0.1987(115) & 0.001434(060) \\
6.7700  & 155.329 &  0.0461 & 0.006998(78) & 0.051354(19) & 0.504(42) & 0.6748(110) & 0.504(37) & 0.2362(072) & 0.001814(075) \\
6.8000  & 159.804 &  0.0448 & 0.006157(63) & 0.049359(18) & 0.543(51) & 0.6731(095) & 0.520(23) & 0.2606(072) & 0.002152(030) \\
6.8400  & 165.949 &  0.0430 & 0.004685(76) & 0.046456(35) & 0.491(27) & 0.6921(068) & 0.657(24) & 0.3571(072) & 0.002821(101) \\
6.8800  & 172.302 &  0.0412 & 0.003745(59) & 0.043759(28) & 0.391(39) & 0.6592(065) & 0.649(19) & 0.3816(101) & 0.003449(067) \\
6.9100  & 177.207 &  0.0401 & 0.003099(47) & 0.041876(36) & 0.253(30) & 0.6421(121) & 0.757(14) & 0.4752(072) & 0.004293(106) \\
6.9500  & 183.938 &  0.0386 & 0.002680(32) & 0.039783(26) & 0.164(13) & 0.6150(067) & 0.743(21) & 0.5069(086) & 0.005024(075) \\
6.9900  & 190.893 &  0.0370 & 0.002307(36) & 0.037522(28) & 0.104(16) & 0.5743(034) & 0.814(17) & 0.5832(086) & 0.006139(092) \\
7.0300  & 198.078 &  0.0356 & 0.002036(14) & 0.035543(24) & 0.062(07) & 0.5318(046) & 0.842(12) & 0.6322(101) & 0.007371(036) \\
7.1000  & 211.225 &  0.0333 & 0.001738(10) & 0.032396(16) & 0.030(04) & 0.5052(046) & 0.855(09) & 0.6998(101) & 0.009511(077) \\
7.1500  & 221.080 &  0.0320 & 0.001611(08) & 0.030768(08) & 0.016(03) & 0.4955(017) & 0.868(05) & 0.7387(043) & 0.011144(095) \\
7.2800  & 248.626 &  0.0285 & 0.001361(07) & 0.026477(08) & 0.012(03) & 0.4710(007) & 0.891(04) & 0.8064(043) & 0.016284(077) \\
\hline
\end{tabular}
\caption{Data for the light and strange quark chiral condensate,
  disconnected and connected chiral susceptibility, the light and
  strange quark number susceptibility and the bare Polyakov loop for
  the HISQ/tree action on $N_\tau=12$ lattices and the LCP defined by
  $m_l/m_s = 0.05$. Two hundred trajectories were discarded for
  thermalization.}
\label{tab:hisq_0.05ms_NT12_obs}
\end{table}

For completeness, we also give the data for the light and strange
quark chiral condensate on zero temperature ensembles for the asqtad
action on $N_\tau=8$ and $12$ lattices in
Table~\ref{tab:asqtad_T0_obs}.  The corresponding data for the
HISQ/tree action are given in Table~\ref{tab:hisq_T0_obs}.

\begin{table}[thbp]
\begin{tabular}{|c|c||c|c|}
\hline
$\beta$ &  $m_s$  & $2 \times \langle \bar \psi \psi \rangle_l$ & $2 \times \langle \bar \psi \psi \rangle_s$ \\
\hline
\multicolumn{4}{|c|}{$32^4$} \\
\hline
6.600 & 0.0650 & 0.04102(5)  &  0.13908(4)  \\
6.625 & 0.0624 & 0.03716(6)  &  0.13127(5)  \\
6.650 & 0.0599 & 0.03389(5)  &  0.12406(4)  \\
6.658 & 0.0590 & 0.03266(7)  &  0.12147(5)  \\
6.666 & 0.0583 & 0.03179(6)  &  0.11944(3)  \\
6.675 & 0.0575 & 0.03086(4)  &  0.11717(3)  \\
6.688 & 0.0563 & 0.02936(4)  &  0.11375(3)  \\
6.700 & 0.0552 & 0.02811(5)  &  0.11071(3)  \\
6.730 & 0.0525 & 0.02506(7)  &  0.10331(4)  \\
6.760 & 0.0500 & 0.02260(5)  &  0.09672(3)  \\
6.800 & 0.0471 & 0.01976(5)  &  0.08896(2)  \\
\hline
\multicolumn{4}{|c|}{$48^4$} \\
\hline
6.800 & 0.0471 & 0.01970(3)  &  0.08886(2)  \\
6.850 & 0.0436 & 0.01672(4)  &  0.07997(3)  \\
6.875 & 0.0420 & 0.01539(3)  &  0.07594(2)  \\
6.900 & 0.0404 & 0.01425(2)  &  0.07213(2)  \\
6.925 & 0.0389 & 0.01319(3)  &  0.06855(2)  \\
6.950 & 0.0375 & 0.01218(2)  &  0.06521(2)  \\
7.000 & 0.0347 & 0.01041(2)  &  0.05888(1)  \\
7.050 & 0.0322 & 0.00898(3)  &  0.05339(2)  \\
7.085 & 0.0305 & 0.00815(3)  &  0.04982(2)  \\
7.125 & 0.0289 & 0.00730(2)  &  0.04640(1)  \\
7.200 & 0.0256 & 0.00583(2)  &  0.03989(1)  \\
7.300 & 0.0220 & 0.00447(3)  &  0.03307(2)  \\
7.400 & 0.0189 & 0.00341(3)  &  0.02748(1)  \\
7.550 & 0.0151 & 0.00223(5)  &  0.02100(2)  \\
\hline
\end{tabular}
\caption{Data for the light and strange quark chiral condensate for
  the asqtad action on $N_\tau=8$ and $12$ zero temperature ensembles
  with the line of constant physics defined by $m_l/m_s = 0.05$.  
The data for the light and strange quark condensate are presented using the 2-flavor normalization.
}
\label{tab:asqtad_T0_obs}
\end{table}

\begin{table}[thbp]
\begin{tabular}{|c|c||c|c|}
\hline
$\beta$ &  $m_s$  & $2 \times \langle \bar \psi \psi \rangle_l$ & $2 \times \langle \bar \psi \psi \rangle_s$ \\
\hline
\multicolumn{4}{|c|}{$24^3 \times 32$} \\
\hline
5.900 & 0.1320 &  0.09833(6)  & 0.21164(3)     \\
6.000 & 0.1138 &  0.07801(22) & 0.18024(10)    \\
6.050 & 0.1064 &  0.06890(8)  & 0.16621(6)     \\
\hline
\multicolumn{4}{|c|}{$28^3 \times 32$} \\
\hline
6.100 & 0.0998 &  0.06070(7)  & 0.15300(6)     \\
\hline
\multicolumn{4}{|c|}{$32^4$} \\                           
\hline
6.195 & 0.0880 & 0.04700(14)  & 0.12983(8)  \\
6.285 & 0.0790 & 0.03660(11)  & 0.11134(3)  \\
6.341 & 0.0740 & 0.03128(8)   & 0.10134(4)  \\
6.354 & 0.0728 & 0.03014(8)   & 0.09900(5)  \\
6.390 & 0.0695 & 0.02723(7)   & 0.09295(4)  \\
6.423 & 0.0670 & 0.02499(7)   & 0.08817(3)  \\
6.488 & 0.0620 & 0.02069(9)   & 0.07906(6)  \\
6.515 & 0.0603 & 0.01920(10)  & 0.07588(5)  \\
6.550 & 0.0582 & 0.01744(9)   & 0.07203(4)  \\
6.575 & 0.0564 & 0.01635(4)   & 0.06908(3)  \\
6.608 & 0.0542 & 0.01485(6)   & 0.06548(5)  \\
6.664 & 0.0514 & 0.01281(6)   & 0.06061(3)  \\
6.800 & 0.0448 & 0.00883(6)   & 0.05008(4)  \\
6.950 & 0.0386 & 0.00611(9)   & 0.04130(3)  \\
7.150 & 0.0320 & 0.00366(8)   & 0.03259(5)  \\
\hline
\multicolumn{4}{|c|}{$32^3 \times 64$} \\       
\hline
6.460 & 0.0642 & 0.02238(5)   &  0.08269(2) \\
\hline
\multicolumn{4}{|c|}{$48^4$} \\ 
\hline
6.740 & 0.0476 & 0.01080(4)   &  0.05445(2)  \\
6.880 & 0.0412 & 0.00765(3)   &  0.04500(1)  \\
7.030 & 0.0356 & 0.00544(4)   &  0.03731(1)  \\
\hline
\multicolumn{4}{|c|}{$48^3 \times 64$} \\       
\hline
7.150 & 0.0320 & 0.00420(5)   &  0.03255(2)  \\
7.280 & 0.0285 & 0.00322(2)   &  0.02815(1)  \\
\hline
\end{tabular}
\caption{Data for the light and strange quark chiral condensate for
  the HISQ/tree action on $N_\tau=6$, $8$ and $12$ zero temperature
  ensembles with the line of constant physics defined by $m_l/m_s =
  0.05$. 
  The data for the light and strange quark condensate are presented using the 2-flavor normalization.
 }
\label{tab:hisq_T0_obs}
\end{table}

\section{Setting the lattice spacing in asqtad and HISQ/tree simulations}
\label{sec:appendix2}

Three observables, $r_0$, $r_1$ and $f_K$, were studied to set the
lattice scale as discussed in Sec.~\ref{sec:parameters}. In
simulations with the asqtad action, we used the values of $r_1$
published in Ref.~\cite{Bazavov:2009bb} and performed an interpolation using a
renormalization group inspired Ansatz \cite{Bazavov:2009bb}
\begin{eqnarray}
& 
\displaystyle\label{asq_r0_fit}
\frac{a}{r_1}=\frac{c_0 f(\beta)+c_2 (10/\beta) f^3(\beta)+ c_4 (10/\beta)^2 f^3(\beta)}{
1+d_2 (10/\beta) f^2(\beta)+ d_4 (10/\beta)^2 f^4(\beta)},\label{ar1_fit} \\[3mm]
&
\displaystyle
f(\beta)=(b_0 (10/\beta))^{-b_1/(2 b_0^2)} \exp(-\beta/(20 b_0)),\label{fbeta}\\[3mm]
&
c_0=c_{00}+(c_{01u} m_l + c_{01s} m_s + c_{02} (2 m_l+m_s))/f(\beta)\\[3mm]
&
c_2=c_{20}+c_{21} (2 m_l+m_s)/f(\beta).
\end{eqnarray}
Here $b_0$ and $b_1$ are the coefficients of the universal 2-loop beta functions and for
the numerical values of the other coefficients we get
\begin{eqnarray}
&
c_{00}  = 4.574615 \cdot 10^{1},
c_{01u} = 6.081198 \cdot 10^{-1},
c_{01s} = 2.689340 \cdot 10^{-1},\nonumber\\
&
c_{02}  = -3.591183 \cdot 10^{-3},
c_{20}  = -5.368781 \cdot 10^{5},
c_{21}  = 8.756186 \cdot 10^{2}, \nonumber\\
&
c_{4}  = 2.930565 \cdot 10^{5},        
d_2  = -3.786570 \cdot 10^{3},
d_4  = 7.385881 \cdot 10^{6}.
\end{eqnarray}

To set the lattice scale using $f_K$, one needs to measure $a f_K$ on
lattices with large volumes and on high statistics ensembles.  The zero
temperature asqtad ensembles generated in this study (see
Tables~\ref{tab:asqtad8_runs} and \ref{tab:asqtad12_runs}) were
primarily for performing subtractions of UV divergences. We,
therefore, used the MILC collaboration study of the light meson decay
constants and masses \cite{Bernard:2007ps}. A systematic fit of $f_K$
data, obtained on a large set of ensembles with similar parameters
values as in this study, using staggered chiral perturbation theory gives values for
$f_K r_1$ at nine $\beta$ values in the range $[6.458, 7.47]$.  We
interpolated these results to obtain $f_K r_1$ at the physical quark
mass at the $\beta$ values used in the $0.05 m_s$ LCP study and given in
Tables~\ref{tab:asqtad8_runs} and \ref{tab:asqtad12_runs}.  With $f_K r_1$ and 
our parametrization of $r_1/a$ in hand we calculated the temperature associated 
with each $\beta$. The uncertainty in these $T$ values is less
than 1\% over the range of $\beta$ values investigated.

In simulations with the HISQ/tree action, we used $r_0$, $r_1$ and
$f_K$ to set the lattice spacing. We determined $r_0/a$ and $r_1/a$
values from the static quark anti-quark potential. These values are
summarized in Tables~\ref{tab:T0_0.2ms_r0} and \ref{tab:r0} for the
$m_l=0.2m_s$ and the $m_l=0.05m_s$ LCP, respectively. The last column
contains the values of the additive renormalization constant $c(\beta)$ defined 
in Eq.~(\ref{eq:Lrenorm}) and used in the calculation of the renormalized Polyakov
loop.  
\begin{table}
\begin{tabular}{|r|r|r|r|r|}
\hline
$\beta$ &  $r_0/a$     &  $r_1/a$    & $r_0^{fit}/a$ & $c(\beta) \cdot r_0$       \\
\hline
6.000   & 2.037(12) & 1.410(13) &  2.052       & -1.622(17) \\
6.038   & 2.141(12) & 1.473(12) &  2.128       & -1.678(28) \\
6.100   & 2.250(12) & 1.544(19) &  2.256       & -1.834(44) \\
6.167   & 2.413(12) & 1.659(19) &  2.403       & -1.965(66) \\
6.200   & 2.501(17) & 1.722(20) &  2.478       & -2.008(26) \\
6.227   & 2.537(12) & 1.745(12) &  2.542       & -2.094(52) \\
6.256   & 2.603(14) & 1.798(20) &  2.611       & -2.124(53) \\
6.285   & 2.715(25) & 1.813(60) &  2.683       & -2.167(32) \\
6.313   & 2.742(20) & 1.848(12) &  2.753       & -2.274(39) \\
6.341   & 2.802(32) & 1.910(10) &  2.826       & -2.299(54) \\
6.369   & 2.916(39) & 1.983(14) &  2.900       & -2.331(66) \\
6.396   & 2.937(30) & 2.016(20) &  2.973       & -2.420(61) \\
6.450   & 3.110(30) & 2.132(14) &  3.124       & -2.546(84) \\
6.800   & 4.330(65) & 2.962(26) &  4.278       & -3.458(73) \\
\hline
\end{tabular}
\caption{Estimates of the scale setting parameters $r_0$, $r_1$ and
  the additive renormalization constant $c(\beta)$ in the
  determination of the potential for the HISQ/tree calculations along the 
  $m_l=0.2m_s$ LCP.  See text for the definition of $r_0^{fit}$.
}
\label{tab:T0_0.2ms_r0}
\end{table}
 
\begin{table}
\begin{tabular}{|r|r|r|r|r|}
\hline
$\beta$ & $r_0/a$     & $r_1/a$     &  $r_1^{fit}/a$ &  $c(\beta) \cdot r_0$       \\
\hline
5.900   & 1.909(11) & 1.23(13)  &  1.263       & -1.441(15) \\
6.000   & 2.094(21) & 1.386(80) &  1.391       & -1.639(28) \\
6.050   & 2.194(22) & 1.440(31) &  1.460       & -1.748(28) \\
6.100   & 2.289(21) & 1.522(30) &  1.533       & -1.828(27) \\
6.195   & 2.531(24) & 1.670(30) &  1.682       & -2.072(31) \\
6.285   & 2.750(30) & 1.822(30) &  1.836       & -2.257(36) \\
6.341   & 2.939(11) & 1.935(30) &  1.940       & -2.440(14) \\
6.354   & 2.986(41) & 1.959(30) &  1.964       & -2.498(49) \\
6.423   & 3.189(22) & 2.096(21) &  2.101       & -2.653(27) \\
6.460   & 3.282(32) & 2.165(20) &  2.178       & -2.706(36) \\
6.488   & 3.395(31) & 2.235(21) &  2.238       & -2.808(37) \\
6.550   & 3.585(14) & 2.369(21) &  2.377       & -2.946(17) \\
6.608   & 3.774(20) & 2.518(21) &  2.513       & -3.070(27) \\
6.664   & 3.994(14) & 2.644(23) &  2.652       & -3.251(16) \\
6.800   & 4.541(30) & 3.025(22) &  3.019       & -3.675(31) \\
6.880   & 4.901(18) & 3.246(22) &  3.255       & -3.896(18) \\
6.950   & 5.249(20) & 3.478(23) &  3.475       & -4.077(40) \\
7.030   & 5.668(49) & 3.728(26) &  3.742       & -4.439(47) \\
7.150   & 6.275(39) & 4.177(31) &  4.176       & -4.791(37) \\
7.280   & 6.991(72) & 4.705(26) &  4.697       & -5.210(89) \\
\hline
\end{tabular}
\caption{Estimates of the scale setting parameters $r_0$, $r_1$ and
  the additive renormalization constant $c(\beta)$ in the
  determination of the potential for the HISQ/tree calculations along the 
  $m_l=0.05m_s$ LCP.  See the text for the definition of $r_1^{fit}/a$ and fit 
  details. 
}
\label{tab:r0}
\end{table}

For the $m_l=0.2m_s$ LCP, we fit $r_0/a$ values with a form similar
to Eq.~(\ref{asq_r0_fit}):
\begin{equation}
\displaystyle\label{hisq_r0_fit_0.2ms}
\frac{a}{r_0}(\beta)_{m_l=0.2m_s}=
\frac{c_0 f(\beta)+c_2 (10/\beta) f^3(\beta)}{
1+d_2 (10/\beta) f^2(\beta)},
\end{equation}
where $c_0=32.83$, $c_2=81127$, $d_2=1778$ and $\chi^2/{\rm dof}=1.01$.
Then we use $r_0=0.468$ fm obtained in Sec.~\ref{sec:parameters}
to calculate the lattice spacing in units of fm.

\begin{figure}
\includegraphics[width=0.5\textwidth]{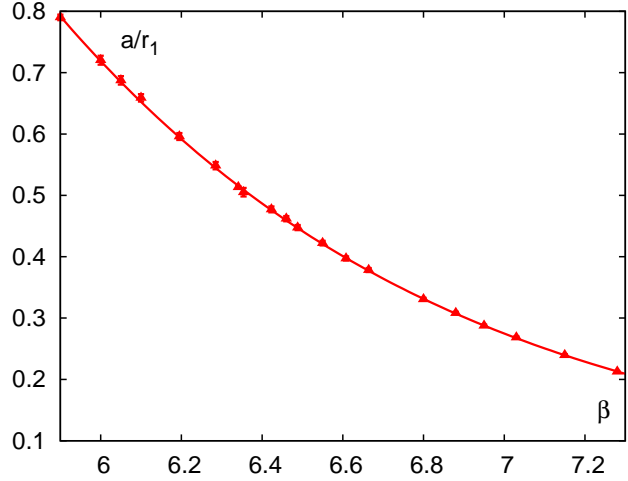}
\caption{The $a/r_1$ data, defined in Eq.~(\ref{ar1_def}), together with the 
smoothing fit for the HISQ/tree
action, $m_l=0.05m_s$ LCP.}
\label{fig_r1scale}
\end{figure}

For the $0.05m_s$ ensembles, we preferred to use the $r_1$ scale, since the
value of $r_1$ in fm is accurately determined by the MILC 
collaboration~\cite{Bazavov:2010hj}
and serves as an external input for this work.
As one can see from Table~\ref{tab:r0}, our data set includes coarse
lattices where $r_1/a<2$. Extracting the value of $r_1/a$ at such short
distances is problematic due to significant systematic errors.
Therefore, we used the following strategy: for coarser lattices
we converted $r_0/a$ values into $r_1/a$ using the continuum
ratio $(r_0/r_1)_{cont}$ derived in Sec.~\ref{sec:parameters} and
defined as
\begin{equation}
\label{ar1_def}
\left.\frac{r_1}{a}\right|_{m_l=0.05m_s}\equiv\left\{
\begin{array}{c}
r_0/a/(r_0/r_1)_{cont},\,\,\,\,\beta<\beta_{01}, \\
r_1/a,\,\,\,\,\beta\geqslant\beta_{01},
\end{array}\right.,
\end{equation}
where the optimal value of $\beta_{01}$ is determined as follows. We fit
$a/r_1$ to the Ansatz:
\begin{equation}
\displaystyle\label{hisq_r1_fit_0.05ms}
\frac{a}{r_1}(\beta)_{m_l=0.05m_s}=
\frac{c_0 f(\beta)+c_2 (10/\beta) f^3(\beta)}{
1+d_2 (10/\beta) f^2(\beta)} \,
\end{equation}
and varied $\beta_{01}$ in the range $[6.423,6.608]$ examining the
$\chi^2$ of the fit. The minimum of $\chi^2$ is achieved at
$\beta_{01}=6.423$. With this choice of $\beta_{01}$ the coefficients
of the fit are $c_0=44.06$, $c_2=272102$, $d_2=4281$ and
$\chi^2/{\rm dof}=0.31$.  The data and the fit are shown in
Fig.~\ref{fig_r1scale}.

\begin{figure}
\includegraphics[width=0.5\textwidth]{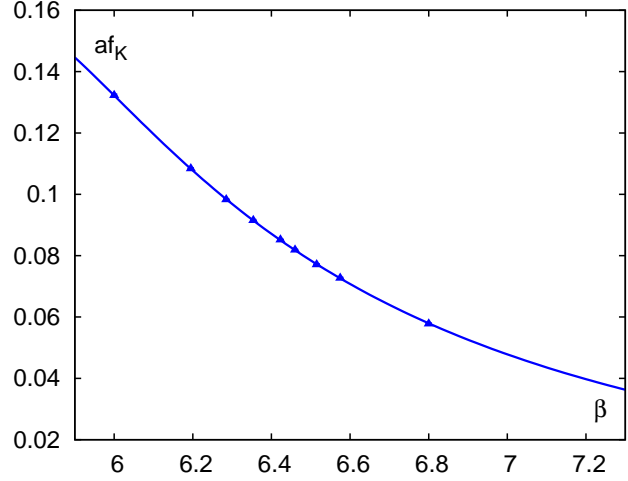}
\caption{The $af_K$ data together with the smoothing fit for the
  HISQ/tree action along the LCP defined by $m_l=0.05m_s$.}
\label{fig_fKscale}
\end{figure}

We also use $f_K$ as an independent observable for scale setting.
We fit $af_K$ data to a form similar to (\ref{hisq_r1_fit_0.05ms}):
\begin{equation}
\displaystyle\label{hisq_fK_fit_0.05ms}
af_K(\beta)_{m_l=0.05m_s}=
\frac{c^K_0 f(\beta)+c^K_2 (10/\beta) f^3(\beta)}{1+d^K_2 (10/\beta) f^2(\beta)}.
\end{equation}
In the continuum limit, the product $r_1f_K$ is fixed:
\begin{equation}
  r_1f_K=\frac{0.3106\mbox{ fm}\cdot 156.1/\sqrt{2}\mbox{ MeV}}
  {197.3\mbox{ fm}\cdot\mbox{MeV}}
  \simeq 0.1738.
\end{equation}
Taking the ratio of Eqs.~(\ref{hisq_fK_fit_0.05ms}) and (\ref{hisq_r1_fit_0.05ms}) one
finds
\begin{equation}
  r_1f_K=\frac{c_0^K}{c_0}\,\,\,\,\,\Rightarrow\,\,\,\,\, c_0^K=7.66.
\end{equation}
We, therefore, varied only $c_2^K$ and $d_2^K$ in the
Ansatz~(\ref{hisq_fK_fit_0.05ms}) and got $c_2^K=32911$, $d_2^K=2388$,
$\chi^2/{\rm dof}=3.5$. Presumably, the errors on $af_K$ are somewhat
underestimated and this results in a high $\chi^2/{\rm dof}$. However,
the fluctuations of the data around the fit are, at worst, 1\%;
therefore we feel comfortable using the Ansatz in 
Eq.~(\ref{hisq_fK_fit_0.05ms}) for setting the temperature scale. This
fit is shown in Fig.~\ref{fig_fKscale}.

\begin{figure}
\includegraphics[width=0.5\textwidth]{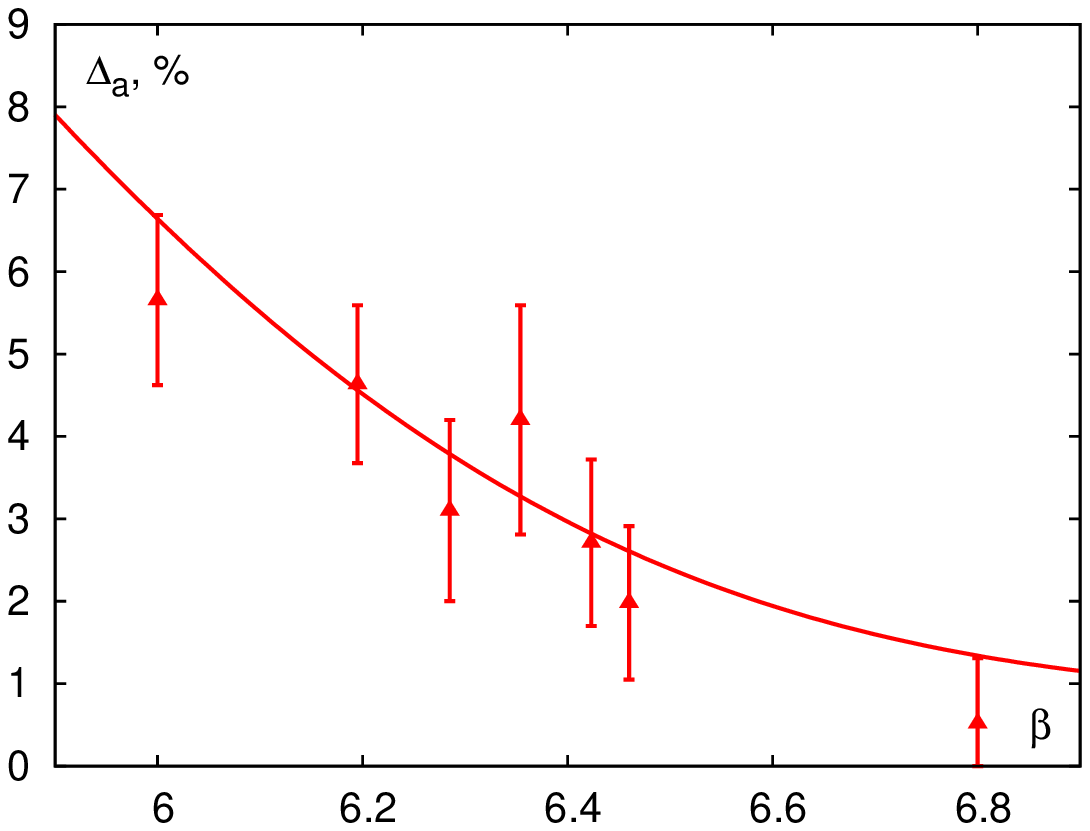}
\caption{The percentage difference in the lattice spacing determined
  from $r_1$ and $f_K$ $vs.$ the gauge coupling $\beta$ for the
  HISQ/tree action. The line corresponds to the difference calculated
  from fits to Eqs.~(\ref{hisq_r1_fit_0.05ms}) and (\ref{hisq_fK_fit_0.05ms}).}
\label{fig_a_diff}
\end{figure}

For those ensembles where $a/r_1$ and $af_K$ measurements are available, we
can compare the difference between the lattice spacing defined from $r_1$
and $f_K$. We define
\begin{equation}
a_{r_1}\equiv\frac{1}{\left.r_1/a\right|_{m_l=0.05m_s}} \times 0.3106\mbox{ fm},
\end{equation}
\begin{equation}
a_{f_K}\equiv(af_K)\cdot\frac{197.3\mbox{ Mev$\cdot$fm}}{156.1/\sqrt{2}\mbox{ MeV}},
\end{equation}
and
\begin{equation}
\Delta_a\equiv\frac{\left|a_{f_K}-a_{r_1}\right|}{a_{r_1}}\cdot 100\%.
\end{equation}
The difference $\Delta_a$ is shown in Fig.~\ref{fig_a_diff}. It is about 8\% in the
lattice spacings corresponding to the coarsest ensemble and decreases to about 1\% for the finest
$N_\tau=12$ ensembles on which $f_K$ was measured. The temperature corresponding to 
these fine lattices is about 240 MeV.

\section{Hadron spectrum with the HISQ/tree action}
\label{sec:appendix3}

This appendix contains results of the calculations of the hadron
correlation functions and some details on the extraction of the masses
and decay constants.  Fits to the correlation functions for mesons and
baryons are performed choosing either the maximum allowed separation,
typically $N_t/2=16$, or up to a distance where the fractional error
exceeds 30\%. The minimum distance is varied to estimate the
systematic effects due to contamination from excited states.  In most
cases, the fit function includes the ground state and the opposite
parity state that contributes with an alternating sign, and their contribution
due to back-propagation. In some cases, the first excited state is also
included.  To reduce the uncertainty due to auto-correlations in the
estimation of statistical errors, we block the data by 10.

The calculation of hadron correlators was done using wall sources. The
pseudoscalar meson masses were calculated on most of our zero
temperature lattices in order to establish the LCP. The data for the
$m_l = 0.2m_s$ LCP are given in Table~\ref{tab:T0_0.2ms_mps}, and for
$0.05m_s$ in Table~\ref{tab:mps}.  We also calculated the pseudoscalar
meson decay constants for different combinations of the quark
masses. For these calculations, we used both wall and point sources and
performed a simultaneous fit to the two correlators. In most
cases, we used one wall source per lattice, but in some cases we used
two or four sources.  In Table~\ref{tab:fpi}, we give the pseudoscalar
meson decay constants in lattice units and specify the number of
sources per lattice used in the calculations.

\begin{table}
\begin{tabular}{|c|c|c|c|}
\hline
$\beta$ & $a M_{\pi}$   & $a M_K$       & $a M_{\eta_{s \bar s}}$  \\
\hline
6.000   & 0.36320(15) & 0.61988(19) & 0.79730(14)            \\
6.038   & 0.34838(14) & 0.59472(18) & 0.76515(15)            \\
6.100   & 0.32895(16) & 0.56187(20) & 0.72335(17)            \\
6.167   & 0.30725(19) & 0.52423(23) & 0.67508(18)            \\
6.200   & 0.29711(34) & 0.50662(36) & 0.65257(32)            \\
6.227   & 0.28871(21) & 0.49280(23) & 0.63519(26)            \\
6.256   & 0.28044(23) & 0.47834(26) & 0.61653(26)            \\
6.285   & 0.27332(25) & 0.46636(33) & 0.60156(33)            \\
6.313   & 0.26543(26) & 0.45313(34) & 0.58415(31)            \\
6.341   & 0.25970(27) & 0.44252(31) & 0.57037(30)            \\
6.369   & 0.25313(32) & 0.43176(43) & 0.55628(35)            \\
6.396   & 0.24563(32) & 0.41907(35) & 0.54098(31)            \\
6.450   & 0.23790(30) & 0.40540(50) & 0.52240(40)            \\
\hline
\end{tabular}
\caption{
The pseudoscalar meson masses for the HISQ/tree action along the 
$m_l=0.2m_s$ LCP.}
\label{tab:T0_0.2ms_mps}
\end{table}
 
\begin{table}
\begin{tabular}{|c|c|c|c|}
\hline
$\beta$ & $a M_{\pi}$   & $a M_K$       & $a M_{\eta_{s \bar s}}$ \\
\hline
5.900   & 0.20162(09) & 0.63407(17) & 0.86972(11)  \\
6.000   & 0.18381(37) & 0.57532(51) & 0.79046(27)  \\
6.195   & 0.15143(14) & 0.47596(16) & 0.65506(11)  \\
6.285   & 0.13823(50) & 0.43501(47) & 0.59951(28)  \\
6.354   & 0.12923(15) & 0.40628(20) & 0.55982(17)  \\
6.423   & 0.12022(12) & 0.37829(19) & 0.52161(17)  \\
6.460   & 0.11528(21) & 0.36272(34) & 0.50137(32)  \\
6.488   & 0.11245(15) & 0.35313(27) & 0.48716(17)  \\
6.515   & 0.10975(12) & 0.34453(29) & 0.47516(29)  \\
6.550   & 0.10629(16) & 0.33322(38) & 0.45989(24)  \\
6.575   & 0.10469(68) & 0.32521(55) & 0.44869(50)  \\
6.608   & 0.10001(17) & 0.31333(28) & 0.43286(29)  \\
6.664   & 0.09572(18) & 0.29837(37) & 0.41178(32)  \\
6.800   & 0.0849(18)  & 0.26387(99) & 0.36257(68)  \\
7.280   & 0.05399(43) & 0.17128(44) & 0.23766(30)  \\
\hline
\end{tabular}
\caption{
The pseudoscalar meson masses for the HISQ/tree action along the 
$m_l=0.05m_s$ LCP.}
\label{tab:mps}
\end{table}

\begin{table}
\begin{tabular}{|c|c|c|c|c|}
\hline
$\beta$ & $a f_{\pi}$   & $a f_K$       & $a f_{s \bar s}$  & \# sources\\
\hline
6.000   & 0.11243(21) & 0.13224(31) & 0.15290(22)     &  1\\
6.195   & 0.09179(21) & 0.10835(13) & 0.12525(13)     &  1\\
6.285   & 0.08366(22) & 0.09826(13) & 0.11390(13)     &  1\\
6.354   & 0.07825(40) & 0.09146(19) & 0.10598(11)     &  1\\
6.423   & 0.07241(18) & 0.08515(11) & 0.09854(07)     &  2\\
6.460   & 0.06885(11) & 0.08185(09) & 0.09454(08)     &  4\\
6.515   & 0.06534(18) & 0.07707(15) & 0.08946(09)     &  4\\
6.575   & 0.06104(49) & 0.07265(19) & 0.08405(14)     &  2\\
6.800   & 0.04883(83) & 0.05774(18) & 0.06717(14)     &  1\\
\hline
\end{tabular}
\caption{Estimates of decay constants of the pseudoscalar mesons in
  lattice units for the HISQ/tree action along the $m_l=0.05 m_s$
  LCP. We use the normalization in which $f_\pi\sim 90$ MeV. In the
  last column, we list the number of source points used on each configuration
  to increase the statistics.}
\label{tab:fpi}
\end{table}

We also calculated the vector meson and the baryon (nucleon and
$\Omega$) masses on some of the $T=0$ ensembles. The results 
are given in Table \ref{tab:mVN_0.2ms} for the $m_l = 0.2m_s$ LCP
and in Table \ref{tab:mVB_0.05ms} for the $m_l = 0.05m_s$ LCP and discussed 
in Sec.~\ref{sec:parameters}. 
To improve the statistics we used 4 sources per lattice for the calculation of vector mesons and nucleon correlation
functions. It was sufficient to use one source for the ÿÿ baryon as it is composed of three heavier strange quarks.

\begin{table}
\begin{tabular}{|c|c|c|c|c|}
\hline
$\beta$  & $a M_{\rho}$  & $a M_{K^*}$  & $a M_{\phi}$ & $a M_N$ \\
\hline
6.000   & 1.0292(199) & 1.1263(82) & 1.2363(24) & 1.3557(100) \\
6.038   & 0.9763(67)  & 1.0867(40) & 1.1907(22) & 1.3277(77)  \\
6.100   & 0.9160(117) & 1.0230(34) & 1.1211(38) & 1.2418(116) \\
6.167   & 0.8499(52)  & 0.9570(41) & 1.0491(18) & 1.1560(55)  \\
6.200   & 0.8300(83)  & 0.9288(48) & 1.0133(24) & 1.1273(73)  \\
6.227   & 0.8002(31)  & 0.9045(37) & 0.9907(25) & 1.0662(92)  \\
6.256   & 0.7822(27)  & 0.8728(48) & 0.9548(14) & 1.0563(52)  \\
6.285   & 0.7700(36)  & 0.8602(33) & 0.9386(22) & 1.0311(25)  \\
6.313   & 0.7462(40)  & 0.8344(35) & 0.9089(21) & 1.0050(29)  \\
6.341   & 0.7170(24)  & 0.8019(30) & 0.8780(17) & 0.9705(36)  \\
6.369   & 0.7006(21)  & 0.7818(23) & 0.8572(26) & 0.9524(25)  \\
6.396   & 0.6843(32)  & 0.7605(20) & 0.8311(29) & 0.9246(37)  \\
6.450   & 0.6505(55)  & 0.7206(43) & 0.7962(18) & 0.8637(106) \\
\hline
\end{tabular}
\caption{The vector meson and the nucleon masses for the HISQ/tree
  action along the $m_l = 0.2m_s$ LCP.}
\label{tab:mVN_0.2ms}
\end{table}

\begin{table}
\begin{tabular}{|c|c|c|c|c|c|}
\hline
$\beta$  & $a M_{\rho}$  & $a M_{K^*}$  & $a M_{\phi}$ & $a M_N$       & $a M_{\Omega}$ \\
\hline
6.000   &             &            &             &             & 1.994(13) \\
6.195   & 0.7562(36)  & 0.8842(18) & 1.00500(93) & 1.0114(56)  & 1.665(17) \\
6.285   &             &            &             &             & 1.4873(67)\\
6.354   & 0.6375(35)  & 0.7499(26) & 0.85234(77) & 0.8315(95)  & 1.374(60) \\
6.423   & 0.6047(43)  & 0.6950(22) & 0.79246(83) & 0.7899(58)  &           \\
6.460   & 0.578(25)   & 0.6709(43) & 0.7644(22)  &             &           \\
6.488   & 0.5647(24)  & 0.6478(22) & 0.73630(68) & 0.7452(51)  &           \\
6.515   & 0.5452(59)  & 0.6358(40) & 0.7167(10)  & 0.7247(55)  &           \\
6.550   & 0.5324(24)  & 0.6118(20) & 0.6929(14)  & 0.7010(90)  &           \\
6.608   & 0.5072(39)  & 0.57572(84)& 0.65227(98) & 0.6442(69)  &           \\
6.664   & 0.4732(43)  & 0.5501(26) & 0.6180(10)  & 0.6048(111) &           \\
6.800   &             &            &             &             & 0.8725(42)\\
\hline
\end{tabular}
\caption{Estimates of the vector mesons, the nucleon and the $\Omega$
  baryon masses in lattice units for the HISQ action along the 
  $m_l=0.05m_s$ LCP.}
\label{tab:mVB_0.05ms}
\end{table}

\section{Autocorrelations in HISQ/tree simulations}
\label{sec:appendix4}

An analysis of autocorrelations in zero-temperature calculations with
the HISQ action at the light quark mass $m_l=0.2m_s$ has been
presented in Ref.~\cite{Bazavov:2010ru}. Note that the HISQ/tree
action used in this paper differs from the HISQ action in
Ref.~\cite{Bazavov:2010ru} in that it does not include a dynamical charm
quark and uses only the tree-level instead of the one-loop improved coefficient for the 
Symanzik gauge action. We calculate the dimensionless autocorrelation
coefficient defined as 
\begin{equation}\label{Cdeltat}
C_{\Delta t}=\frac{\langle x_i x_{i+\Delta t}\rangle
 -\langle x_i\rangle^2}{\langle x_i^2 \rangle 
 -\langle x_i\rangle^2}
\end{equation}
for several representative ensembles both at zero and finite
temperature at $m_l=0.05m_s$. 

The autocorrelation coefficients for the plaquette, light and strange
quark chiral condensate and the topological charge are shown in
Tables~\ref{tab_hisq_corr_T0} and \ref{tab_hisq_corr_T}. At zero
temperature, the autocorrelation coefficient is given at time
separation $\Delta t=5$ for $\beta=6.000$, 6.285 and 6.460, and $\Delta t=6$ for
$\beta=7.280$. At finite temperature, the autocorrelation coefficient
is given at time separation $\Delta t=10$.  The chiral condensate was measured using one
source on zero temperature lattices and with ten sources on finite
temperature lattices. In several cases (mostly for zero temperature
ensembles), the autocorrelation function for the light quark chiral
condensate is too noisy to extract $C_{\Delta t}$ reliably.  

The integrated autocorrelation time, 
\begin{equation}
\tau_{int}=1+2\sum_{\Delta t=1}^{\infty} C(\Delta t) \, ,
\end{equation}
provides an estimate of the time separation at which measurements can
be considered statistically independent. To calculate $\tau_{int}$
reliably requires time series substantially longer than generated in
this study, so we provide a rough estimate.  We assume that the
autocorrelation function is dominated by a single exponential, in
which case
\begin{equation}
  C(\Delta t) = \exp (-{\Delta t}/\tau_1) \, .
\label{eq:Cdelta}
\end{equation}
and 
\begin{equation}
  \tau_1 = -\frac{\Delta t}{\ln C_{\Delta t}}.
\end{equation}
The integrated autocorrelation time is then given by
\begin{equation}
  \tau_{int} \approx 1+2\sum_{\Delta t=1}^{\infty}\exp( -{\Delta t}/\tau_1 )
  =1+2\,\frac{\exp(-1/\tau_1)}{1-\exp(-1/\tau_1)}
  \approx 1+2\tau_1 \, .
\end{equation}

Our data indicate that Eq.~(\ref{eq:Cdelta}) is a reasonable
approximation for the plaquette but not for the chiral condensate and
the topological susceptibility. To get a rough estimate of
$\tau_{int}$ we consider the data for the plaquette given in Table
\ref{tab_hisq_corr_T0}. Estimates of $\tau_1$ vary between 3.4--5.1
and, consequently, for $\tau_{int}$ between 7--10.  Estimates of
$\tau_{int}$ for the chiral condensate and the topological
susceptibility could be larger due to multiple exponentials
contributing long tails even though the $C_{\Delta t}$ are smaller.
Based on these rough estimates for the plaquette, we typically save
every tenth lattice for further measurements, for example, the quark
number susceptibility. We also check for the statistical significance
of the data for a given observable by binning to the extent justified
by the statistics.

\begin{figure}[thbp]
\includegraphics[width=0.48\textwidth]{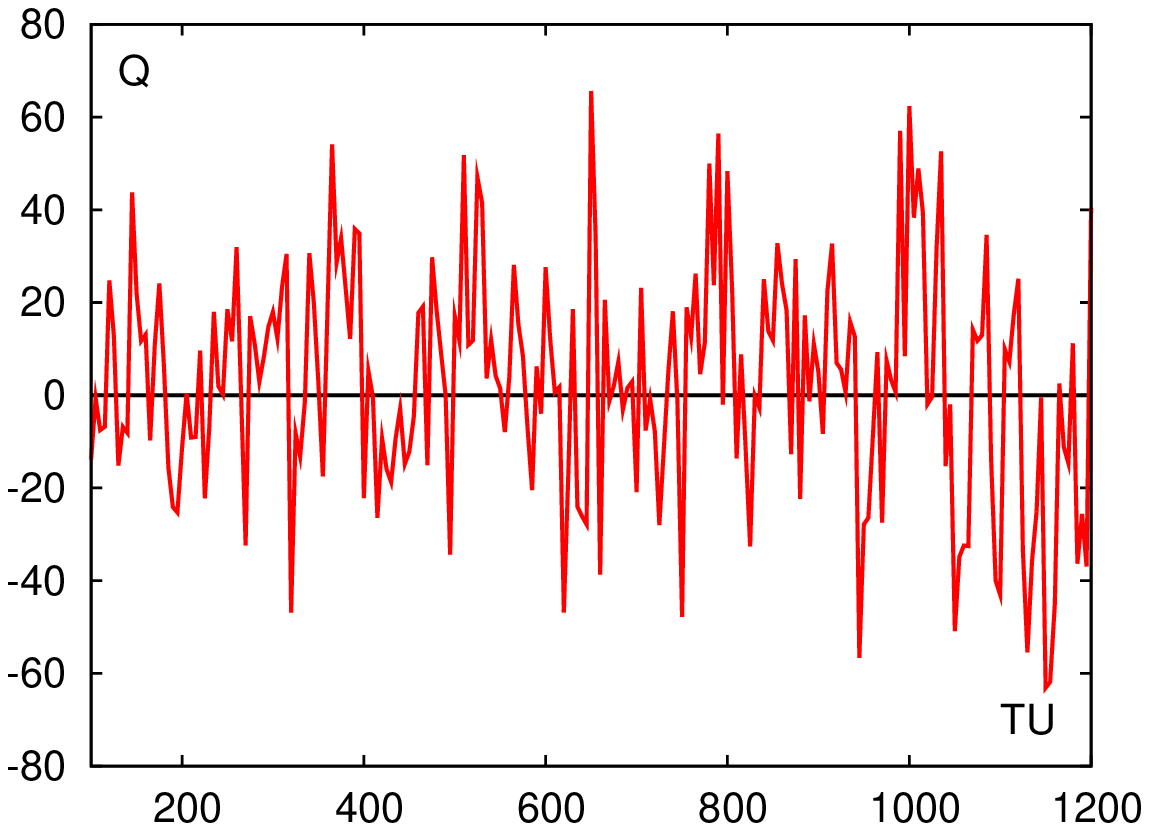}
\caption{Time history of the topological charge measured every 5 time units
on $24^3\times32$ configurations at $\beta=6.000$ with the HISQ/tree action.}
\label{fig_topo2432b6000}
\end{figure}

\begin{figure}[thbp]
\includegraphics[width=0.48\textwidth]{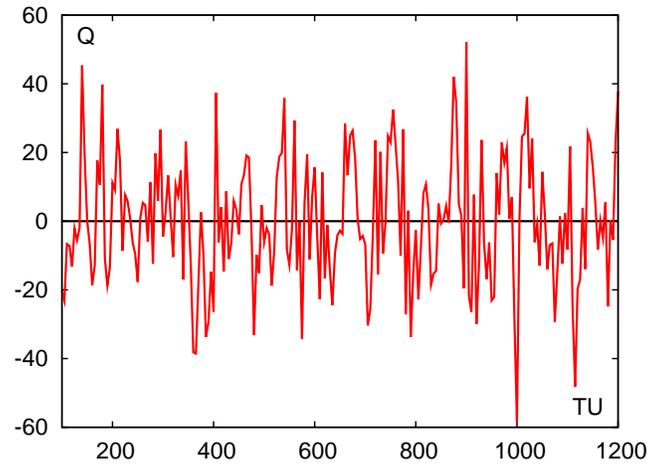}
\caption{Time history of the topological charge measured every 5 time units
on $32^4$ configurations at $\beta=6.285$ with the HISQ/tree action.}
\label{fig_topo3232b6285}
\end{figure}

\begin{figure}[thbp]
\includegraphics[width=0.48\textwidth]{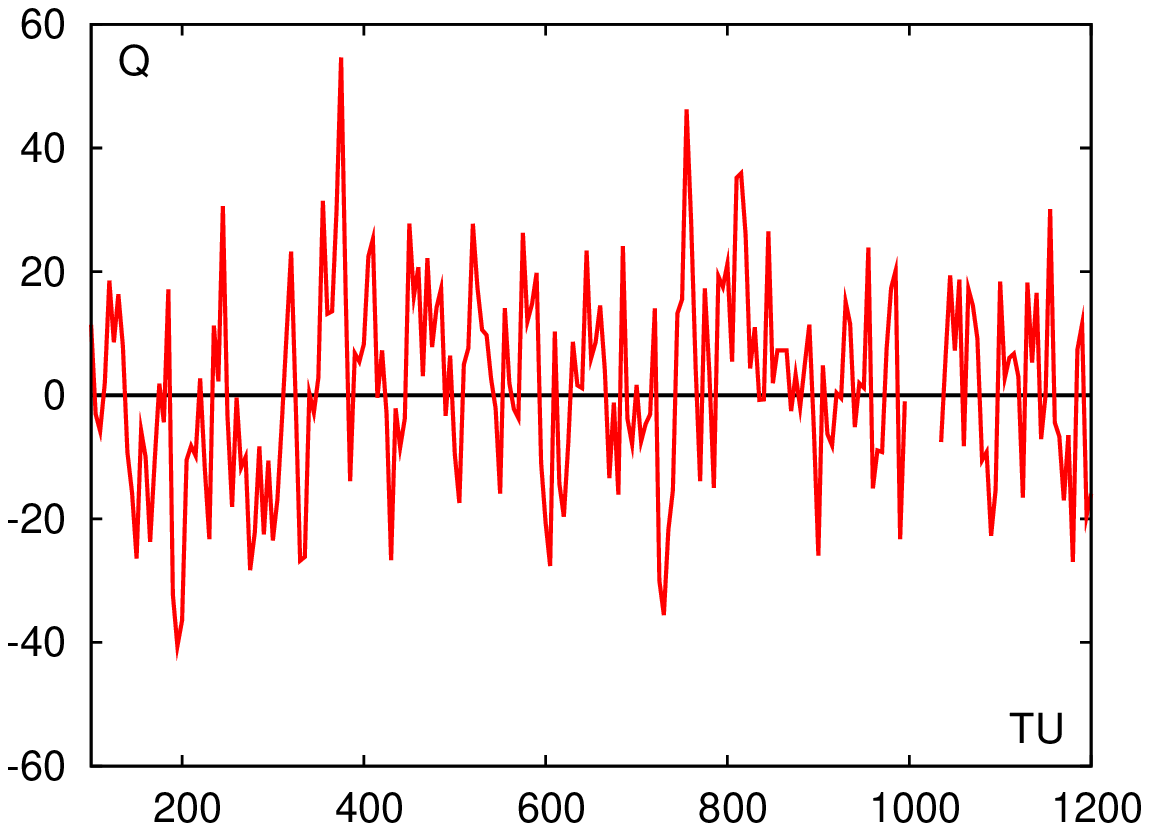}
\caption{Time history of the topological charge measured every 5 time units
on $32^3\times64$ configurations on $\beta=6.460$ with the HISQ/tree action.}
\label{fig_topo3264b6460}
\end{figure}

\begin{figure}[thbp]
\includegraphics[width=0.48\textwidth]{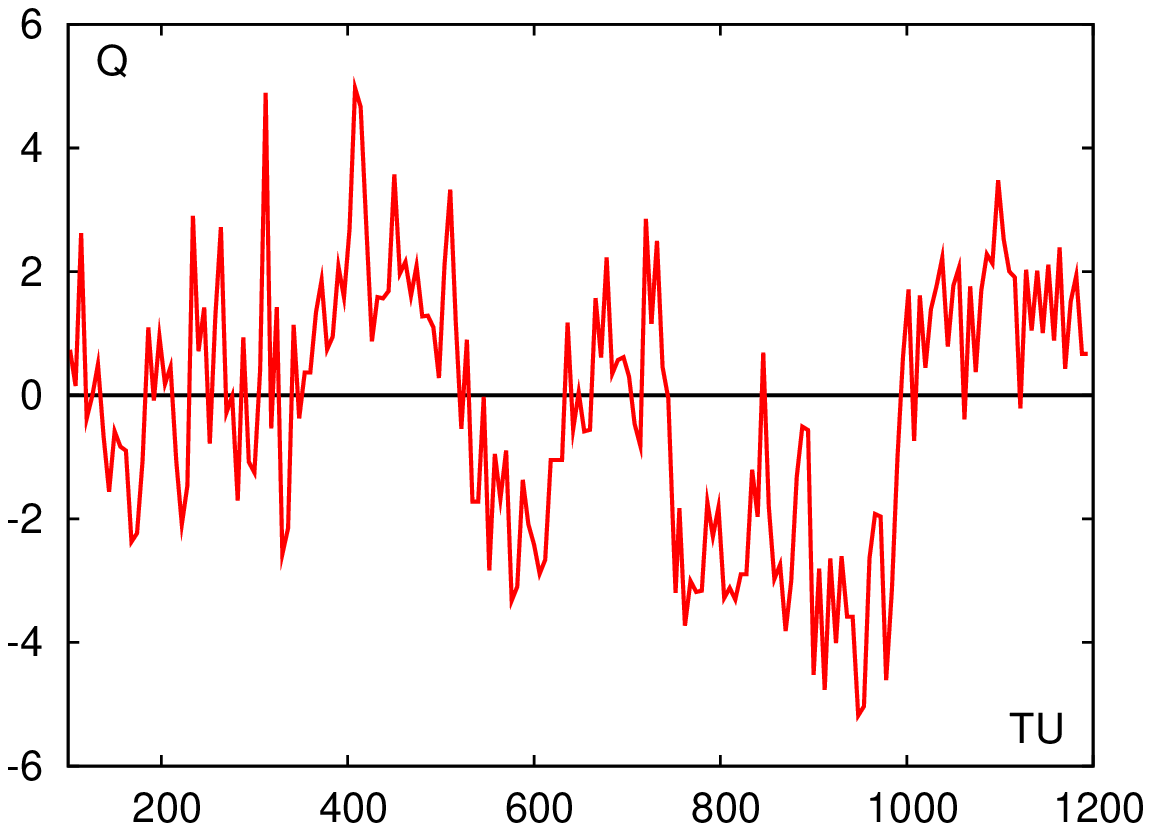}
\caption{Time history of the topological charge measured every 6 time units
on $48^3\times64$ configurations at  $\beta=7.280$ with the HISQ/tree action.}
\label{fig_topo4864b7280}
\end{figure}

The topological charge $Q$ is calculated on zero-temperature ensembles
following the prescription in Ref.~\cite{Bazavov:2010xr}.  The
corresponding time histories are shown in
Figs.~\ref{fig_topo2432b6000}--\ref{fig_topo4864b7280}.  In these
figures, the number of TUs examined is set to 1200
(corresponding to the shortest time series at $\beta=7.280$) to
make comparison easier. 

\begin{table}[thbp]
\begin{tabular}{|c|c|c|c|c|}
\hline
$\beta$ & $\Delta t$ &    $\Box$  & $\langle\bar\psi\psi\rangle_s$ & $Q$ \\\hline
6.000   & 5 &  0.378(42) & 0.073(14) & 0.361(57) \\
6.285   & 5 &  0.369(29) & 0.076(19) & 0.262(39) \\
6.460   & 5 &  0.336(28) & 0.120(21) & 0.397(51) \\
7.280   & 6 &  0.168(34) & 0.025(40) &  -        \\
\hline
\end{tabular}
\caption{Estimates of the autocorrelation coefficient $C_{\Delta t}$
  with $\Delta t$ in $TU$ for a representative set of the HISQ/tree
  zero-temperature lattices measured using the plaquette, strange
  quark chiral condensate and the topological charge.}
\label{tab_hisq_corr_T0}
\end{table}

\begin{table}[thbp]
\begin{tabular}{|c||c|c|c||c|c|c||c|c|c|}
\hline
 & \multicolumn{3}{c||}{$N_\tau=6$} 
 & \multicolumn{3}{c||}{$N_\tau=8$} 
 & \multicolumn{3}{c|}{$N_\tau=12$} \\ \hline
$\beta$ & $\Box$  & $\langle\bar\psi\psi\rangle_l$ & $\langle\bar\psi\psi\rangle_s$ 
        & $\Box$  & $\langle\bar\psi\psi\rangle_l$ & $\langle\bar\psi\psi\rangle_s$ 
        & $\Box$  & $\langle\bar\psi\psi\rangle_l$ & $\langle\bar\psi\psi\rangle_s$ \\\hline
6.000 & 0.172(18) & 0.119(14) & 0.219(22) & --        & --        & --        & --        & --        & --        \\
6.100 & 0.158(09) & 0.219(06) & 0.270(11) & --        & --        & --        & --        & --        & --        \\
6.195 & 0.121(16) & 0.313(29) & 0.371(23) & 0.156(19) & 0.096(18) & 0.226(29) & --        & --        & --        \\
6.285 & --        & --        & --        & 0.116(09) & 0.163(12) & 0.218(15) & --        & --        & --        \\
6.354 & 0.085(12) & 0.068(18) & 0.295(39) & 0.168(34) & 0.262(33) & 0.328(29) & --        & --        & --        \\
6.460 & --        & --        & --        & 0.155(09) & 0.353(21) & 0.413(30) & --        & --        & --        \\
6.550 & 0.129(16) & N/A       & 0.172(21) & 0.108(17) & 0.238(37) & 0.413(36) & --        & --        & --        \\
6.664 & --        & --        & --        & 0.132(29) & 0.078(20) & 0.337(32) & 0.117(37) & 0.285(34) & 0.342(42) \\
6.800 & 0.027(17) & N/A       & N/A       & 0.117(13) & 0.043(10) & 0.209(10) & 0.091(22) & 0.392(46) & 0.298(42) \\
6.950 & --        & --        & --        & 0.063(28) & N/A       & 0.085(14) & 0.121(12) & 0.355(31) & 0.385(41) \\
7.030 & --        & --        & --        & --        & --        & --        & 0.116(21) & 0.278(32) & 0.292(34) \\
7.150 & --        & --        & --        & --        & --        & --        & 0.091(04) & 0.132(29) & 0.177(15) \\
\hline
\end{tabular}
\caption{Estimates of the autocorrelation coefficient $C_{\Delta t}$ with $\Delta t=10$ 
  for a representative set of HISQ/tree finite temperature lattices
  measured using the plaquette, and the light and strange quark chiral
  condensates. The correlations decrease away from the transition region, therefore in
  some cases, we were unable to extract the autocorrelation coefficient. Note 
  that the chiral condensate is calculated with stochastic estimators 
  that introduce additional noise in the correlation function.
}
\label{tab_hisq_corr_T}
\end{table}

\bibliography{Tc_HotQCD}

\begin{thebibliography}{82}
\expandafter\ifx\csname natexlab\endcsname\relax\def\natexlab#1{#1}\fi
\expandafter\ifx\csname bibnamefont\endcsname\relax
  \def\bibnamefont#1{#1}\fi
\expandafter\ifx\csname bibfnamefont\endcsname\relax
  \def\bibfnamefont#1{#1}\fi
\expandafter\ifx\csname citenamefont\endcsname\relax
  \def\citenamefont#1{#1}\fi
\expandafter\ifx\csname url\endcsname\relax
  \def\url#1{\texttt{#1}}\fi
\expandafter\ifx\csname urlprefix\endcsname\relax\def\urlprefix{URL }\fi
\providecommand{\bibinfo}[2]{#2}
\providecommand{\eprint}[2][]{\url{#2}}

\bibitem[{\citenamefont{Hagedorn}(1965)}]{Hagedorn:1965st}
\bibinfo{author}{\bibfnamefont{R.}~\bibnamefont{Hagedorn}},
  \bibinfo{journal}{Nuovo Cim. Suppl.} \textbf{\bibinfo{volume}{3}},
  \bibinfo{pages}{147} (\bibinfo{year}{1965}).

\bibitem[{\citenamefont{Cabibbo and Parisi}(1975)}]{Cabibbo:1975ig}
\bibinfo{author}{\bibfnamefont{N.}~\bibnamefont{Cabibbo}} \bibnamefont{and}
  \bibinfo{author}{\bibfnamefont{G.}~\bibnamefont{Parisi}},
  \bibinfo{journal}{Phys. Lett.} \textbf{\bibinfo{volume}{B59}},
  \bibinfo{pages}{67} (\bibinfo{year}{1975}).

\bibitem[{\citenamefont{Shuryak}(1978)}]{Shuryak:1977ut}
\bibinfo{author}{\bibfnamefont{E.~V.} \bibnamefont{Shuryak}},
  \bibinfo{journal}{Sov. Phys. JETP} \textbf{\bibinfo{volume}{47}},
  \bibinfo{pages}{212} (\bibinfo{year}{1978}).

\bibitem[{\citenamefont{McLerran and Svetitsky}(1981)}]{McLerran:1981pb}
\bibinfo{author}{\bibfnamefont{L.~D.} \bibnamefont{McLerran}} \bibnamefont{and}
  \bibinfo{author}{\bibfnamefont{B.}~\bibnamefont{Svetitsky}},
  \bibinfo{journal}{Phys. Rev.} \textbf{\bibinfo{volume}{D24}},
  \bibinfo{pages}{450} (\bibinfo{year}{1981}).

\bibitem[{\citenamefont{Kuti et~al.}(1981)\citenamefont{Kuti, Polonyi, and
  Szlachanyi}}]{Kuti:1980gh}
\bibinfo{author}{\bibfnamefont{J.}~\bibnamefont{Kuti}},
  \bibinfo{author}{\bibfnamefont{J.}~\bibnamefont{Polonyi}}, \bibnamefont{and}
  \bibinfo{author}{\bibfnamefont{K.}~\bibnamefont{Szlachanyi}},
  \bibinfo{journal}{Phys. Lett.} \textbf{\bibinfo{volume}{B98}},
  \bibinfo{pages}{199} (\bibinfo{year}{1981}).

\bibitem[{\citenamefont{Engels et~al.}(1981)\citenamefont{Engels, Karsch, Satz,
  and Montvay}}]{Engels:1980ty}
\bibinfo{author}{\bibfnamefont{J.}~\bibnamefont{Engels}},
  \bibinfo{author}{\bibfnamefont{F.}~\bibnamefont{Karsch}},
  \bibinfo{author}{\bibfnamefont{H.}~\bibnamefont{Satz}}, \bibnamefont{and}
  \bibinfo{author}{\bibfnamefont{I.}~\bibnamefont{Montvay}},
  \bibinfo{journal}{Phys. Lett.} \textbf{\bibinfo{volume}{B101}},
  \bibinfo{pages}{89} (\bibinfo{year}{1981}).

\bibitem[{\citenamefont{Kogut et~al.}(1982)\citenamefont{Kogut, Stone, Wyld,
  Shigemitsu, Shenker et~al.}}]{Kogut:1982fn}
\bibinfo{author}{\bibfnamefont{J.~B.} \bibnamefont{Kogut}},
  \bibinfo{author}{\bibfnamefont{M.}~\bibnamefont{Stone}},
  \bibinfo{author}{\bibfnamefont{H.}~\bibnamefont{Wyld}},
  \bibinfo{author}{\bibfnamefont{J.}~\bibnamefont{Shigemitsu}},
  \bibinfo{author}{\bibfnamefont{S.}~\bibnamefont{Shenker}},
  \bibnamefont{et~al.}, \bibinfo{journal}{Phys. Rev. Lett.}
  \textbf{\bibinfo{volume}{48}}, \bibinfo{pages}{1140} (\bibinfo{year}{1982}),
  \bibinfo{note}{revised version}.

\bibitem[{\citenamefont{Kogut et~al.}(1983)\citenamefont{Kogut, Stone, Wyld,
  Gibbs, Shigemitsu et~al.}}]{Kogut:1982rt}
\bibinfo{author}{\bibfnamefont{J.~B.} \bibnamefont{Kogut}},
  \bibinfo{author}{\bibfnamefont{M.}~\bibnamefont{Stone}},
  \bibinfo{author}{\bibfnamefont{H.}~\bibnamefont{Wyld}},
  \bibinfo{author}{\bibfnamefont{W.}~\bibnamefont{Gibbs}},
  \bibinfo{author}{\bibfnamefont{J.}~\bibnamefont{Shigemitsu}},
  \bibnamefont{et~al.}, \bibinfo{journal}{Phys. Rev. Lett.}
  \textbf{\bibinfo{volume}{50}}, \bibinfo{pages}{393} (\bibinfo{year}{1983}).

\bibitem[{\citenamefont{Pisarski and Wilczek}(1984)}]{Pisarski:1983ms}
\bibinfo{author}{\bibfnamefont{R.~D.} \bibnamefont{Pisarski}} \bibnamefont{and}
  \bibinfo{author}{\bibfnamefont{F.}~\bibnamefont{Wilczek}},
  \bibinfo{journal}{Phys. Rev.} \textbf{\bibinfo{volume}{D29}},
  \bibinfo{pages}{338} (\bibinfo{year}{1984}).

\bibitem[{\citenamefont{Bernard et~al.}(2005)}]{Bernard:2004je}
\bibinfo{author}{\bibfnamefont{C.}~\bibnamefont{Bernard}} \bibnamefont{et~al.}
  (\bibinfo{collaboration}{MILC Collaboration}), \bibinfo{journal}{Phys. Rev.}
  \textbf{\bibinfo{volume}{D71}}, \bibinfo{pages}{034504}
  (\bibinfo{year}{2005}), \eprint{hep-lat/0405029}.

\bibitem[{\citenamefont{Cheng et~al.}(2006)\citenamefont{Cheng, Christ, Datta,
  van~der Heide, Jung et~al.}}]{Cheng:2006qk}
\bibinfo{author}{\bibfnamefont{M.}~\bibnamefont{Cheng}},
  \bibinfo{author}{\bibfnamefont{N.}~\bibnamefont{Christ}},
  \bibinfo{author}{\bibfnamefont{S.}~\bibnamefont{Datta}},
  \bibinfo{author}{\bibfnamefont{J.}~\bibnamefont{van~der Heide}},
  \bibinfo{author}{\bibfnamefont{C.}~\bibnamefont{Jung}}, \bibnamefont{et~al.},
  \bibinfo{journal}{Phys. Rev.} \textbf{\bibinfo{volume}{D74}},
  \bibinfo{pages}{054507} (\bibinfo{year}{2006}), \eprint{hep-lat/0608013}.

\bibitem[{\citenamefont{Aoki et~al.}(2006{\natexlab{a}})\citenamefont{Aoki,
  Endrodi, Fodor, Katz, and Szabo}}]{Aoki:2006we}
\bibinfo{author}{\bibfnamefont{Y.}~\bibnamefont{Aoki}},
  \bibinfo{author}{\bibfnamefont{G.}~\bibnamefont{Endrodi}},
  \bibinfo{author}{\bibfnamefont{Z.}~\bibnamefont{Fodor}},
  \bibinfo{author}{\bibfnamefont{S.}~\bibnamefont{Katz}}, \bibnamefont{and}
  \bibinfo{author}{\bibfnamefont{K.}~\bibnamefont{Szabo}},
  \bibinfo{journal}{Nature} \textbf{\bibinfo{volume}{443}},
  \bibinfo{pages}{675} (\bibinfo{year}{2006}{\natexlab{a}}),
  \eprint{hep-lat/0611014}.

\bibitem[{\citenamefont{Braun-Munzinger
  et~al.}(2003)\citenamefont{Braun-Munzinger, Redlich, and
  Stachel}}]{BraunMunzinger:2003zd}
\bibinfo{author}{\bibfnamefont{P.}~\bibnamefont{Braun-Munzinger}},
  \bibinfo{author}{\bibfnamefont{K.}~\bibnamefont{Redlich}}, \bibnamefont{and}
  \bibinfo{author}{\bibfnamefont{J.}~\bibnamefont{Stachel}}
  (\bibinfo{year}{2003}), \bibinfo{note}{to appear in Quark Gluon Plasma 3,
  eds. R.C. Hwa and Xin-Nian Wang, World Scientific Publishing},
  \eprint{nucl-th/0304013}.

\bibitem[{\citenamefont{Ejiri et~al.}(2010)}]{Ejiri:2009hq}
\bibinfo{author}{\bibfnamefont{S.}~\bibnamefont{Ejiri}} \bibnamefont{et~al.}
  (\bibinfo{collaboration}{WHOT-QCD Collaboration}), \bibinfo{journal}{Phys.
  Rev.} \textbf{\bibinfo{volume}{D82}}, \bibinfo{pages}{014508}
  (\bibinfo{year}{2010}), \eprint{0909.2121}.

\bibitem[{\citenamefont{Bornyakov et~al.}(2010)\citenamefont{Bornyakov,
  Horsley, Morozov, Nakamura, Polikarpov et~al.}}]{Bornyakov:2009qh}
\bibinfo{author}{\bibfnamefont{V.}~\bibnamefont{Bornyakov}},
  \bibinfo{author}{\bibfnamefont{R.}~\bibnamefont{Horsley}},
  \bibinfo{author}{\bibfnamefont{S.}~\bibnamefont{Morozov}},
  \bibinfo{author}{\bibfnamefont{Y.}~\bibnamefont{Nakamura}},
  \bibinfo{author}{\bibfnamefont{M.}~\bibnamefont{Polikarpov}},
  \bibnamefont{et~al.}, \bibinfo{journal}{Phys. Rev.}
  \textbf{\bibinfo{volume}{D82}}, \bibinfo{pages}{014504}
  (\bibinfo{year}{2010}), \eprint{0910.2392}.

\bibitem[{\citenamefont{Cheng et~al.}(2010{\natexlab{a}})\citenamefont{Cheng,
  Christ, Li, Mawhinney, Renfrew et~al.}}]{Cheng:2009be}
\bibinfo{author}{\bibfnamefont{M.}~\bibnamefont{Cheng}},
  \bibinfo{author}{\bibfnamefont{N.~H.} \bibnamefont{Christ}},
  \bibinfo{author}{\bibfnamefont{M.}~\bibnamefont{Li}},
  \bibinfo{author}{\bibfnamefont{R.~D.} \bibnamefont{Mawhinney}},
  \bibinfo{author}{\bibfnamefont{D.}~\bibnamefont{Renfrew}},
  \bibnamefont{et~al.}, \bibinfo{journal}{Phys. Rev.}
  \textbf{\bibinfo{volume}{D81}}, \bibinfo{pages}{054510}
  (\bibinfo{year}{2010}{\natexlab{a}}), \eprint{0911.3450}.

\bibitem[{\citenamefont{Aoki et~al.}(2006{\natexlab{b}})\citenamefont{Aoki,
  Fodor, Katz, and Szabo}}]{Aoki:2005vt}
\bibinfo{author}{\bibfnamefont{Y.}~\bibnamefont{Aoki}},
  \bibinfo{author}{\bibfnamefont{Z.}~\bibnamefont{Fodor}},
  \bibinfo{author}{\bibfnamefont{S.}~\bibnamefont{Katz}}, \bibnamefont{and}
  \bibinfo{author}{\bibfnamefont{K.}~\bibnamefont{Szabo}},
  \bibinfo{journal}{JHEP} \textbf{\bibinfo{volume}{0601}}, \bibinfo{pages}{089}
  (\bibinfo{year}{2006}{\natexlab{b}}), \eprint{hep-lat/0510084}.

\bibitem[{\citenamefont{Bernard
  et~al.}(2007{\natexlab{a}})\citenamefont{Bernard, Burch, DeTar, Gottlieb,
  Levkova et~al.}}]{Bernard:2006nj}
\bibinfo{author}{\bibfnamefont{C.}~\bibnamefont{Bernard}},
  \bibinfo{author}{\bibfnamefont{T.}~\bibnamefont{Burch}},
  \bibinfo{author}{\bibfnamefont{C.~E.} \bibnamefont{DeTar}},
  \bibinfo{author}{\bibfnamefont{S.}~\bibnamefont{Gottlieb}},
  \bibinfo{author}{\bibfnamefont{L.}~\bibnamefont{Levkova}},
  \bibnamefont{et~al.}, \bibinfo{journal}{Phys. Rev.}
  \textbf{\bibinfo{volume}{D75}}, \bibinfo{pages}{094505}
  (\bibinfo{year}{2007}{\natexlab{a}}), \eprint{hep-lat/0611031}.

\bibitem[{\citenamefont{Cheng et~al.}(2008)\citenamefont{Cheng, Christ, Datta,
  van~der Heide, Jung et~al.}}]{Cheng:2007jq}
\bibinfo{author}{\bibfnamefont{M.}~\bibnamefont{Cheng}},
  \bibinfo{author}{\bibfnamefont{N.}~\bibnamefont{Christ}},
  \bibinfo{author}{\bibfnamefont{S.}~\bibnamefont{Datta}},
  \bibinfo{author}{\bibfnamefont{J.}~\bibnamefont{van~der Heide}},
  \bibinfo{author}{\bibfnamefont{C.}~\bibnamefont{Jung}}, \bibnamefont{et~al.},
  \bibinfo{journal}{Phys. Rev.} \textbf{\bibinfo{volume}{D77}},
  \bibinfo{pages}{014511} (\bibinfo{year}{2008}), \eprint{0710.0354}.

\bibitem[{\citenamefont{Bazavov
  et~al.}(2009{\natexlab{a}})\citenamefont{Bazavov, Bhattacharya, Cheng,
  Christ, DeTar et~al.}}]{Bazavov:2009zn}
\bibinfo{author}{\bibfnamefont{A.}~\bibnamefont{Bazavov}},
  \bibinfo{author}{\bibfnamefont{T.}~\bibnamefont{Bhattacharya}},
  \bibinfo{author}{\bibfnamefont{M.}~\bibnamefont{Cheng}},
  \bibinfo{author}{\bibfnamefont{N.}~\bibnamefont{Christ}},
  \bibinfo{author}{\bibfnamefont{C.}~\bibnamefont{DeTar}},
  \bibnamefont{et~al.}, \bibinfo{journal}{Phys. Rev.}
  \textbf{\bibinfo{volume}{D80}}, \bibinfo{pages}{014504}
  (\bibinfo{year}{2009}{\natexlab{a}}), \eprint{0903.4379}.

\bibitem[{\citenamefont{Cheng et~al.}(2010{\natexlab{b}})\citenamefont{Cheng,
  Ejiri, Hegde, Karsch, Kaczmarek et~al.}}]{Cheng:2009zi}
\bibinfo{author}{\bibfnamefont{M.}~\bibnamefont{Cheng}},
  \bibinfo{author}{\bibfnamefont{S.}~\bibnamefont{Ejiri}},
  \bibinfo{author}{\bibfnamefont{P.}~\bibnamefont{Hegde}},
  \bibinfo{author}{\bibfnamefont{F.}~\bibnamefont{Karsch}},
  \bibinfo{author}{\bibfnamefont{O.}~\bibnamefont{Kaczmarek}},
  \bibnamefont{et~al.}, \bibinfo{journal}{Phys. Rev.}
  \textbf{\bibinfo{volume}{D81}}, \bibinfo{pages}{054504}
  (\bibinfo{year}{2010}{\natexlab{b}}), \eprint{0911.2215}.

\bibitem[{\citenamefont{Aoki et~al.}(2006{\natexlab{c}})\citenamefont{Aoki,
  Fodor, Katz, and Szabo}}]{Aoki:2006br}
\bibinfo{author}{\bibfnamefont{Y.}~\bibnamefont{Aoki}},
  \bibinfo{author}{\bibfnamefont{Z.}~\bibnamefont{Fodor}},
  \bibinfo{author}{\bibfnamefont{S.}~\bibnamefont{Katz}}, \bibnamefont{and}
  \bibinfo{author}{\bibfnamefont{K.}~\bibnamefont{Szabo}},
  \bibinfo{journal}{Phys. Lett.} \textbf{\bibinfo{volume}{B643}},
  \bibinfo{pages}{46} (\bibinfo{year}{2006}{\natexlab{c}}),
  \eprint{hep-lat/0609068}.

\bibitem[{\citenamefont{Aoki et~al.}(2009)\citenamefont{Aoki, Borsanyi, Durr,
  Fodor, Katz et~al.}}]{Aoki:2009sc}
\bibinfo{author}{\bibfnamefont{Y.}~\bibnamefont{Aoki}},
  \bibinfo{author}{\bibfnamefont{S.}~\bibnamefont{Borsanyi}},
  \bibinfo{author}{\bibfnamefont{S.}~\bibnamefont{Durr}},
  \bibinfo{author}{\bibfnamefont{Z.}~\bibnamefont{Fodor}},
  \bibinfo{author}{\bibfnamefont{S.~D.} \bibnamefont{Katz}},
  \bibnamefont{et~al.}, \bibinfo{journal}{JHEP}
  \textbf{\bibinfo{volume}{0906}}, \bibinfo{pages}{088} (\bibinfo{year}{2009}),
  \eprint{0903.4155}.

\bibitem[{\citenamefont{Borsanyi et~al.}(2010)}]{Borsanyi:2010bp}
\bibinfo{author}{\bibfnamefont{S.}~\bibnamefont{Borsanyi}} \bibnamefont{et~al.}
  (\bibinfo{collaboration}{Wuppertal-Budapest Collaboration}),
  \bibinfo{journal}{JHEP} \textbf{\bibinfo{volume}{1009}}, \bibinfo{pages}{073}
  (\bibinfo{year}{2010}), \eprint{1005.3508}.

\bibitem[{\citenamefont{Follana et~al.}(2007)}]{Follana:2006rc}
\bibinfo{author}{\bibfnamefont{E.}~\bibnamefont{Follana}} \bibnamefont{et~al.}
  (\bibinfo{collaboration}{HPQCD Collaboration, UKQCD Collaboration}),
  \bibinfo{journal}{Phys. Rev.} \textbf{\bibinfo{volume}{D75}},
  \bibinfo{pages}{054502} (\bibinfo{year}{2007}), \eprint{hep-lat/0610092}.

\bibitem[{\citenamefont{Detar and Gupta}(2007)}]{Detar:2007as}
\bibinfo{author}{\bibfnamefont{C.~E.} \bibnamefont{Detar}} \bibnamefont{and}
  \bibinfo{author}{\bibfnamefont{R.}~\bibnamefont{Gupta}}
  (\bibinfo{collaboration}{HotQCD Collaboration}), \bibinfo{journal}{PoS}
  \textbf{\bibinfo{volume}{LAT2007}}, \bibinfo{pages}{179}
  (\bibinfo{year}{2007}), \eprint{0710.1655}.

\bibitem[{\citenamefont{Gupta}(2009)}]{Gupta:2009tv}
\bibinfo{author}{\bibfnamefont{R.}~\bibnamefont{Gupta}}
  (\bibinfo{collaboration}{HotQCD}) (\bibinfo{year}{2009}), \eprint{0912.1374}.

\bibitem[{\citenamefont{Bazavov and Petreczky}(2009)}]{Bazavov:2009mi}
\bibinfo{author}{\bibfnamefont{A.}~\bibnamefont{Bazavov}} \bibnamefont{and}
  \bibinfo{author}{\bibfnamefont{P.}~\bibnamefont{Petreczky}}
  (\bibinfo{collaboration}{HotQCD Collaboration}), \bibinfo{journal}{PoS}
  \textbf{\bibinfo{volume}{LAT2009}}, \bibinfo{pages}{163}
  (\bibinfo{year}{2009}), \eprint{0912.5421}.

\bibitem[{\citenamefont{Bazavov and
  Petreczky}(2010{\natexlab{a}})}]{Bazavov:2010sb}
\bibinfo{author}{\bibfnamefont{A.}~\bibnamefont{Bazavov}} \bibnamefont{and}
  \bibinfo{author}{\bibfnamefont{P.}~\bibnamefont{Petreczky}}
  (\bibinfo{collaboration}{HotQCD collaboration}),
  \bibinfo{journal}{J.Phys.Conf.Ser.} \textbf{\bibinfo{volume}{230}},
  \bibinfo{pages}{012014} (\bibinfo{year}{2010}{\natexlab{a}}),
  \eprint{1005.1131}.

\bibitem[{\citenamefont{Bazavov and
  Petreczky}(2010{\natexlab{b}})}]{Bazavov:2010bx}
\bibinfo{author}{\bibfnamefont{A.}~\bibnamefont{Bazavov}} \bibnamefont{and}
  \bibinfo{author}{\bibfnamefont{P.}~\bibnamefont{Petreczky}}
  (\bibinfo{collaboration}{HotQCD Collaboration})
  (\bibinfo{year}{2010}{\natexlab{b}}), \eprint{1009.4914}.

\bibitem[{\citenamefont{Soldner}(2010)}]{Soldner:2010xk}
\bibinfo{author}{\bibfnamefont{W.}~\bibnamefont{Soldner}}
  (\bibinfo{collaboration}{HotQCD collaboration}), \bibinfo{journal}{PoS}
  \textbf{\bibinfo{volume}{LATTICE2010}}, \bibinfo{pages}{215}
  (\bibinfo{year}{2010}), \eprint{1012.4484}.

\bibitem[{\citenamefont{Bazavov and
  Petreczky}(2010{\natexlab{c}})}]{Bazavov:2010pg}
\bibinfo{author}{\bibfnamefont{A.}~\bibnamefont{Bazavov}} \bibnamefont{and}
  \bibinfo{author}{\bibfnamefont{P.}~\bibnamefont{Petreczky}}
  (\bibinfo{collaboration}{for the HotQCD Collaboration}),
  \bibinfo{journal}{PoS} \textbf{\bibinfo{volume}{LATTICE2010}},
  \bibinfo{pages}{169} (\bibinfo{year}{2010}{\natexlab{c}}),
  \eprint{1012.1257}.

\bibitem[{\citenamefont{Bazavov and
  Petreczky}(2011{\natexlab{a}})}]{Bazavov:2011sd}
\bibinfo{author}{\bibfnamefont{A.}~\bibnamefont{Bazavov}} \bibnamefont{and}
  \bibinfo{author}{\bibfnamefont{P.}~\bibnamefont{Petreczky}}
  (\bibinfo{collaboration}{HotQCD Collaboration})
  (\bibinfo{year}{2011}{\natexlab{a}}), \eprint{1107.5027}.

\bibitem[{\citenamefont{Bazavov and
  Petreczky}(2011{\natexlab{b}})}]{Bazavov:2011jx}
\bibinfo{author}{\bibfnamefont{A.}~\bibnamefont{Bazavov}} \bibnamefont{and}
  \bibinfo{author}{\bibfnamefont{P.}~\bibnamefont{Petreczky}}
  (\bibinfo{year}{2011}{\natexlab{b}}), \bibinfo{note}{* Temporary entry *},
  \eprint{1110.2160}.

\bibitem[{\citenamefont{Karsch et~al.}(2003)\citenamefont{Karsch, Redlich, and
  Tawfik}}]{Karsch:2003vd}
\bibinfo{author}{\bibfnamefont{F.}~\bibnamefont{Karsch}},
  \bibinfo{author}{\bibfnamefont{K.}~\bibnamefont{Redlich}}, \bibnamefont{and}
  \bibinfo{author}{\bibfnamefont{A.}~\bibnamefont{Tawfik}},
  \bibinfo{journal}{Eur. Phys. J.} \textbf{\bibinfo{volume}{C29}},
  \bibinfo{pages}{549} (\bibinfo{year}{2003}), \bibinfo{note}{dedicated to Rolf
  Hagedorn}, \eprint{hep-ph/0303108}.

\bibitem[{\citenamefont{Huovinen and
  Petreczky}(2010{\natexlab{a}})}]{Huovinen:2009yb}
\bibinfo{author}{\bibfnamefont{P.}~\bibnamefont{Huovinen}} \bibnamefont{and}
  \bibinfo{author}{\bibfnamefont{P.}~\bibnamefont{Petreczky}},
  \bibinfo{journal}{Nucl. Phys.} \textbf{\bibinfo{volume}{A837}},
  \bibinfo{pages}{26} (\bibinfo{year}{2010}{\natexlab{a}}), \eprint{0912.2541}.

\bibitem[{\citenamefont{Huovinen and
  Petreczky}(2010{\natexlab{b}})}]{Huovinen:2010tv}
\bibinfo{author}{\bibfnamefont{P.}~\bibnamefont{Huovinen}} \bibnamefont{and}
  \bibinfo{author}{\bibfnamefont{P.}~\bibnamefont{Petreczky}},
  \bibinfo{journal}{J.Phys.Conf.Ser.} \textbf{\bibinfo{volume}{230}},
  \bibinfo{pages}{012012} (\bibinfo{year}{2010}{\natexlab{b}}),
  \eprint{1005.0324}.

\bibitem[{\citenamefont{Huovinen and Petreczky}(2011)}]{Huovinen:2011xc}
\bibinfo{author}{\bibfnamefont{P.}~\bibnamefont{Huovinen}} \bibnamefont{and}
  \bibinfo{author}{\bibfnamefont{P.}~\bibnamefont{Petreczky}},
  \bibinfo{journal}{J.Phys.G} \textbf{\bibinfo{volume}{G38}},
  \bibinfo{pages}{124103} (\bibinfo{year}{2011}), \eprint{1106.6227}.

\bibitem[{\citenamefont{Blum et~al.}(1997)\citenamefont{Blum, Detar, Gottlieb,
  Rummukainen, Heller et~al.}}]{Blum:1996uf}
\bibinfo{author}{\bibfnamefont{T.}~\bibnamefont{Blum}},
  \bibinfo{author}{\bibfnamefont{C.~E.} \bibnamefont{Detar}},
  \bibinfo{author}{\bibfnamefont{S.~A.} \bibnamefont{Gottlieb}},
  \bibinfo{author}{\bibfnamefont{K.}~\bibnamefont{Rummukainen}},
  \bibinfo{author}{\bibfnamefont{U.~M.} \bibnamefont{Heller}},
  \bibnamefont{et~al.}, \bibinfo{journal}{Phys. Rev.}
  \textbf{\bibinfo{volume}{D55}}, \bibinfo{pages}{1133} (\bibinfo{year}{1997}),
  \eprint{hep-lat/9609036}.

\bibitem[{\citenamefont{Orginos et~al.}(1999)\citenamefont{Orginos, Toussaint,
  and Sugar}}]{Orginos:1999cr}
\bibinfo{author}{\bibfnamefont{K.}~\bibnamefont{Orginos}},
  \bibinfo{author}{\bibfnamefont{D.}~\bibnamefont{Toussaint}},
  \bibnamefont{and} \bibinfo{author}{\bibfnamefont{R.}~\bibnamefont{Sugar}}
  (\bibinfo{collaboration}{MILC Collaboration}), \bibinfo{journal}{Phys. Rev.}
  \textbf{\bibinfo{volume}{D60}}, \bibinfo{pages}{054503}
  (\bibinfo{year}{1999}), \eprint{hep-lat/9903032}.

\bibitem[{\citenamefont{Bazavov
  et~al.}(2010{\natexlab{a}})\citenamefont{Bazavov, Toussaint, Bernard, Laiho,
  DeTar et~al.}}]{Bazavov:2009bb}
\bibinfo{author}{\bibfnamefont{A.}~\bibnamefont{Bazavov}},
  \bibinfo{author}{\bibfnamefont{D.}~\bibnamefont{Toussaint}},
  \bibinfo{author}{\bibfnamefont{C.}~\bibnamefont{Bernard}},
  \bibinfo{author}{\bibfnamefont{J.}~\bibnamefont{Laiho}},
  \bibinfo{author}{\bibfnamefont{C.}~\bibnamefont{DeTar}},
  \bibnamefont{et~al.}, \bibinfo{journal}{Rev. Mod. Phys.}
  \textbf{\bibinfo{volume}{82}}, \bibinfo{pages}{1349}
  (\bibinfo{year}{2010}{\natexlab{a}}), \eprint{0903.3598}.

\bibitem[{\citenamefont{Hasenfratz and Knechtli}(2001)}]{Hasenfratz:2001hp}
\bibinfo{author}{\bibfnamefont{A.}~\bibnamefont{Hasenfratz}} \bibnamefont{and}
  \bibinfo{author}{\bibfnamefont{F.}~\bibnamefont{Knechtli}},
  \bibinfo{journal}{Phys. Rev.} \textbf{\bibinfo{volume}{D64}},
  \bibinfo{pages}{034504} (\bibinfo{year}{2001}), \eprint{hep-lat/0103029}.

\bibitem[{\citenamefont{Hasenfratz}(2003)}]{Hasenfratz:2002vv}
\bibinfo{author}{\bibfnamefont{A.}~\bibnamefont{Hasenfratz}},
  \bibinfo{journal}{Nucl. Phys. Proc. Suppl.} \textbf{\bibinfo{volume}{119}},
  \bibinfo{pages}{131} (\bibinfo{year}{2003}), \eprint{hep-lat/0211007}.

\bibitem[{\citenamefont{Hasenfratz et~al.}(2007)\citenamefont{Hasenfratz,
  Hoffmann, and Schaefer}}]{Hasenfratz:2007rf}
\bibinfo{author}{\bibfnamefont{A.}~\bibnamefont{Hasenfratz}},
  \bibinfo{author}{\bibfnamefont{R.}~\bibnamefont{Hoffmann}}, \bibnamefont{and}
  \bibinfo{author}{\bibfnamefont{S.}~\bibnamefont{Schaefer}},
  \bibinfo{journal}{JHEP} \textbf{\bibinfo{volume}{0705}}, \bibinfo{pages}{029}
  (\bibinfo{year}{2007}), \eprint{hep-lat/0702028}.

\bibitem[{\citenamefont{Bazavov et~al.}(2009{\natexlab{b}})}]{Bazavov:2009wm}
\bibinfo{author}{\bibfnamefont{A.}~\bibnamefont{Bazavov}} \bibnamefont{et~al.}
  (\bibinfo{collaboration}{MILC Collaboration}), \bibinfo{journal}{PoS}
  \textbf{\bibinfo{volume}{LAT2009}}, \bibinfo{pages}{123}
  (\bibinfo{year}{2009}{\natexlab{b}}), \eprint{0911.0869}.

\bibitem[{\citenamefont{Bazavov et~al.}(2010{\natexlab{b}})}]{Bazavov:2010ru}
\bibinfo{author}{\bibfnamefont{A.}~\bibnamefont{Bazavov}} \bibnamefont{et~al.}
  (\bibinfo{collaboration}{MILC collaboration}), \bibinfo{journal}{Phys. Rev.}
  \textbf{\bibinfo{volume}{D82}}, \bibinfo{pages}{074501}
  (\bibinfo{year}{2010}{\natexlab{b}}), \eprint{1004.0342}.

\bibitem[{\citenamefont{Sharpe}(2006)}]{Sharpe:2006re}
\bibinfo{author}{\bibfnamefont{S.~R.} \bibnamefont{Sharpe}},
  \bibinfo{journal}{PoS} \textbf{\bibinfo{volume}{LAT2006}},
  \bibinfo{pages}{022} (\bibinfo{year}{2006}), \eprint{hep-lat/0610094}.

\bibitem[{\citenamefont{Creutz}(2007)}]{Creutz:2007rk}
\bibinfo{author}{\bibfnamefont{M.}~\bibnamefont{Creutz}},
  \bibinfo{journal}{PoS} \textbf{\bibinfo{volume}{LAT2007}},
  \bibinfo{pages}{007} (\bibinfo{year}{2007}), \eprint{0708.1295}.

\bibitem[{\citenamefont{Heller et~al.}(1999)\citenamefont{Heller, Karsch, and
  Sturm}}]{Heller:1999xz}
\bibinfo{author}{\bibfnamefont{U.~M.} \bibnamefont{Heller}},
  \bibinfo{author}{\bibfnamefont{F.}~\bibnamefont{Karsch}}, \bibnamefont{and}
  \bibinfo{author}{\bibfnamefont{B.}~\bibnamefont{Sturm}},
  \bibinfo{journal}{Phys. Rev.} \textbf{\bibinfo{volume}{D60}},
  \bibinfo{pages}{114502} (\bibinfo{year}{1999}), \eprint{hep-lat/9901010}.

\bibitem[{\citenamefont{Hegde et~al.}(2008)\citenamefont{Hegde, Karsch,
  Laermann, and Shcheredin}}]{Hegde:2008nx}
\bibinfo{author}{\bibfnamefont{P.}~\bibnamefont{Hegde}},
  \bibinfo{author}{\bibfnamefont{F.}~\bibnamefont{Karsch}},
  \bibinfo{author}{\bibfnamefont{E.}~\bibnamefont{Laermann}}, \bibnamefont{and}
  \bibinfo{author}{\bibfnamefont{S.}~\bibnamefont{Shcheredin}},
  \bibinfo{journal}{Eur. Phys. J.} \textbf{\bibinfo{volume}{C55}},
  \bibinfo{pages}{423} (\bibinfo{year}{2008}), \eprint{0801.4883}.

\bibitem[{\citenamefont{Clark et~al.}(2005)\citenamefont{Clark, Kennedy, and
  Sroczynski}}]{Clark:2004cp}
\bibinfo{author}{\bibfnamefont{M.}~\bibnamefont{Clark}},
  \bibinfo{author}{\bibfnamefont{A.}~\bibnamefont{Kennedy}}, \bibnamefont{and}
  \bibinfo{author}{\bibfnamefont{Z.}~\bibnamefont{Sroczynski}},
  \bibinfo{journal}{Nucl. Phys. Proc. Suppl.} \textbf{\bibinfo{volume}{140}},
  \bibinfo{pages}{835} (\bibinfo{year}{2005}), \eprint{hep-lat/0409133}.

\bibitem[{\citenamefont{Clark et~al.}(2006)\citenamefont{Clark, de~Forcrand,
  and Kennedy}}]{Clark:2005sq}
\bibinfo{author}{\bibfnamefont{M.}~\bibnamefont{Clark}},
  \bibinfo{author}{\bibfnamefont{P.}~\bibnamefont{de~Forcrand}},
  \bibnamefont{and} \bibinfo{author}{\bibfnamefont{A.}~\bibnamefont{Kennedy}},
  \bibinfo{journal}{PoS} \textbf{\bibinfo{volume}{LAT2005}},
  \bibinfo{pages}{115} (\bibinfo{year}{2006}), \eprint{hep-lat/0510004}.

\bibitem[{\citenamefont{Bazavov et~al.}(2008)}]{Bazavov:2009jc}
\bibinfo{author}{\bibfnamefont{A.}~\bibnamefont{Bazavov}} \bibnamefont{et~al.}
  (\bibinfo{collaboration}{MILC Collaboration}), \bibinfo{journal}{PoS}
  \textbf{\bibinfo{volume}{LATTICE2008}}, \bibinfo{pages}{033}
  (\bibinfo{year}{2008}), \eprint{0903.0874}.

\bibitem[{\citenamefont{Sommer}(1994)}]{Sommer:1993ce}
\bibinfo{author}{\bibfnamefont{R.}~\bibnamefont{Sommer}},
  \bibinfo{journal}{Nucl. Phys.} \textbf{\bibinfo{volume}{B411}},
  \bibinfo{pages}{839} (\bibinfo{year}{1994}), \eprint{hep-lat/9310022}.

\bibitem[{\citenamefont{Aubin et~al.}(2004)\citenamefont{Aubin, Bernard, DeTar,
  Osborn, Gottlieb et~al.}}]{Aubin:2004wf}
\bibinfo{author}{\bibfnamefont{C.}~\bibnamefont{Aubin}},
  \bibinfo{author}{\bibfnamefont{C.}~\bibnamefont{Bernard}},
  \bibinfo{author}{\bibfnamefont{C.}~\bibnamefont{DeTar}},
  \bibinfo{author}{\bibfnamefont{J.}~\bibnamefont{Osborn}},
  \bibinfo{author}{\bibfnamefont{S.}~\bibnamefont{Gottlieb}},
  \bibnamefont{et~al.}, \bibinfo{journal}{Phys. Rev.}
  \textbf{\bibinfo{volume}{D70}}, \bibinfo{pages}{094505}
  (\bibinfo{year}{2004}), \eprint{hep-lat/0402030}.

\bibitem[{\citenamefont{Necco and Sommer}(2002)}]{Necco:2001xg}
\bibinfo{author}{\bibfnamefont{S.}~\bibnamefont{Necco}} \bibnamefont{and}
  \bibinfo{author}{\bibfnamefont{R.}~\bibnamefont{Sommer}},
  \bibinfo{journal}{Nucl. Phys.} \textbf{\bibinfo{volume}{B622}},
  \bibinfo{pages}{328} (\bibinfo{year}{2002}), \eprint{hep-lat/0108008}.

\bibitem[{Tou()}]{Toussaint:private}
\bibinfo{note}{{D. Toussaint}, private communications}.

\bibitem[{\citenamefont{Bazavov et~al.}(2010{\natexlab{c}})}]{Bazavov:2010hj}
\bibinfo{author}{\bibfnamefont{A.}~\bibnamefont{Bazavov}} \bibnamefont{et~al.}
  (\bibinfo{collaboration}{MILC Collaboration}), \bibinfo{journal}{PoS}
  \textbf{\bibinfo{volume}{LATTICE2010}}, \bibinfo{pages}{074}
  (\bibinfo{year}{2010}{\natexlab{c}}), \eprint{1012.0868}.

\bibitem[{\citenamefont{Davies et~al.}(2010)\citenamefont{Davies, Follana,
  Kendall, Lepage, and McNeile}}]{Davies:2009tsa}
\bibinfo{author}{\bibfnamefont{C.}~\bibnamefont{Davies}},
  \bibinfo{author}{\bibfnamefont{E.}~\bibnamefont{Follana}},
  \bibinfo{author}{\bibfnamefont{I.}~\bibnamefont{Kendall}},
  \bibinfo{author}{\bibfnamefont{G.}~\bibnamefont{Lepage}}, \bibnamefont{and}
  \bibinfo{author}{\bibfnamefont{C.}~\bibnamefont{McNeile}}
  (\bibinfo{collaboration}{HPQCD Collaboration}), \bibinfo{journal}{Phys. Rev.}
  \textbf{\bibinfo{volume}{D81}}, \bibinfo{pages}{034506}
  (\bibinfo{year}{2010}), \eprint{0910.1229}.

\bibitem[{\citenamefont{Gray et~al.}(2005)\citenamefont{Gray, Allison, Davies,
  Dalgic, Lepage et~al.}}]{Gray:2005ur}
\bibinfo{author}{\bibfnamefont{A.}~\bibnamefont{Gray}},
  \bibinfo{author}{\bibfnamefont{I.}~\bibnamefont{Allison}},
  \bibinfo{author}{\bibfnamefont{C.}~\bibnamefont{Davies}},
  \bibinfo{author}{\bibfnamefont{E.}~\bibnamefont{Dalgic}},
  \bibinfo{author}{\bibfnamefont{G.}~\bibnamefont{Lepage}},
  \bibnamefont{et~al.}, \bibinfo{journal}{Phys. Rev.}
  \textbf{\bibinfo{volume}{D72}}, \bibinfo{pages}{094507}
  (\bibinfo{year}{2005}), \eprint{hep-lat/0507013}.

\bibitem[{\citenamefont{Bernard et~al.}(2001)\citenamefont{Bernard, Burch,
  Orginos, Toussaint, DeGrand et~al.}}]{Bernard:2001av}
\bibinfo{author}{\bibfnamefont{C.~W.} \bibnamefont{Bernard}},
  \bibinfo{author}{\bibfnamefont{T.}~\bibnamefont{Burch}},
  \bibinfo{author}{\bibfnamefont{K.}~\bibnamefont{Orginos}},
  \bibinfo{author}{\bibfnamefont{D.}~\bibnamefont{Toussaint}},
  \bibinfo{author}{\bibfnamefont{T.~A.} \bibnamefont{DeGrand}},
  \bibnamefont{et~al.}, \bibinfo{journal}{Phys. Rev.}
  \textbf{\bibinfo{volume}{D64}}, \bibinfo{pages}{054506}
  (\bibinfo{year}{2001}), \eprint{hep-lat/0104002}.

\bibitem[{\citenamefont{Lee and Sharpe}(1999)}]{Lee:1999zxa}
\bibinfo{author}{\bibfnamefont{W.-J.} \bibnamefont{Lee}} \bibnamefont{and}
  \bibinfo{author}{\bibfnamefont{S.~R.} \bibnamefont{Sharpe}},
  \bibinfo{journal}{Phys. Rev.} \textbf{\bibinfo{volume}{D60}},
  \bibinfo{pages}{114503} (\bibinfo{year}{1999}), \eprint{hep-lat/9905023}.

\bibitem[{\citenamefont{Cheng et~al.}(2007)\citenamefont{Cheng, Christ, Jung,
  Karsch, Mawhinney et~al.}}]{Cheng:2006wj}
\bibinfo{author}{\bibfnamefont{M.}~\bibnamefont{Cheng}},
  \bibinfo{author}{\bibfnamefont{N.}~\bibnamefont{Christ}},
  \bibinfo{author}{\bibfnamefont{C.}~\bibnamefont{Jung}},
  \bibinfo{author}{\bibfnamefont{F.}~\bibnamefont{Karsch}},
  \bibinfo{author}{\bibfnamefont{R.}~\bibnamefont{Mawhinney}},
  \bibnamefont{et~al.}, \bibinfo{journal}{Eur. Phys. J.}
  \textbf{\bibinfo{volume}{C51}}, \bibinfo{pages}{875} (\bibinfo{year}{2007}),
  \eprint{hep-lat/0612030}.

\bibitem[{\citenamefont{Ejiri et~al.}(2009)\citenamefont{Ejiri, Karsch,
  Laermann, Miao, Mukherjee et~al.}}]{Ejiri:2009ac}
\bibinfo{author}{\bibfnamefont{S.}~\bibnamefont{Ejiri}},
  \bibinfo{author}{\bibfnamefont{F.}~\bibnamefont{Karsch}},
  \bibinfo{author}{\bibfnamefont{E.}~\bibnamefont{Laermann}},
  \bibinfo{author}{\bibfnamefont{C.}~\bibnamefont{Miao}},
  \bibinfo{author}{\bibfnamefont{S.}~\bibnamefont{Mukherjee}},
  \bibnamefont{et~al.}, \bibinfo{journal}{Phys. Rev.}
  \textbf{\bibinfo{volume}{D80}}, \bibinfo{pages}{094505}
  (\bibinfo{year}{2009}), \eprint{0909.5122}.

\bibitem[{\citenamefont{Engels et~al.}(2000)\citenamefont{Engels, Holtmann,
  Mendes, and Schulze}}]{Engels:2000xw}
\bibinfo{author}{\bibfnamefont{J.}~\bibnamefont{Engels}},
  \bibinfo{author}{\bibfnamefont{S.}~\bibnamefont{Holtmann}},
  \bibinfo{author}{\bibfnamefont{T.}~\bibnamefont{Mendes}}, \bibnamefont{and}
  \bibinfo{author}{\bibfnamefont{T.}~\bibnamefont{Schulze}},
  \bibinfo{journal}{Phys. Lett.} \textbf{\bibinfo{volume}{B492}},
  \bibinfo{pages}{219} (\bibinfo{year}{2000}), \eprint{hep-lat/0006023}.

\bibitem[{\citenamefont{Toussaint}(1997)}]{Toussaint:1996qr}
\bibinfo{author}{\bibfnamefont{D.}~\bibnamefont{Toussaint}},
  \bibinfo{journal}{Phys. Rev.} \textbf{\bibinfo{volume}{D55}},
  \bibinfo{pages}{362} (\bibinfo{year}{1997}), \eprint{hep-lat/9607084}.

\bibitem[{\citenamefont{Engels and Mendes}(2000)}]{Engels:1999wf}
\bibinfo{author}{\bibfnamefont{J.}~\bibnamefont{Engels}} \bibnamefont{and}
  \bibinfo{author}{\bibfnamefont{T.}~\bibnamefont{Mendes}},
  \bibinfo{journal}{Nucl. Phys.} \textbf{\bibinfo{volume}{B572}},
  \bibinfo{pages}{289} (\bibinfo{year}{2000}), \eprint{hep-lat/9911028}.

\bibitem[{\citenamefont{Engels et~al.}(2001)\citenamefont{Engels, Holtmann,
  Mendes, and Schulze}}]{Engels:2001bq}
\bibinfo{author}{\bibfnamefont{J.}~\bibnamefont{Engels}},
  \bibinfo{author}{\bibfnamefont{S.}~\bibnamefont{Holtmann}},
  \bibinfo{author}{\bibfnamefont{T.}~\bibnamefont{Mendes}}, \bibnamefont{and}
  \bibinfo{author}{\bibfnamefont{T.}~\bibnamefont{Schulze}},
  \bibinfo{journal}{Phys. Lett.} \textbf{\bibinfo{volume}{B514}},
  \bibinfo{pages}{299} (\bibinfo{year}{2001}), \eprint{hep-lat/0105028}.

\bibitem[{\citenamefont{Kaczmarek et~al.}(2011)\citenamefont{Kaczmarek, Karsch,
  Laermann, Miao, Mukherjee et~al.}}]{Kaczmarek:2011zz}
\bibinfo{author}{\bibfnamefont{O.}~\bibnamefont{Kaczmarek}},
  \bibinfo{author}{\bibfnamefont{F.}~\bibnamefont{Karsch}},
  \bibinfo{author}{\bibfnamefont{E.}~\bibnamefont{Laermann}},
  \bibinfo{author}{\bibfnamefont{C.}~\bibnamefont{Miao}},
  \bibinfo{author}{\bibfnamefont{S.}~\bibnamefont{Mukherjee}},
  \bibnamefont{et~al.}, \bibinfo{journal}{Phys. Rev.}
  \textbf{\bibinfo{volume}{D83}}, \bibinfo{pages}{014504}
  (\bibinfo{year}{2011}), \eprint{1011.3130}.

\bibitem[{\citenamefont{Cucchieri et~al.}(2002)\citenamefont{Cucchieri, Engels,
  Holtmann, Mendes, and Schulze}}]{Cucchieri:2002hu}
\bibinfo{author}{\bibfnamefont{A.}~\bibnamefont{Cucchieri}},
  \bibinfo{author}{\bibfnamefont{J.}~\bibnamefont{Engels}},
  \bibinfo{author}{\bibfnamefont{S.}~\bibnamefont{Holtmann}},
  \bibinfo{author}{\bibfnamefont{T.}~\bibnamefont{Mendes}}, \bibnamefont{and}
  \bibinfo{author}{\bibfnamefont{T.}~\bibnamefont{Schulze}},
  \bibinfo{journal}{J. Phys.} \textbf{\bibinfo{volume}{A35}},
  \bibinfo{pages}{6517} (\bibinfo{year}{2002}), \eprint{cond-mat/0202017}.

\bibitem[{\citenamefont{Wallace and Zia}(1975)}]{Wallace:1975vi}
\bibinfo{author}{\bibfnamefont{D.}~\bibnamefont{Wallace}} \bibnamefont{and}
  \bibinfo{author}{\bibfnamefont{R.}~\bibnamefont{Zia}},
  \bibinfo{journal}{Phys. Rev.} \textbf{\bibinfo{volume}{B12}},
  \bibinfo{pages}{5340} (\bibinfo{year}{1975}).

\bibitem[{\citenamefont{Hasenfratz and Leutwyler}(1990)}]{Hasenfratz:1989pk}
\bibinfo{author}{\bibfnamefont{P.}~\bibnamefont{Hasenfratz}} \bibnamefont{and}
  \bibinfo{author}{\bibfnamefont{H.}~\bibnamefont{Leutwyler}},
  \bibinfo{journal}{Nucl. Phys.} \textbf{\bibinfo{volume}{B343}},
  \bibinfo{pages}{241} (\bibinfo{year}{1990}).

\bibitem[{\citenamefont{Smilga and Stern}(1993)}]{Smilga:1993in}
\bibinfo{author}{\bibfnamefont{A.~V.} \bibnamefont{Smilga}} \bibnamefont{and}
  \bibinfo{author}{\bibfnamefont{J.}~\bibnamefont{Stern}},
  \bibinfo{journal}{Phys. Lett.} \textbf{\bibinfo{volume}{B318}},
  \bibinfo{pages}{531} (\bibinfo{year}{1993}).

\bibitem[{\citenamefont{Smilga and Verbaarschot}(1996)}]{Smilga:1995qf}
\bibinfo{author}{\bibfnamefont{A.~V.} \bibnamefont{Smilga}} \bibnamefont{and}
  \bibinfo{author}{\bibfnamefont{J.}~\bibnamefont{Verbaarschot}},
  \bibinfo{journal}{Phys. Rev.} \textbf{\bibinfo{volume}{D54}},
  \bibinfo{pages}{1087} (\bibinfo{year}{1996}), \eprint{hep-ph/9511471}.

\bibitem[{\citenamefont{Cheng et~al.}(2011)\citenamefont{Cheng, Datta, Francis,
  van~der Heide, Jung et~al.}}]{Cheng:2010fe}
\bibinfo{author}{\bibfnamefont{M.}~\bibnamefont{Cheng}},
  \bibinfo{author}{\bibfnamefont{S.}~\bibnamefont{Datta}},
  \bibinfo{author}{\bibfnamefont{A.}~\bibnamefont{Francis}},
  \bibinfo{author}{\bibfnamefont{J.}~\bibnamefont{van~der Heide}},
  \bibinfo{author}{\bibfnamefont{C.}~\bibnamefont{Jung}}, \bibnamefont{et~al.},
  \bibinfo{journal}{Eur.Phys.J.} \textbf{\bibinfo{volume}{C71}},
  \bibinfo{pages}{1564} (\bibinfo{year}{2011}), \eprint{1010.1216}.

\bibitem[{rbc()}]{rbcbi_future}
\bibinfo{note}{{RBC-Bielefeld Collaboration} work in progress}.

\bibitem[{\citenamefont{Kaczmarek et~al.}(2002)\citenamefont{Kaczmarek, Karsch,
  Petreczky, and Zantow}}]{Kaczmarek:2002mc}
\bibinfo{author}{\bibfnamefont{O.}~\bibnamefont{Kaczmarek}},
  \bibinfo{author}{\bibfnamefont{F.}~\bibnamefont{Karsch}},
  \bibinfo{author}{\bibfnamefont{P.}~\bibnamefont{Petreczky}},
  \bibnamefont{and} \bibinfo{author}{\bibfnamefont{F.}~\bibnamefont{Zantow}},
  \bibinfo{journal}{Phys. Lett.} \textbf{\bibinfo{volume}{B543}},
  \bibinfo{pages}{41} (\bibinfo{year}{2002}), \eprint{hep-lat/0207002}.

\bibitem[{\citenamefont{Digal et~al.}(2003)\citenamefont{Digal, Fortunato, and
  Petreczky}}]{Digal:2003jc}
\bibinfo{author}{\bibfnamefont{S.}~\bibnamefont{Digal}},
  \bibinfo{author}{\bibfnamefont{S.}~\bibnamefont{Fortunato}},
  \bibnamefont{and}
  \bibinfo{author}{\bibfnamefont{P.}~\bibnamefont{Petreczky}},
  \bibinfo{journal}{Phys. Rev.} \textbf{\bibinfo{volume}{D68}},
  \bibinfo{pages}{034008} (\bibinfo{year}{2003}), \eprint{hep-lat/0304017}.

\bibitem[{\citenamefont{Kaczmarek and Zantow}(2005)}]{Kaczmarek:2005ui}
\bibinfo{author}{\bibfnamefont{O.}~\bibnamefont{Kaczmarek}} \bibnamefont{and}
  \bibinfo{author}{\bibfnamefont{F.}~\bibnamefont{Zantow}},
  \bibinfo{journal}{Phys. Rev.} \textbf{\bibinfo{volume}{D71}},
  \bibinfo{pages}{114510} (\bibinfo{year}{2005}), \eprint{hep-lat/0503017}.

\bibitem[{\citenamefont{Petreczky and Petrov}(2004)}]{Petreczky:2004pz}
\bibinfo{author}{\bibfnamefont{P.}~\bibnamefont{Petreczky}} \bibnamefont{and}
  \bibinfo{author}{\bibfnamefont{K.}~\bibnamefont{Petrov}},
  \bibinfo{journal}{Phys. Rev.} \textbf{\bibinfo{volume}{D70}},
  \bibinfo{pages}{054503} (\bibinfo{year}{2004}), \eprint{hep-lat/0405009}.

\bibitem[{\citenamefont{Bernard
  et~al.}(2007{\natexlab{b}})\citenamefont{Bernard, DeTar, Levkova, Gottlieb,
  Heller et~al.}}]{Bernard:2007ps}
\bibinfo{author}{\bibfnamefont{C.}~\bibnamefont{Bernard}},
  \bibinfo{author}{\bibfnamefont{C.~E.} \bibnamefont{DeTar}},
  \bibinfo{author}{\bibfnamefont{L.}~\bibnamefont{Levkova}},
  \bibinfo{author}{\bibfnamefont{S.}~\bibnamefont{Gottlieb}},
  \bibinfo{author}{\bibfnamefont{U.}~\bibnamefont{Heller}},
  \bibnamefont{et~al.}, \bibinfo{journal}{PoS}
  \textbf{\bibinfo{volume}{LAT2007}}, \bibinfo{pages}{090}
  (\bibinfo{year}{2007}{\natexlab{b}}), \eprint{0710.1118}.

\bibitem[{\citenamefont{Bazavov et~al.}(2010{\natexlab{d}})}]{Bazavov:2010xr}
\bibinfo{author}{\bibfnamefont{A.}~\bibnamefont{Bazavov}} \bibnamefont{et~al.}
  (\bibinfo{collaboration}{MILC collaboration}), \bibinfo{journal}{Phys. Rev.}
  \textbf{\bibinfo{volume}{D81}}, \bibinfo{pages}{114501}
  (\bibinfo{year}{2010}{\natexlab{d}}), \eprint{1003.5695}.

\end{thebibliography}

\end{document}